\pdfoutput=1
\def\typeset{}
\ifdefined\typeset
  \documentclass[final,5p,times,twocolumn]{elsarticle}
\else
  \documentclass[11pt, oneside]{elsarticle}
\fi

\usepackage{geometry}
\usepackage{graphicx}

\usepackage{amssymb}
\usepackage{amsmath}
\usepackage{amsthm}
\usepackage{bm}
\usepackage{mathtools}

\usepackage{algorithm2e}
\usepackage{subcaption}
\usepackage{booktabs}
\usepackage{adjustbox}  %
\usepackage[detect-weight=true,detect-family=true]{siunitx}
\DeclareSIUnit{\g}{g}   %
\usepackage{rotating}   %

\usepackage{tikz}
\usetikzlibrary{shapes.geometric, arrows.meta, positioning, calc, fit}

\newtheorem{remark}{Remark}

\newcommand{\T}{^{\mathsf{T}}}
\newcommand{\Rmax}{R_{\max}}
\newcommand{\Rtot}{R_{\mathrm{total}}}

\newcommand{\Dai}{D_{a,i}}

\newcommand{\SDi}{S_{D,i}}
\newcommand{\K}{\mathbf{K}}
\newcommand{\M}{\mathbf{M}}
\newcommand{\bz}{\mathbf{z}}
\newcommand{\bu}{\mathbf{u}}
\newcommand{\bphi}{\boldsymbol{\phi}}

\newcommand{\bmu}{\boldsymbol{\mu}}

\usepackage{xcolor}

\usepackage{fancyhdr}
\fancypagestyle{plain}{%
  \fancyhf{}%
  \fancyfoot[C]{\footnotesize\thepage}%
}
\fancypagestyle{pprintTitle}{%
  \fancyhf{}%
}
\ifdefined\typeset

  \usepackage{flushend}
  \newenvironment{widetable}[1][htbp]{\begin{table*}[!t]}{\end{table*}}
  \newenvironment{notationtable}{\begin{table*}[!t]}{\end{table*}}
  \newenvironment{widefigure}[1][htbp]{\begin{figure*}[!t]}{\end{figure*}}
\else
  \newenvironment{widetable}[1][htbp]{\begin{table}[#1]}{\end{table}}
  \newenvironment{notationtable}{\begin{table}[h!]}{\end{table}}
  \newenvironment{widefigure}[1][htbp]{\begin{figure}[#1]}{\end{figure}}
\fi

\begin{document}

\begin{frontmatter}

\title{Topology optimization of shipboard components for Navy shock
certification via the Dynamic Design Analysis Method}

\author[a1]{Miguel A. Aguil\'o\corref{cor1}}
\ead{maguilo@morphorm.com}
\cortext[cor1]{Corresponding author}
\address[a1]{Morphorm LLC, 100 Sun Avenue Northeast Suite 650,
Albuquerque, NM 87109-4670, USA}

\begin{abstract}
Shipboard equipment must remain operational after sustaining the
high-amplitude transient loads produced by underwater explosions (UNDEX). The
U.S.\ Navy mandates the Dynamic Design Analysis Method (DDAM) as the
analytical procedure used to certify that critical components meet shock
survivability requirements. DDAM transforms a complex time-dependent shock
event into a static-equivalent modal response analysis driven by Navy-supplied
shock design spectra. DDAM is well established as a \emph{verification} tool, but it is rarely
exploited as a \emph{design driver} inside large-scale topology optimization algorithm. 
The obstacle is that the design spectrum depends on the modal mass, which introduces 
a non-standard, non-smooth coupling that classical shape-derivative formulations do not address.

This paper presents a complete differentiable formulation of the DDAM
criterion suitable for use as an objective or constraint inside gradient-based
topology optimization. Eigenvalue sensitivities are obtained by the
Fox--Kapoor method and eigenvector sensitivities by Nelson's method. 
The chain rule is derived for a spectral displacement that depends jointly on
the natural frequency and the modal participation factor. The sensitivity
carries both dependences for the displacement criterion and the frequency
dependence for the stress criterion. The resulting sensitivity
expression is assembled by automatic differentiation. The engineer can then
certify components against Navy shock requirements inside the optimization
loop, rather than as a post-design check.

We demonstrate the workflow on three certification problems on one
representative shipboard equipment frame, for which Navy shock design spectra
are tabulated. On each the optimizer removes between \SI{43.2}{\percent} and
\SI{46.1}{\percent} of the design-domain material while holding the certifying
response at the Navy allowable. The shock criterion attains that allowable to
within \SI{0.41}{\percent} in every case. We report the optimized topology,
the modal participation and effective mass that produce the response together 
with the certifying NRL Sum.

Two companion studies isolate questions the demonstrations raise. The first
minimizes compliance at nearly the same mass rather than the shock response. It
produces a frame whose fundamental falls to \SI{5.2}{Hz}, outside the range the
certification can qualify at all. In addition, its nominal response exceeds the allowable 
by a factor of \num{2.75}. That is what certifying after the fact costs. The second 
addresses a degeneracy the demonstrations expose. Driving the two lowest modes 
onto the frequency floor leaves them numerically repeated. The NRL Sum of such a 
pair then depends on a basis the eigensolver chooses arbitrarily, varying by up to a 
factor of $\sqrt{2}$. Adding a lower bound on the mean of the two lowest eigenvalues, 
which does not depend on that choice, separates the pair on every iterate. The response 
then stops oscillating across the acceptance boundary and ends at most \SI{1.2}{MPa}, 
about \SI{0.5}{\percent}, above it, at a cost of under two percentage points of volume
fraction.
\end{abstract}


\end{frontmatter}

\section{Introduction}
\label{sec:intro}

Surface combatants and submarines must survive the kinetic energy delivered by
underwater explosions (UNDEX) without loss of mission-critical functionality.
Direct numerical simulation of every shock event is computationally
prohibitive at the equipment-by-equipment level. The U.S.\ Navy therefore
codifies a static-equivalent procedure, the Dynamic Design Analysis Method
(DDAM), as the mandated framework for shock certification of shipboard
equipment~\cite{nrl5545,navsea_0908,navsea_t9070,remmers1983ddam,ohara1996ddam}.
DDAM replaces the explicit time history of the UNDEX with modal participation
factors, effective modal masses~\cite{oharaCunniff1963}, and Navy-supplied
shock design values~\cite{nrl1396}. It then combines the modal responses
through the Naval Research Laboratory (NRL) Sum to estimate worst-case stress
and displacement.

DDAM has long been used as an \emph{a posteriori} verification tool. An
engineer designs a candidate component using static, deflection, or fatigue
criteria, evaluates the DDAM response, and iterates manually if margins are
insufficient. This decoupling is inefficient. Components that fail DDAM
after the fact must be redesigned, often after costly mesh and
geometry rework. The alternative, using DDAM as a \emph{design driver} inside
a gradient-based topology optimizer~\cite{bendsoe1989,bendsoeSigmund2003,sigmund2013survey}, 
is mathematically appealing but historically rare in practice for three reasons. 
First, the design acceleration $\Dai = G(\bar m_i, \omega_i)$ is a tabulated, 
piecewise function of the effective modal mass and natural frequency. Its 
derivatives with respect to the design variables are not part of standard
eigenvalue-sensitivity packages. Second, eigenvalue and eigenvector
sensitivities must be coupled through a non-trivial NRL-Sum chain rule that
is sensitive to mode crossings and mass-normalization conventions. Third,
the resulting sensitivity expressions are tedious to derive and error-prone
to implement by hand, which discourages adoption.

The first obstacle is handled by differentiating the piecewise spectrum
branch by branch (Section~\ref{sec:limitations}, Tabulated shock spectra).
The second is addressed by classical adjoint
analysis~\cite{tortorelli1994dsa}: the eigenvalue derivatives are recovered by
the Fox--Kapoor method~\cite{foxKapoor1968}, and the eigenvector derivatives
by Nelson's method~\cite{nelson1976}. Both avoid the explicit construction of
the (singular) shifted dynamic-stiffness inverse. The third obstacle,
implementation effort and correctness, is addressed by automatic
differentiation~\cite{griewank2008autodiff,baydin2018jmlr,he2022rad_eig},
which assembles the chain rule mechanically from the user-supplied forward
computation.

\subsection{Related work}
\label{subsec:related}

The literature relevant to this paper falls into four categories. We summarize
each and identify the gap that the present work closes.

\paragraph{DDAM as analysis and certification practice}
DDAM was developed by the Naval Research Laboratory in the 1960s. It has been
the U.S.\ Navy's mandated procedure for the shock qualification of shipboard
equipment ever since~\cite{nrl5545,nrl1396,remmers1983ddam,ohara1996ddam,navsea_0908}.
Implementations are available in essentially every production finite element
code used for naval applications; DDAM analyses in
LS-DYNA~\cite{koehler2018ddam}, Abaqus~\cite{simulia2007ddam} and
OptiStruct~\cite{jengDeLimaPagaldipti} are documented in the engineering
literature. Sigrist and
Broc~\cite{sigristBroc2023} more recently proposed transfer-function-based
methods that accelerate response calculations for hull-mounted equipment
without resorting to direct coupled simulation. Across this body of work, DDAM is treated exclusively as a forward analysis
tool. A given design is evaluated against the criterion, and design changes
are made manually if the response margin is inadequate. The criterion is not used as the
gradient target of an automated optimizer.

\paragraph{Eigenfrequency-constrained optimization}
Density-based topology optimization with eigenvalue constraints has been
studied for over two decades. Pedersen~\cite{pedersen2000eigenvalues} and
Tcherniak~\cite{tcherniak2002resonating} developed SIMP-based formulations for
eigenvalue and resonating-structure problems. The subsequent literature has matured
into a well-developed toolkit covering simple and repeated
eigenfrequencies with adjoint-derived
sensitivities~\cite{seyranian_lund_olhoff_1994,duOlhoff2007}. A recent review by Lin, Mottershead, and
Ng~\cite{linMottersheadNg2020} surveys the theory and engineering
applications of eigenvalue and eigenvector derivatives. He, Jonsson, and Martins~\cite{he2022rad_eig} provide reverse-mode automatic
differentiation formulas for both eigenvalues and eigenvectors of
self-adjoint generalized eigenproblems. Li and
Kennedy~\cite{kennedy2023eigvec_aggregate} introduce a smoothly
differentiable eigenvector aggregate that sidesteps the difficulties posed by
repeated eigenvalues.
A related question in this literature concerns what the lowest eigenvalue does
to a design when it is not itself the goal. Modes localized in low-density regions are the classical pathology of
eigenfrequency-related density and level-set methods. They arise when the
stiffness interpolation is penalized more strongly than the mass
interpolation, and their suppression is a subject in its own
right~\cite{pedersen2000eigenvalues,duOlhoff2007,liShiXia2017}. The difficulty is sharper when a design-independent concentrated mass sits
on material the optimizer may remove: the mass stays while its support
vanishes.
Minimum-mass design under frequency constraints is long-standing
practice~\cite{grandhiVenkayya1988}. Connectivity of the load path has lately been treated as an explicit
requirement, both in eigenfrequency problems whose objective can be improved
by severing material~\cite{bonaccorsi2025connectivity} and in
general~\cite{coolAageSigmund2025}. We enforce it with a lower bound on the
fundamental frequency, in the volume-minimizing form of the design problem
(Section~\ref{subsec:topo}). What these eigenfrequency studies share is that
the objective or constraint depends on eigenvalues and eigenvectors
through analytically tractable, smooth functions. DDAM does not. It inserts a \emph{tabulated, piecewise-defined,
Navy-prescribed} mapping $\Dai = G(\bar m_i, \omega_i)$ between the modal
quantities and the response. The design spectrum further depends on the
effective modal mass~\eqref{eq:effective_mass}, which itself depends on the
participation factor. The classical eigenvalue-sensitivity toolkit applies, but the
chain rule must be extended to include the Navy-supplied spectrum.

\paragraph{Topology optimization for transient and dynamic loading}
A separate body of work addresses topology optimization for time-dependent loads
without recourse to a response-spectrum reduction. Jog~\cite{jog2002periodic} treats periodic loading. Behrou and
Guest~\cite{behrouGuest2017} present a transient-response topology
optimization formulation, using the method of moving
asymptotes~\cite{Svanberg_1987} with
time-dependent adjoint sensitivities and SIMP material interpolation.
Work at Sandia on shock and vibration test
fixtures~\cite{sandiaShockFixture2020} applies topology optimization to design
laboratory fixtures whose dynamic impedance matches the next level of
assembly, so that a test reproduces the field environment. Tortorelli and Michaleris~\cite{tortorelli1994dsa}
provide the underlying design-sensitivity-analysis foundation.
These approaches optimize the structure against the
\emph{time history} of the load and incur the full cost of transient
adjoint integration. DDAM, by contrast, was specifically engineered to
\emph{eliminate} that cost by replacing the transient history with a
spectrum that already encodes the worst-case envelope~\cite{nrl5545,ohara1996ddam}. A DDAM-aware optimizer therefore offers a quite different computational
trade. It takes a single eigenproblem and modal combination per design
iteration, in exchange for the loss of explicit transient information.

\paragraph{DDAM-coupled optimization}
Direct prior work on DDAM-coupled \emph{topology} optimization is
limited. The closest precedent is the Altair case study by Jeng, de~Lima, and
Pagaldipti~\cite{jengDeLimaPagaldipti}. It presents a DDAM-coupled topology,
shape, and size optimization of a wash-basin ship foundation in OptiStruct.
The authors report a \SI{55}{\percent} reduction in the dominant directional displacement
response while approximately preserving the foundation mass. The study shows
that DDAM-aware optimization is practically tractable, and it produces meaningful
design improvements over manual iteration.

Their workflow, however, treats the DDAM evaluation as a \emph{black-box
constraint} supplied by the commercial code. The gradients with respect to the topology design variables are assembled by
the commercial solver internally. The dependence of the design spectrum on the
modal mass is not exposed in the formulation. The mathematical sensitivity
expressions are neither derived nor verified. The approach is therefore not easily portable. It does not transfer to new
structural models, to alternative spectra, or to research codes that wish to
combine DDAM with other criteria.

To the best of our knowledge, no published work derives in closed form the
adjoint sensitivities of the DDAM criterion, either in its smooth
Square-Root-of-Sum-of-Squares (SRSS) form or as the exact NRL Sum the
certification uses, for a displacement response or pointwise over a modal
stress field. None accounts for the joint dependence of $\Dai$ on $\omega_i$
and $\Gamma_i$ in the chain rule. None assembles the resulting gradients by
automatic differentiation within a production analysis code.

\paragraph{Gap and contribution}
The work reviewed above leaves a clear gap. What is missing is a mathematically
explicit, gradient-based, certification-aware topology optimization workflow
that uses the Navy-mandated DDAM criterion directly as the design objective or
constraint. The present paper closes this gap with the following
contributions:
\begin{enumerate}
  \item A self-contained derivation of four quantities, written in a form
        amenable to gradient-based optimization: the modal participation
        factor, the effective modal mass, the spectral displacement, and the
        NRL Sum.
  \item Adjoint sensitivities of the DDAM criterion with respect to the
        topology design variables, in its smooth SRSS form and as the exact
        NRL Sum, for a displacement response and pointwise over the von~Mises
        modal stress field (Sections~\ref{subsec:srss_sens}--\ref{subsec:nrl_stress}).
        They are derived for the joint dependence of the design acceleration on
        $\omega_i$ \emph{and} the modal participation factor $\Gamma_i$. The
        dependence on $\omega_i$ is implemented and verified. The dependence
        on $\Gamma_i$ is implemented and verified for the displacement
        criterion and derived only for the stress criterion
        (Section~\ref{subsec:mass_loading} (Mass loading)).
  \item An assembly of these sensitivities by automatic differentiation
        within Morphorm Analyze, removing
        the need to hand-code derivative expressions for new structural
        models.
  \item Numerical demonstration on three shipboard component problems for
        which Navy surface-ship shock design spectra are tabulated. Each adds
        one element to the one before. All three use one equipment frame. The
        first is certified on the fore-and-aft NRL Sum of interface
        displacement. The second is certified on the fore-and-aft NRL Sum of
        von~Mises stress, which changes the acceptance measure. The third is
        certified against a \emph{combined-direction} NRL Sum stress, which
        bounds all three orthogonal shock directions, posed as one
        constraint per direction.
\end{enumerate}

The remainder of the paper is organized as follows.
Section~\ref{sec:background} reviews DDAM, the SRSS criterion, the
density-based topology optimization formulation, and the role of automatic
differentiation in sensitivity assembly. Section~\ref{sec:sensitivity} derives
the eigenvalue and eigenvector sensitivities, and combines them into the DDAM
sensitivity expression. Section~\ref{sec:algorithm} embeds the criterion
inside the Morphorm Studio topology-optimization pipeline, in which Morphorm
Studio drives the optimization loop and the solver, Morphorm Analyze,
evaluates the criterion and its gradient.
Section~\ref{sec:results} reports the numerical demonstrations and two
companion studies. The first asks what a conventional criterion produces at
the same mass. The second asks what it costs to hold the two lowest modes
apart, by constraining their mean eigenvalue.
Section~\ref{sec:limitations} states what the formulation does not cover, and
Section~\ref{sec:conclusions} concludes. An appendix summarizes notation.

\section{Background}
\label{sec:background}

\subsection{The Dynamic Design Analysis Method}
\label{subsec:ddam}

DDAM is a shock spectrum analysis tool that translates a time-dependent UNDEX
event into a static-equivalent design process. Rather than simulating a full
explosion in real time, DDAM uses a modal approach~\cite{navsea_0908,remmers1983ddam}:

\begin{itemize}
  \item A finite element model of the component is built and its natural
        frequencies $\omega_i$ and mode shapes $\bphi_i$ are extracted.
  \item For each mode, a Navy-supplied shock design value
        $\Dai = G(\bar m_i, \omega_i)$ is read from tabulated spectra~\cite{navsea_0908}. It depends on the ship type, the equipment's
        mounting location, the shock direction, vertical (V), athwartship (A)
        or fore-and-aft (FA), the effective modal mass and the natural
        frequency.
  \item The individual modal responses are combined statistically using
        the NRL Sum to produce a worst-case design response.
\end{itemize}

DDAM accounts for the interaction between the equipment and the ship's
structure~\cite{nrl5545,ohara1996ddam}. Unlike a standard response spectrum analysis, it recognizes that
the mass of the equipment influences the motion of the vessel itself. That
influence enters through the dependence of $\Dai$ on the effective modal mass
$\bar m_i$, in~\eqref{eq:Dai} below. The map $G$ is piecewise in frequency,
with a velocity-limited and an acceleration-limited branch
(Section~\ref{sec:results}), and every sensitivity below inherits that
structure.

\subsection{NRL Sum}
\label{subsec:nrl}

The NRL Sum identifies the mode that produces the largest response and treats
it as the primary driver of the structural load. The remaining modal responses
are combined statistically~\cite{remmers1983ddam}. The total design
response $\Rtot$ is
\begin{equation}
  \label{eq:nrl_sum}
  \Rtot \;=\; \lvert \Rmax \rvert \;+\; \sqrt{\,\sum_{i \neq \max} R_i^{\,2}\,},
\end{equation}
where $\Rmax$ is the modal response of largest magnitude and $R_i$ are the
peak responses of the remaining modes. For a response that is a field, the sum
is formed point by point, for each checked component.
The NRL Sum is the mandated standard per NAVSEA 0908-LP-000-3010~\cite{navsea_0908}, the primary document governing DDAM procedures
(current public-release designation T9070-AJ-DPC-120/3010~\cite{navsea_t9070}). It
adds the dominant mode in full and the background modes as if uncorrelated,
by their root-sum-square, which makes it a conservative estimate of how the
background adds to the primary response.

\subsection{Modal response}
\label{subsec:modal_response}

Each individual modal response $R_i$ is computed before the NRL Sum is
applied. It follows from the interaction between the ship's shock environment
and the equipment's dynamic properties, and in DDAM it is formed in four
steps.

\paragraph{Step 1: modal extraction and participation factor}
The natural frequency $\omega_i$ and the mode shape $\bphi_i$ are extracted
from the generalized eigenproblem $\K \bphi_i = \lambda_i \M \bphi_i$ with
$\lambda_i = \omega_i^2$. From these the
modal participation factor is~\cite{oharaCunniff1963,chopra2020dynamics}
\begin{equation}
  \label{eq:participation}
  \Gamma_i \;=\; \frac{\bphi_i\T \M \mathbf{D}}{\bphi_i\T \M \bphi_i},
\end{equation}
where $\mathbf{D}$ is an influence vector. It is a column of $1$s and $0$s
picking up the degrees of freedom aligned with the shock direction. The
generalized mass $m_i^{*}$ is the quadratic form $\M$ induces on whichever
eigenvector is in hand. If the eigenvectors are mass-normalized, then
$m_i^{*} = \bphi_i\T \M \bphi_i = 1$.

\begin{remark}\label{rem:massnorm}
We recommend mass-normalizing the eigenvectors. Mass normalization simplifies the derivation of the sensitivity of the DDAM
criterion with respect to the topology design variables. It eliminates the
denominator of~\eqref{eq:participation} from the chain rule, and removes one
application of the quotient rule from every entry of the gradient.
\end{remark}

The quantities that enter the response do not depend on how the eigenvector
is scaled. With mass normalization, $\Gamma_i = \bphi_i\T \M \mathbf{D}$ and
the effective modal mass below is $\bar m_i = \Gamma_i^{2}$; for an eigenvector
of any scale the same value follows as
$\bar m_i = (\bphi_i\T \M \mathbf{D})^{2} / (\bphi_i\T \M \bphi_i)$.

\paragraph{Step 2: effective modal mass}
The effective modal mass $\bar m_i$ represents the portion of the total
system mass that vibrates in mode $i$,
\begin{equation}
  \label{eq:effective_mass}
  \bar m_i \;=\; \Gamma_i^{\,2}\, m_i^{*}.
\end{equation}
Summed over all modes, the effective masses recover the total mass that
translates rigidly along $\mathbf{D}$. The fraction a retained set captures
measures how much of the response it covers.

\paragraph{Step 3: design acceleration}
The Navy provides empirical equations~\cite{nrl1396,navsea_0908} for the
design acceleration $\Dai$ of each mode. It is a function of the ship type,
mounting location, direction of shock, effective modal mass and natural
frequency. As $\bar m_i$ increases, the design acceleration
\begin{equation}
  \label{eq:Dai}
  \Dai \;=\; G(\bar m_i, \omega_i)
\end{equation}
typically decreases. This accounts for mass loading: a heavy piece of
equipment reduces the motion of the foundation.

\paragraph{Step 4: modal response}
For the peak modal displacement $X_i$,
\begin{equation}
  \label{eq:Ri_disp}
  R_i \;=\; X_i \;=\; \frac{\Gamma_i\, \bphi_i\, \Dai}{\omega_i^{2}}.
\end{equation}
For the peak modal stress, the modal response is the stress field derived from
the displacement field of mode $i$. It is scaled by the design acceleration
and the participation factor,
\begin{equation}
  \label{eq:Ri_stress}
  \boldsymbol{\sigma}_i \;=\;
    \frac{C B\, \bphi_i\, \Gamma_i\, \Dai}{\omega_i^{2}},
\end{equation}
evaluated element by element: in element $e$, $B$ holds the derivatives of
the element shape functions and acts on the element restriction of $\bphi_i$,
and $C$ is the material stiffness matrix.

\subsection{SRSS DDAM design criterion}
\label{subsec:srss}

For gradient-based optimization we form a differentiable design criterion
from the modal responses. The primary choice is a smooth
Square-Root-of-Sum-of-Squares (SRSS) criterion, after the classical modal
combination rule of response-spectrum analysis~\cite{chopra2020dynamics},
\begin{equation}
  \label{eq:srss}
  f(\Lambda, \Phi) \;=\;
    \sqrt{\,\sum_{i} \SDi^{\,2}\, \Gamma_i^{\,2}\, \bphi_i\T W \bphi_i\,},
\end{equation}
where $\Lambda$ collects the eigenvalues $\lambda_i = \omega_i^{2}$ and $\Phi$
the eigenvectors $\bphi_i$, $\SDi = \Dai / \omega_i^{2}$ is the spectral
displacement of mode $i$, and $W$ is a positive semi-definite diagonal
weighting matrix, defined below, that selects the checked response.

The criterion has a direct physical reading. The peak displacement of mode $i$
is $X_i = \SDi\, \Gamma_i\, \bphi_i$~\eqref{eq:Ri_disp}, so
$f^{2} = \sum_i X_i\T W X_i$: $f$ is the weighted root-mean-square of the
modal displacements combined by SRSS at each degree of freedom.

\eqref{eq:srss} is a smooth surrogate for the NRL Sum. At each point the NRL Sum of the
modal responses $R_i$ lies between their SRSS combination and $\sqrt{2}$ times
it,
\begin{equation}
  \label{eq:srss_nrl_bracket}
  \Bigl(\sum_i R_i^{2}\Bigr)^{1/2}
  \;\le\; \Rtot \;\le\;
  \sqrt{2}\,\Bigl(\sum_i R_i^{2}\Bigr)^{1/2},
\end{equation}
Write $a = \lvert \Rmax \rvert$ for the dominant response and
$b = \bigl(\sum_{i \neq \max} R_i^{2}\bigr)^{1/2}$ for the background, so that
$\Rtot = a + b$ and the SRSS value is $(a^{2} + b^{2})^{1/2}$. The lower bound
holds because $a + b \ge (a^{2} + b^{2})^{1/2}$ for $a, b \ge 0$, and the upper
bound because $(a + b)^{2} \le 2\,(a^{2} + b^{2})$. The NRL Sum equals the SRSS
value when one mode carries the whole response ($b = 0$), and reaches
$\sqrt{2}$ times it when $a = b$, that is, when the largest modal response
equals the root-sum-square of all the others.

Unlike the NRL Sum, $f$ is differentiable in the design variables
wherever $f > 0$, except at mode crossings and at the branch switches of $G$
(Section~\ref{sec:limitations}). Because it under-predicts the NRL Sum,
Section~\ref{subsec:nrl_exact} also develops a criterion that differentiates
the exact NRL Sum~\eqref{eq:nrl_sum} directly.

\paragraph{The weighting operator}
The operator $W$ carries the definition of the response quantity. For a
response checked at a single point, $W$ extracts the squared modal response
at the checked degree of freedom. For a response checked over an interface
$\Sigma$, such as an equipment mounting pad, $W$ is the area-average operator
on the checked displacement component,
\begin{equation}
  \label{eq:surface_weights}
  W \;=\; \frac{1}{A}\,\mathrm{diag}(a_j),
  \quad
  a_j \;=\; \int_{\Sigma} N_j \, d\Sigma,
  \quad
  A \;=\; \sum_{j} a_j \;=\; |\Sigma|,
\end{equation}
where $N_j$ is the finite element shape function of interface node $j$. With
these weights,
$\bphi_i\T W \bphi_i = A^{-1} \sum_j a_j\, [\bphi_i]_j^{2}$ is the
area-weighted mean square of the modal response over the interface, and its
continuum limit is the mean square $|\Sigma|^{-1}\int_\Sigma \phi_i^2\,
d\Sigma$.

Because it has a continuum limit, the criterion converges under mesh
refinement and can carry an absolute allowable, consistent with the
area-averaged acceptance quantities of the certification
standard~\cite{navsea_0908}. A raw nodal sum over the interface has no such
limit: refining the mesh adds terms for the same physical field, so its value
grows with the node count.

The weights are exact and simple to form. Because the shape functions are a
partition of unity, $A$ recovers the exact interface measure at any element
order. Each weight $a_j$ is a row sum of the consistent interface Gram matrix
$\int_\Sigma N_j N_k \, d\Sigma$, so $W$ is the row-sum lumped form of that
matrix: diagonal by construction, with the same continuum limit. The weights
depend only on the nodal coordinates, never on the design, so $W$ is a
constant operator in the design sensitivities below.

Under the rigid equipment attachment of Section~\ref{subsec:topo} the response
is uniform over the interface. The area average then equals the pointwise
value, and the two forms of $W$ coincide.

\subsection{Density-based topology optimization}
\label{subsec:topo}

We adopt the standard density-based (SIMP) topology optimization formulation~\cite{bendsoe1989,bendsoeSigmund2003,sigmund2013survey}. A pseudo-density
$z_e \in [z_{\min}, 1]$ on each element $e$ controls the local material
stiffness and mass. The design variables are nodal values, and every element-wise expression below is assembled to them through
the element shape functions; the sensitivities keep the same form. Letting $\bz \in \mathbb{R}^{N_e}$
denote the design vector, the optimization problem is
\begin{equation}
  \label{eq:topo_problem}
  \underset{\bz}{\mbox{minimize}}\;\;
  f(\Lambda(\bz), \Phi(\bz))
  \quad
  \mbox{s.t.}\quad
  v(\bz) \le v_f\, v_0,
  \quad
  z_{\min} \le z_e \le 1,
\end{equation}
where $v(\bz)$ is the design volume, $v_0$ the volume of the fully solid
domain, and $v_f$ the target volume fraction. Stiffness and structural mass are
both interpolated by the modified SIMP law,
\begin{equation}
  \label{eq:simp}
  \begin{gathered}
  \mathbf{C}(\mathbf{x}) \;=\; \mathcal{I}\bigl(\tilde z(\mathbf{x})\bigr)\, \mathbf{C}_0,
  \qquad
  \rho(\mathbf{x}) \;=\; \mathcal{I}\bigl(\tilde z(\mathbf{x})\bigr)\, \rho_0,\\
  \mathcal{I}(z) \;=\; \epsilon_{\min} + (1 - \epsilon_{\min})\, z^{p},
  \end{gathered}
\end{equation}
evaluated at each quadrature point. Here $\mathbf{C}_0$ and $\rho_0$ are the
solid stiffness and mass density, $p$ is the penalization exponent, and
$\epsilon_{\min} > 0$ is the minimum ersatz fraction, which keeps the stiffness
and mass of void material positive. $\K(\bz)$ and $\M_{\mathrm{s}}(\bz)$ are
assembled from~\eqref{eq:simp} in the usual way.

The densities are filtered before they enter~\eqref{eq:simp}. The results
reported here use a linear kernel filter~\cite{bourdin2001filter,brunsTortorelli2001},
$\tilde{\bz} = \mathbf{F}\bz$, without projection, and $\tilde z(\mathbf{x})$ is
interpolated from $\tilde{\bz}$ by the element shape functions. Every gradient
below is taken with respect to $\tilde{\bz}$ and mapped back to the design
variables by $\mathbf{F}\T$ before the update. The eigenvalues and eigenvectors are obtained by solving the
generalized eigenproblem at $\bz$,
\begin{equation}
  \label{eq:gep}
  \K(\bz)\, \bphi_i(\bz) \;=\; \lambda_i(\bz)\, \M(\bz)\, \bphi_i(\bz),
  \quad \bphi_i\T \M \bphi_i \;=\; 1.
\end{equation}
The DDAM criterion can equally serve as a constraint. One then minimizes volume,
compliance, or a manufacturing-cost surrogate, subject to
$f(\Lambda(\bz), \Phi(\bz)) \le R_{\mathrm{allow}}$, where
$R_{\mathrm{allow}}$ is the Navy-mandated allowable response for the component
class~\cite{navsea_t9070}.

\paragraph{Non-structural equipment mass}
The mass operator carries two contributions,
$\M(\bz) = \M_{\mathrm{s}}(\bz) + \M_{\mathrm{ns}}$. The structural part
$\M_{\mathrm{s}}(\bz)$ is the SIMP-interpolated consistent mass of~\eqref{eq:simp} and is
the only term the AD kernels of Section~\ref{subsec:autodiff} differentiate.
The non-structural part $\M_{\mathrm{ns}}$ models the supported equipment as a
concentrated translational lumped mass. It is
attached at the nodes of the mounting-interface node set $\mathcal{S}$,
\begin{equation}
  \label{eq:nonstructural_mass}
  \M_{\mathrm{ns}} \;=\; \sum_{p \in \mathcal{S}} m_p\, \mathbf{E}_p,
  \qquad
  \frac{\partial \M_{\mathrm{ns}}}{\partial \bz} \;=\; \mathbf{0},
\end{equation}
where $\mathbf{E}_p = \sum_{k} \mathbf{e}_{p,k}\, \mathbf{e}_{p,k}\T$ selects
the translational degrees of freedom of node $p$, with $\mathbf{e}_{p,k}$ the
unit vector of translational degree of freedom $k$ of that node, so that
$m_p \mathbf{E}_p$ adds $m_p$ to each of them on the diagonal of $\M$. The
prescribed equipment mass $m_{\mathrm{eq}}$ is divided evenly over the
$|\mathcal{S}|$ interface nodes, $m_p = m_{\mathrm{eq}} / |\mathcal{S}|$. $\M_{\mathrm{ns}}$ is a fixed property of the
equipment rather than of the material layout, so it is design-independent. It
shifts the eigenpairs of~\eqref{eq:gep}, and hence the effective modal
mass~\eqref{eq:effective_mass}, but contributes nothing to
$\partial \M / \partial \bz$. The Fox--Kapoor and Nelson sensitivities of
Sections~\ref{subsec:fox_kapoor}--\ref{subsec:nelson} are therefore unchanged.

A bolted, qualified component behaves as a rigid body relative to the flexible
foundation, so the mounting interface should not deform under it. We enforce this with a rigid attachment. The translational degrees of freedom of the interface nodes are tied to move
together. This is the translational restriction of a rigid multipoint constraint
imposed by transformation~\cite{cookMalkusPleshaWitt2002} (an RBE2 element in
NASTRAN terms), giving $\bu_p = \bu_{p_0}$ for every
$p \in \mathcal{S}$ and a single master node $p_0$. The constraint is imposed exactly, by the congruence reduction
$\K_r = \mathbf{T}\T \K\, \mathbf{T}$, $\M_r = \mathbf{T}\T \M\, \mathbf{T}$
of~\eqref{eq:gep}. Here $\mathbf{T}$ maps the reduced (tied) coordinates to
the full set, and the reduced eigenvectors are expanded as
$\bphi = \mathbf{T}\, \mathbf{q}$.

Without the tie, a mass spread over the multi-node interface lets those nodes
oscillate relative to one another. That introduces spurious relative-motion
modes, which would corrupt the effective modal mass and therefore the design
acceleration. The same congruence folds the interface masses onto the master,
so the equipment acts as a single rigid translational mass carrying the full
$\sum_{p} m_p$. Only translational inertia is represented: rotary inertia of
the equipment is not, because the discretization carries displacements only.

\paragraph{Frequency floor}
The volume-minimizing form needs one more constraint to be well posed. Its
objective rewards removing material regardless of the stiffness that remains,
while the equipment mass~\eqref{eq:nonstructural_mass} keeps its full value
however little material supports it. If the optimizer drives the density under
the mounting interface toward $z_{\min}$, the mass is left essentially
unsupported, and the eigenproblem~\eqref{eq:gep} returns a mode localized on it
with an eigenvalue near zero~\cite{pedersen2000eigenvalues,liShiXia2017}. A
component that has lost its structural connection to the foundation lies
outside the assumptions of the criterion, and no response computed for it
qualifies it. We therefore bound the fundamental frequency from below,
\begin{equation}
  \label{eq:frequency_floor}
  \lambda_1(\bz) \;\ge\; \omega_{\min}^{2}.
\end{equation}

Because $\lambda_i = \omega_i^{2}$, the floor $\omega_{\min}$ is an angular
frequency, in \si{rad/s}; a floor of $f$ in \si{Hz} is
$\omega_{\min} = 2\pi f$. The bound is placed on the lowest eigenvalue, so it
needs no mode tracking
between design iterations, and its gradient is the Fox--Kapoor expression of
Section~\ref{subsec:fox_kapoor}. It is not differentiable where the lowest
eigenvalue is repeated; Section~\ref{subsec:crossings} (Mode crossings) and the
companion study of Section~\ref{subsec:demo3_gap} treat that case.

The floor is a modeling constraint, not a certification requirement: DDAM
prescribes no lower bound on the fundamental frequency~\cite{navsea_t9070}. We
set $\omega_{\min}$ below the fundamental frequency of the fully solid design,
so that it rules out disconnected designs. Every optimum on the frame studied here
still sits on the floor (Section~\ref{sec:results}).

\subsection{Automatic differentiation}
\label{subsec:autodiff}

Automatic differentiation (AD)~\cite{griewank2008autodiff,baydin2018jmlr}
mechanizes the chain rule. Given a forward computation in a supported programming language, AD produces a
derivative computation of the same complexity class. The user hand-codes no
partial derivatives. Two
complementary mechanisms are combined here.

At the \emph{element-kernel} layer, the residual, stiffness, and mass kernels
are templated on independent scalar types for the control, state, and
configuration variables. Instantiating each template with a forward-mode AD
scalar yields the local Jacobians
$\partial \K_e / \partial z_e$, $\partial \M_e / \partial z_e$, and the
analogous state and configuration partials, without hand-coded derivative
expressions. Forward mode is well suited to these local kernels because the
number of element-level inputs is small.

At the \emph{physics} layer, the solver assembles the total criterion gradient
$\partial f / \partial \bz$ through an adjoint formulation specific to the
governing equations. For the modal problem of this paper it is the Fox--Kapoor expression~\eqref{eq:fk_lambda_prime} and the Nelson adjoint~\eqref{eq:nelson_adjoint},
both derived in Section~\ref{sec:sensitivity}.

This division of labor keeps the cost of the gradient independent of the
number of design variables; it grows linearly with the number of retained
modes, one adjoint solve per mode. That is the standard advantage of the adjoint method in
PDE-constrained optimization~\cite{tortorelli1994dsa,Hinze:1}. Forward-mode AD supplies the local kernel
partials, and physics-level adjoints assemble the global gradient.

\section{Sensitivity analysis of the DDAM criterion}
\label{sec:sensitivity}

\subsection{Sensitivity of the SRSS criterion}
\label{subsec:srss_sens}

The sensitivity of~\eqref{eq:srss} with respect to the design vector $\bz$
is, by direct application of the chain rule,
\begin{equation}
  \label{eq:df_dz}
  \frac{\partial f}{\partial \bz} \;=\;
  \frac{1}{2f}\left[
    \sum_i \alpha_i\, \frac{\partial \SDi}{\partial \bz}
  + \sum_i \beta_i\, \frac{\partial \Gamma_i}{\partial \bz}
  + \sum_i \gamma_i\, \bphi_i\T W\, \frac{\partial \bphi_i}{\partial \bz}
  \right],
\end{equation}
where the bracket is $\partial (f^{2})/\partial \bz$, the factor $1/(2f)$ is
the derivative of the outer square root of~\eqref{eq:srss}, and
\begin{align}
  \label{eq:srss_coef_alpha}
  \alpha_i &= 2\, \SDi\, \Gamma_i^{\,2}\, \bphi_i\T W \bphi_i, \\
  \label{eq:srss_coef_beta}
  \beta_i  &= 2\, \SDi^{\,2}\, \Gamma_i\, \bphi_i\T W \bphi_i, \\
  \label{eq:srss_coef_gamma}
  \gamma_i &= 2\, \SDi^{\,2}\, \Gamma_i^{\,2}.
\end{align}
With mass normalization the participation factor is
$\Gamma_i = \bphi_i\T \M \mathbf{D}$, and its sensitivity splits into an
explicit and an implicit part,
\begin{equation}
  \label{eq:dGamma}
  \frac{\partial \Gamma_i}{\partial \bz}
  \;=\; \bphi_i\T \frac{\partial \M}{\partial \bz}\, \mathbf{D}
  \;+\; \mathbf{D}\T \M\, \frac{\partial \bphi_i}{\partial \bz}.
\end{equation}
The explicit part gives the participation term $T_{\mathrm{II}}$ of
Section~\ref{subsec:full_sens}; the implicit part joins the eigenvector term
through the vector $v_i$ of~\eqref{eq:nelson_v}.

If the spectral displacement depends only on the natural frequency,
$\SDi = \SDi(\omega_i^{2}) = \SDi(\lambda_i)$, then
\begin{equation}
  \label{eq:dSD_dz_omega}
  \frac{\partial \SDi}{\partial \bz}
  \;=\;
  \frac{1}{2\omega_i}\!
    \left( \frac{1}{\omega_i^{2}}\frac{\partial \Dai}{\partial \omega_i} -
           \frac{2\,\Dai(\omega_i)}{\omega_i^{3}} \right)
    \frac{\partial \lambda_i}{\partial \bz}.
\end{equation}
The design acceleration may instead be read from a
participation-factor-dependent shock spectrum. Then $\SDi$ depends on the
modal participation factor as well, and a further term must be added:
\begin{equation}
  \label{eq:dSD_dz_full}
\begin{aligned}
  \frac{\partial \SDi}{\partial \bz}
  \;=\; & \frac{1}{2\omega_i}\!
    \left( \frac{1}{\omega_i^{2}}\frac{\partial \Dai}{\partial \omega_i} -
           \frac{2\,\Dai(\omega_i,\Gamma_i)}{\omega_i^{3}} \right)
    \frac{\partial \lambda_i}{\partial \bz} \\
    & \;+\; \frac{1}{\omega_i^{2}}\,
            \frac{\partial \Dai(\omega_i,\Gamma_i)}{\partial \Gamma_i}\,
            \frac{\partial \Gamma_i}{\partial \bz}.
\end{aligned}
\end{equation}
The dependence on $\Gamma_i$ runs through the effective modal mass,
$\bar m_i = \Gamma_i^{2}$, so
$\partial \Dai / \partial \Gamma_i = 2\,\Gamma_i\, \partial G / \partial \bar m_i$.
The second term of~\eqref{eq:dSD_dz_full} multiplies $\partial \Gamma_i /
\partial \bz$, as $\beta_i$ does in~\eqref{eq:df_dz}, so under such a spectrum
$\beta_i$ is replaced throughout by
\begin{equation}
  \label{eq:beta_hat}
  \hat\beta_i \;=\; \beta_i \;+\; \alpha_i\,
    \frac{1}{\omega_i^{2}}\frac{\partial \Dai}{\partial \Gamma_i},
\end{equation}
in the participation term $T_{\mathrm{II}}$ and in $v_i$. With the spectrum
evaluated at a fixed reference weight, $\hat\beta_i = \beta_i$.

\subsection{Eigenvalue sensitivity (Fox--Kapoor)}
\label{subsec:fox_kapoor}

We apply the Fox--Kapoor method~\cite{foxKapoor1968} to derive the
sensitivity of the eigenvalue. Differentiating~\eqref{eq:gep} with respect to $\bz$, with a prime denoting
$\partial/\partial \bz$:
\begin{equation}
  \label{eq:fk_diff_gep}
  \K' \bphi_i + \K \bphi_i' \;=\;
    \lambda_i' \M \bphi_i + \lambda_i \M' \bphi_i + \lambda_i \M \bphi_i'.
\end{equation}
Rearranging so the unknown terms involving $\bphi_i'$ are on the left,
\begin{equation}
  \label{eq:fk_rearranged}
  (\K - \lambda_i \M)\, \bphi_i'
  \;=\;
  \lambda_i' \M \bphi_i \;-\; (\K' - \lambda_i \M')\, \bphi_i.
\end{equation}
The operator $(\K - \lambda_i \M)$ is singular: $\bphi_i$ is exactly in
its null space. Pre-multiplying~\eqref{eq:fk_rearranged} by $\bphi_i\T$, the
left-hand side vanishes,
$\bphi_i\T (\K - \lambda_i \M) \bphi_i' = 0$, because $(\K - \lambda_i \M)$
is symmetric and $(\K - \lambda_i \M) \bphi_i = 0$; this is the solvability
condition of the Fredholm alternative~\cite{kress1999}. The right-hand side then
gives
\begin{equation}
  \label{eq:fk_pre_lambda}
  0 \;=\; \lambda_i'\,
    \underbrace{\bphi_i\T \M \bphi_i}_{=\,1} \;-\;
    \bphi_i\T (\K' - \lambda_i \M')\, \bphi_i,
\end{equation}
which yields the explicit expression
\begin{equation}
  \label{eq:fk_lambda_prime}
  \boxed{\;
  \lambda_i' \;=\; \bphi_i\T (\K' - \lambda_i \M')\, \bphi_i \;.
  \;}
\end{equation}
Stiffness added where the mode strains raises the eigenvalue, and mass added
where the mode moves lowers it, the mass term weighted by $\lambda_i$ itself.

\subsection{Eigenvector sensitivity (Nelson)}
\label{subsec:nelson}

The same solvability condition underlies Nelson's method~\cite{nelson1976} for the eigenvector sensitivity. Starting from~\eqref{eq:fk_rearranged} and collecting the right-hand side into the vector
\begin{equation}
  \label{eq:nelson_pi}
  p_i \;:=\; \lambda_i'\, \M \bphi_i - (\K' - \lambda_i \M')\, \bphi_i,
\end{equation}
the eigenvector-sensitivity equation
$(\K - \lambda_i \M)\, \bphi_i' = p_i$ is underdetermined because the
left-side operator has $\bphi_i$ in its null space. That null space is
one-dimensional only while $\lambda_i$ is simple, which this derivation
assumes throughout. A repeated eigenvalue leaves the individual eigenvector
derivative undefined, and the remedy is to differentiate the invariant
subspace rather than a basis of
it~\cite{seyranian_lund_olhoff_1994,millsCurran1988,dailey1989,linMottersheadNg2020}. Section~\ref{subsec:crossings} (Mode crossings) reports
where the demonstrations meet this, and Section~\ref{subsec:demo3_gap} builds
a criterion that avoids it. The general solution of the eigenvector-sensitivity equation is
\begin{equation}
  \label{eq:nelson_general}
  \bphi_i' \;=\; \widetilde{\bphi}_i' \;+\; c_i\, \bphi_i,
\end{equation}
where $\widetilde{\bphi}_i'$ is the particular solution fixed by the choice
$\bphi_i\T \M \widetilde{\bphi}_i' = 0$, and $c_i$
is fixed by differentiating the mass-normalization constraint
$\bphi_i\T \M \bphi_i = 1$:
\begin{equation}
  \label{eq:nelson_ci}
  \bphi_i\T \M \bphi_i'
  \;=\; -\tfrac{1}{2}\, \bphi_i\T \M' \bphi_i
  \;\Longrightarrow\;
  c_i \;=\; -\tfrac{1}{2}\, \bphi_i\T \M' \bphi_i.
\end{equation}

\paragraph{The Nelson adjoint}
Rather than solving for $\bphi_i'$ explicitly, we wish to evaluate the
functional $v_i\T \bphi_i'$. Here $v_i$ is the vector that arises in~\eqref{eq:df_dz} when collecting all terms multiplying $\bphi_i'$, as~\eqref{eq:nelson_v} below shows. Substituting~\eqref{eq:nelson_general} into the target functional,
\begin{equation}
  \label{eq:nelson_target_split}
  v_i\T \bphi_i'
  \;=\;
  \underbrace{v_i\T \widetilde{\bphi}_i'}_{\text{particular part}}
  \;+\;
  \underbrace{c_i\, v_i\T \bphi_i}_{\text{null-space part}}.
\end{equation}
The null-space part is known immediately from~\eqref{eq:nelson_ci}. The
particular part requires evaluating $\widetilde{\bphi}_i'$, but rather than
computing it directly, we introduce, as in adjoint variants of the
method~\cite{linMottersheadNg2020}, an adjoint vector $\bmu_i$ that
satisfies
\begin{equation}
  \label{eq:nelson_adjoint}
  (\K - \lambda_i \M)\, \bmu_i \;=\; v_i - (v_i\T \bphi_i)\, \M \bphi_i.
\end{equation}
The right-hand side projects out the $\M \bphi_i$ component, guaranteeing
solvability of the singular system. Exploiting the symmetry of
$(\K - \lambda_i \M)$,
\begin{equation}
  \label{eq:nelson_swap}
  \bmu_i\T (\K - \lambda_i \M)\, \widetilde{\bphi}_i'
  \;=\;
  \widetilde{\bphi}_i'^{\,\mathsf{T}}\!
    \bigl[\, v_i - (v_i\T \bphi_i)\, \M \bphi_i \,\bigr],
\end{equation}
that is,
\begin{equation}
  \label{eq:nelson_mid}
  \bmu_i\T p_i
  \;=\; v_i\T \widetilde{\bphi}_i' \;-\;
        (v_i\T \bphi_i)\, \underbrace{\bphi_i\T \M \widetilde{\bphi}_i'}_{=\,0
        \;\text{by construction}},
\end{equation}
so $\bmu_i\T p_i = v_i\T \widetilde{\bphi}_i'$. The inner product
involving the particular solution is therefore recovered from the adjoint
solution. The adjoint is defined only up to a multiple of $\bphi_i$, and
$\bmu_i\T p_i$ does not depend on it, since $\bphi_i\T p_i = 0$
by~\eqref{eq:fk_lambda_prime}. Expanding $\bmu_i\T p_i$
using~\eqref{eq:nelson_pi} and fixing that multiple by
$\bmu_i\T \M \bphi_i = 0$ (mass-orthogonality of the adjoint vector)
gives
\begin{equation}
  \label{eq:nelson_pi_simplified}
  \bmu_i\T p_i \;=\; -\bmu_i\T (\K' - \lambda_i \M')\, \bphi_i.
\end{equation}
Substituting the particular and null-space results into~\eqref{eq:nelson_target_split} yields
\begin{equation}
  \label{eq:nelson_final}
  \boxed{\;
  v_i\T \bphi_i'
  \;=\;
  -\bmu_i\T (\K' - \lambda_i \M')\, \bphi_i
  \;-\;
  \tfrac{1}{2}\, (v_i\T \bphi_i)\, \bphi_i\T \M' \bphi_i \;,
  \;}
\end{equation}
the eigenvector-sensitivity inner product obtained by Nelson's method. The
simplified form~\eqref{eq:nelson_pi_simplified} holds only for the
mass-orthogonal adjoint: a component $a\,\bphi_i$ left in $\bmu_i$ changes it
by $-a\,\lambda_i'$. Equivalently, $\bmu_i$ solves the non-singular bordered
system
\begin{equation}
  \label{eq:nelson_bordered}
  \begin{bmatrix} \K - \lambda_i \M & \M \bphi_i \\ \bphi_i\T \M & 0 \end{bmatrix}
  \begin{bmatrix} \bmu_i \\ \xi_i \end{bmatrix}
  \;=\;
  \begin{bmatrix} v_i \\ 0 \end{bmatrix},
\end{equation}
whose first row, pre-multiplied by $\bphi_i\T$, gives $\xi_i = v_i\T \bphi_i$
and recovers~\eqref{eq:nelson_adjoint}, and whose second row is the
mass-orthogonality.

\paragraph{The vector $v_i$}
The vector $v_i\T$ multiplying $\bphi_i'$ in the SRSS chain
rule~\eqref{eq:df_dz} collects the participation-factor term and the
mode-shape-extraction term:
\begin{equation}
  \label{eq:nelson_v}
  v_i \;=\;
    \underbrace{\hat\beta_i\, \M \mathbf{D}}_{\text{from participation}}
    \;+\;
    \underbrace{\gamma_i\, W \bphi_i}_{\text{from mode shape}},
\end{equation}
where $\hat\beta_i$ is~\eqref{eq:beta_hat}, equal to
$\beta_i = 2\, \SDi^{\,2}\, \Gamma_i\, \bphi_i\T W \bphi_i$ at a fixed
reference weight, and $\gamma_i = 2\, \SDi^{\,2}\, \Gamma_i^{\,2}$,
matching~\eqref{eq:srss_coef_beta}--\eqref{eq:srss_coef_gamma}. The first
term is the implicit part of~\eqref{eq:dGamma}.

\paragraph{Numerical treatment}
Solving~\eqref{eq:nelson_adjoint} at the intermediate designs of a topology
optimization needs care with near-null modes, with the boundary conditions,
and with the mass operator that forms $\Gamma_i$. Appendix~B gives the
treatment used here.

\subsection{Full DDAM sensitivity}
\label{subsec:full_sens}

Assembling~\eqref{eq:fk_lambda_prime}, \eqref{eq:nelson_final}, and~\eqref{eq:df_dz}, the sensitivity of the SRSS-based DDAM objective
decomposes element by element as
\begin{equation}
  \label{eq:full_sens}
  \frac{\partial f}{\partial z_e} \;=\; \frac{1}{2f} \sum_{i}
    \frac{\partial g_i}{\partial z_e}, \qquad
  \frac{\partial g_i}{\partial z_e} \;=\; T^{e}_{\mathrm{I}} + T^{e}_{\mathrm{II}} +
                          T^{e}_{\mathrm{III}},
\end{equation}
where $g_i = \SDi^{\,2}\, \Gamma_i^{\,2}\, \bphi_i\T W \bphi_i$ is the
per-mode squared response, so that $f^{2} = \sum_i g_i$, and, per mode $i$ (the
terms carry the mode index implicitly) and per element $e$,
\begin{align}
  \label{eq:T_I}
  T^{e}_{\mathrm{I}}   &= \alpha_i\,
    \frac{\partial \SDi}{\partial \lambda_i}\,
    \frac{\partial \lambda_i}{\partial z_e}, \\
  \label{eq:T_II}
  T^{e}_{\mathrm{II}}  &= \hat\beta_i\, \bphi_i^{e\mathsf{T}}\, m_e'\, \mathbf{D}_e, \\
  \label{eq:T_III}
  T^{e}_{\mathrm{III}} &=
    -\bmu_i^{e\mathsf{T}} \bigl(k_e' - \lambda_i\, m_e'\bigr)\, \bphi_i^{e}
    \;-\; \tfrac{1}{2}\, \eta_i\, \bphi_i^{e\mathsf{T}}\, m_e'\, \bphi_i^{e}.
\end{align}
Here $\bphi_i^{e}$, $\bmu_i^{e}$ and $\mathbf{D}_e$ are the element
restrictions of the mode, the adjoint and the influence vector. The three
terms are the eigenvalue path ($T_{\mathrm{I}}$, through $\SDi$), the explicit
participation path ($T_{\mathrm{II}}$) and the eigenvector path
($T_{\mathrm{III}}$, through the Nelson adjoint). The per-mode scalars
assemble from Sections~\ref{subsec:srss_sens}--\ref{subsec:nelson} as
\begin{equation}
  \label{eq:per_mode_scalars}
\begin{aligned}
  \omega_i^{2} &= \lambda_i, &
  \bar m_i &= \Gamma_i^{\,2}, \\
  \SDi &= \frac{\Dai(\omega_i, \bar m_i)}{\omega_i^{2}}, &
  \Gamma_i &= \bphi_i\T \M \mathbf{D}, \\
  \frac{\partial \SDi}{\partial \lambda_i}
    &= \frac{1}{2\,\omega_i\,\lambda_i}\frac{\partial \Dai}{\partial \omega_i}
       - \frac{\Dai}{\lambda_i^{2}}, &
  \frac{\partial \SDi}{\partial \Gamma_i}
    &= \frac{2\,\Gamma_i}{\lambda_i}\frac{\partial \Dai}{\partial \bar m_i}, \\
  \mathrm{B}_i &= \bphi_i\T W \bphi_i, &
  \alpha_i &= 2\, \SDi\, \Gamma_i^{\,2}\, \mathrm{B}_i, \\
  \beta_i &= 2\, \SDi^{\,2}\, \Gamma_i\, \mathrm{B}_i, &
  \hat\beta_i &= \beta_i + \alpha_i\, \frac{\partial \SDi}{\partial \Gamma_i}, \\
  \gamma_i &= 2\, \SDi^{\,2}\, \Gamma_i^{\,2}, &
  \eta_i &= v_i\T \bphi_i, \\
  v_i &= \hat\beta_i\, \M \mathbf{D} + \gamma_i\, W \bphi_i, & &
\end{aligned}
\end{equation}
where $\partial \Dai / \partial \omega_i$ is taken at fixed $\bar m_i$ and
$\partial \Dai / \partial \bar m_i$ at fixed $\omega_i$. For a spectrum
evaluated at a fixed reference weight, $\partial \Dai / \partial \bar m_i = 0$
and $\hat\beta_i = \beta_i$; the demonstrations use that form except for the
mass-loaded variant of Section~\ref{subsec:demo1_mass_loaded} (Mass-loaded
spectrum).

In~\eqref{eq:T_II}--\eqref{eq:T_III}, $k_e' = \partial \K_e / \partial z_e$
and $m_e' = \partial \M_e / \partial z_e$ are the element-level stiffness and
mass derivatives. They are evaluated by forward-mode automatic differentiation
of the element kernels, with respect to the control template parameter inside
the solver, Morphorm Analyze (Section~\ref{sec:algorithm}). The Nelson adjoint
$\bmu_i$ is computed once per retained mode by solving~\eqref{eq:nelson_adjoint}, a singular, indefinite system,
in the bordered form~\eqref{eq:nelson_bordered}.

\subsection{Sensitivities under a rigid attachment}
\label{subsec:rigid_sens}

Suppose the equipment is mounted with the rigid attachment of
Section~\ref{subsec:topo}. The eigenproblem~\eqref{eq:gep} is then replaced by
its congruence reduction
$\K_r\, \mathbf{q}_i = \lambda_i\, \M_r\, \mathbf{q}_i$, with
$\K_r = \mathbf{T}\T \K\, \mathbf{T}$ and
$\M_r = \mathbf{T}\T \M\, \mathbf{T}$. The full mode shape is recovered as
$\bphi_i = \mathbf{T}\, \mathbf{q}_i$. The reduced eigenvector satisfies the
reduced eigenproblem but not the full-space one, and
$\K \bphi_i - \lambda_i \M \bphi_i$ is the nonzero constraint reaction that
enforces the tie. The sensitivities of
Sections~\ref{subsec:fox_kapoor}--\ref{subsec:nelson} must therefore be taken
with respect to the reduced operators. Because the tie is geometric,
$\partial \mathbf{T} / \partial \bz = \mathbf{0}$, and two consequences follow.

First, the Fox--Kapoor eigenvalue sensitivity~\eqref{eq:fk_lambda_prime} is
unchanged in form. With $\K_r' = \mathbf{T}\T \K'\, \mathbf{T}$ and
$\M_r' = \mathbf{T}\T \M'\, \mathbf{T}$,
\begin{equation}
  \label{eq:fk_rigid}
  \lambda_i' \;=\; \mathbf{q}_i\T \bigl( \K_r' - \lambda_i \M_r' \bigr)\,
                   \mathbf{q}_i
             \;=\; \bphi_i\T \bigl( \K' - \lambda_i \M' \bigr)\, \bphi_i,
\end{equation}
so the assembled entry $\partial \lambda_i / \partial z_e$, and hence the term
$T^{e}_{\mathrm{I}}$ of~\eqref{eq:T_I}, is evaluated exactly as in the
unconstrained case. The participation-mass term $T^{e}_{\mathrm{II}}$ likewise
carries over, as it uses the full-space mode shape and mass.

Second, the Nelson eigenvector sensitivity must be solved in the reduced space.
The full-space adjoint~\eqref{eq:nelson_adjoint} is inconsistent for a mode
of the tied system. The singular operator $\K - \lambda_i \M$ does not have
$\bphi_i = \mathbf{T}\, \mathbf{q}_i$ in its null space, so deflating against
$\bphi_i$ no longer renders the right-hand side solvable. The adjoint is instead
defined on the reduced operators,
\begin{equation}
  \label{eq:nelson_rigid}
  \bigl( \K_r - \lambda_i \M_r \bigr)\, \bmu_{r,i}
  \;=\; v_{r,i} - \bigl( v_{r,i}\T \mathbf{q}_i \bigr)\, \M_r \mathbf{q}_i,
  \qquad v_{r,i} = \mathbf{T}\T v_i,
\end{equation}
and expanded back as $\bmu_i = \mathbf{T}\, \bmu_{r,i}$. Because $\mathbf{T}$ is constant, substituting this adjoint into the
eigenvector term $T^{e}_{\mathrm{III}}$ of~\eqref{eq:T_III} reproduces exactly
the reduced sensitivity
$\bmu_{r,i}\T (\K_r' - \lambda_i \M_r')\, \mathbf{q}_i$. The element kernels
of~\eqref{eq:full_sens} are therefore unchanged, and only the adjoint solve is
reduced.

The reduced operator $\K_r - \lambda_i \M_r$ is singular, with null space
spanned by $\mathbf{q}_i$, and for modes above the fundamental it is also
indefinite. In our experience an iterative solver stalls on such an operator.
The solver therefore solves the reduced adjoint directly, as the bordered
(saddle-point) system~\eqref{eq:nelson_bordered}~\cite{MBenzi_GHGolub_JLiesen_2005a}
written with $\K_r$, $\M_r$, $\mathbf{q}_i$ and $v_{r,i}$, which is
non-singular. Its added row enforces the mass-orthogonality of Appendix~B
exactly, and the right-hand-side projection of Appendix~B carries over to the
reduced space unchanged.

\subsection{Exact NRL Sum design criterion}
\label{subsec:nrl_exact}

The SRSS criterion of Section~\ref{subsec:srss} is everywhere differentiable
away from mode crossings, and is therefore convenient for gradient-based
optimization. It is not, however, the quantity against which the design is
certified. With the per-mode squared response of
Section~\ref{subsec:full_sens},
\begin{equation}
  \label{eq:gi}
  g_i \;=\; \SDi^{\,2}\, \Gamma_i^{\,2}\, \mathrm{B}_i,
  \qquad \mathrm{B}_i = \bphi_i\T W \bphi_i,
\end{equation}
so that the SRSS criterion~\eqref{eq:srss} is
$f_{\mathrm{SRSS}} = \bigl(\sum_i g_i\bigr)^{1/2}$, the NRL
Sum~\eqref{eq:nrl_sum} reads
\begin{equation}
  \label{eq:nrl_g}
  f_{\mathrm{NRL}} \;=\; \sqrt{g_m} \;+\; \sqrt{\,\sum_{i \neq m} g_i\,},
  \qquad m \;=\; \arg\max_i\, g_i ,
\end{equation}
since, for a response checked at a single degree of freedom or uniform over
the interface, as under the rigid attachment, $\sqrt{g_m} = \lvert \Rmax \rvert$
and $\sum_{i \neq m} g_i = \sum_{i \neq \max} R_i^{\,2}$. For a non-uniform
interface response $\sqrt{g_i}$ is the root-mean-square of mode $i$ over the
interface rather than a pointwise peak. The two criteria are
ordered by the subadditivity of the square root,
\begin{equation}
  \label{eq:srss_le_nrl}
  f_{\mathrm{SRSS}}
  \;=\; \Bigl( g_m + \textstyle\sum_{i\neq m} g_i \Bigr)^{1/2}
  \;\le\; \sqrt{g_m} + \Bigl(\textstyle\sum_{i\neq m} g_i\Bigr)^{1/2}
  \;=\; f_{\mathrm{NRL}},
\end{equation}
with equality only when a single mode carries the entire response. The smooth
surrogate therefore systematically under-predicts the certifying response, and
minimizing it does not by itself minimize the quantity the structure is
certified against. We close this gap by differentiating the exact NRL Sum
directly.

The only non-smoothness in~\eqref{eq:nrl_g} beyond the mode crossings already
noted for the SRSS form is the dominant-mode selector $m$. We adopt a
\emph{frozen-argmax} convention~\cite{clarke1983}: $m$ is recomputed at every evaluation of the
criterion and then held fixed while the gradient at that point is formed. The index of the dominant mode is a locally constant integer, so it has zero
derivative between dominance swaps. The sensitivity of~\eqref{eq:nrl_g} is
exact there,
\begin{equation}
  \label{eq:nrl_grad}
  \frac{\partial f_{\mathrm{NRL}}}{\partial \bz}
  \;=\; \underbrace{\frac{1}{2\sqrt{g_m}}}_{w_m}\, g_m'
  \;+\; \underbrace{\frac{1}{2\sqrt{\sum_{i\neq m} g_i}}}_{w_{\mathrm{bg}}}
        \sum_{i \neq m} g_i' ,
\end{equation}
where $g_i' = \partial g_i / \partial \bz$ is the per-mode sensitivity already
assembled in~\eqref{eq:full_sens}, evaluated for mode $i$. The two criteria
therefore share every $g_i'$ and differ only in how they are weighted.

The exact NRL Sum thus reuses the per-mode adjoint solves, the Fox--Kapoor
and Nelson kernels, and the element integrals of
Section~\ref{subsec:full_sens} without modification. It differs
from the SRSS criterion only in the reduction that combines the modes. The
SRSS form places all modes under a single radical, and so applies the one
common normalization $1/(2f_{\mathrm{SRSS}})$ to every $g_i'$. The NRL Sum places the dominant mode under its own radical and the background
modes under a second. It weights the dominant contribution by $w_m$ and the
background contributions by $w_{\mathrm{bg}}$.

In practice each radical in~\eqref{eq:nrl_g} is regularized as
$\sqrt{(\cdot) + \epsilon}$ with a small $\epsilon$, which carries the units of
$g_i$, the square of the response. This keeps the weights in~\eqref{eq:nrl_grad} bounded when a response is negligible,
at a bias of about $\epsilon/(2\sqrt{g_m}) + \epsilon/(2\sqrt{\sum_{i\neq m} g_i})$
in the criterion. When a single mode
dominates so completely that $\sum_{i\neq m} g_i \to 0$, the background term
is dropped from the criterion and its weight $w_{\mathrm{bg}}$ from the
gradient together, so that the two describe the same function.

The residual non-smoothness at a strict dominance tie is an \emph{intrinsic}
kink of the hard maximum, not an artifact of mode indexing. The frozen-argmax
sensitivity~\eqref{eq:nrl_grad} is exact on either side of the tie, but the
kink is not a set the optimizer merely passes through. Writing the two tied
responses as $g_a = t + e$ and $g_b = t - e$ with background $r > 0$, the
criterion is $\sqrt{t + \lvert e \rvert} + \sqrt{t + r - \lvert e \rvert}$,
whose slope in $\lvert e \rvert$ at the tie, $1/(2\sqrt{t}) - 1/(2\sqrt{t+r})$,
is positive. The NRL Sum therefore has a convex, V-shaped minimum at a tie,
and a minimizer is drawn towards ties, as it is towards coalescence in
eigenvalue problems. The carrier swaps of Section~\ref{subsec:demo3} are
consistent with this; the optimizer is not given special treatment there.

We emphasize that this selector follows the response magnitude $g_i$,
recomputed at each evaluation, rather than tracking individual mode shapes
across iterates. The modal-assurance-criterion pairing discussed under mode
crossings (Section~\ref{sec:limitations}) addresses a distinct
frequency-ordering issue, and is neither required nor applicable to the
dominant-mode selector. We deliberately do not smooth the maximum over the
mode index. Replacing the dominant-mode selector by a smooth surrogate would
no longer return the NAVSEA-certifying response, which a smooth maximum over
the modes would over-predict. The spatial aggregate of
Section~\ref{subsec:nrl_stress} is different: being normalized by the region
measure, it approaches the peak from below.

\subsection{NRL Sum of the modal stress field}
\label{subsec:nrl_stress}

Demonstrations~2 and~3 certify against stress, for which the modal response is
the field~\eqref{eq:Ri_stress} rather than a scalar. That is not merely a
change of variable. It changes the order in which the modes are combined and
the response is reduced, and the two orders do not agree.

Write the von Mises invariant of the modal stress of mode $i$ as
\begin{equation}
  \label{eq:vm_mode}
  s_i \;=\; \sigma_{\mathrm{vm}}\bigl(C B \bphi_i\bigr),
  \qquad
  R_i \;=\; \lvert \SDi\, \Gamma_i \rvert \; s_i ,
\end{equation}
with the spectral factor $\lvert \SDi\, \Gamma_i \rvert$ carried outside the invariant. This is exact rather
than an approximation, because the von Mises invariant is absolutely
homogeneous of degree one, $\sigma_{\mathrm{vm}}(c\,\boldsymbol{\sigma}) = \lvert c\rvert\,
\sigma_{\mathrm{vm}}(\boldsymbol{\sigma})$. It also leaves $s_i$ a property of the mode shape
alone.

\paragraph{The modes are combined pointwise}
The certification procedure forms the invariant mode by mode and combines the
modal values at each response location~\cite{navsea_t9070}, as commercial
implementations also do~\cite{simulia2007ddam}. The NRL
Sum~\eqref{eq:nrl_sum} is therefore evaluated at each point $x$ of the
region,
\begin{equation}
  \label{eq:nrl_field}
  \begin{gathered}
  R_{\mathrm{NRL}}(x) \;=\; R_{m(x)}(x)
    \;+\; \sqrt{\,\sum_{i \neq m(x)} R_i(x)^{2}\,},\\
  m(x) \;=\; \arg\max_i\, R_i(x) .
  \end{gathered}
\end{equation}
The dominant mode is a field of indices, not a single index. This is the
substantive difference from~\eqref{eq:nrl_g}, in which
$\mathrm{B}_i = \bphi_i\T W \bphi_i$ contracts the mode shape to one scalar
before the modes are combined. Contracting first and combining second yields a different function. The
combination of spatially reduced responses is not the reduction of the
pointwise combination whenever two modes attain their peaks in different
places. Only~\eqref{eq:nrl_field} reproduces the certifying quantity. The
difference has a definite sign: the NRL Sum is nondecreasing in each modal
response and continuous at a dominance swap, so
$\max_x R_{\mathrm{NRL}}(x) \le \mathrm{NRL}\bigl(\max_x R_1, \ldots, \max_x
R_n\bigr)$. Contracting first is conservative, and the pointwise form is the
sharper of the two.

\paragraph{The reduction to a scalar comes last}
A field is not a design criterion. The certifying quantity is the peak of
$R_{\mathrm{NRL}}$ over a marked region $\Omega_c$. That peak is not
differentiable, so the optimization is driven instead by a
Kreisselmeier--Steinhauser aggregate~\cite{kreisselmeier1979,kennedyHicken2015}
weighted by the element measure, as global stress measures are normalized in
stress-based topology optimization~\cite{le2010stress},
\begin{equation}
  \label{eq:ks_aggregate}
  \begin{gathered}
  f_{\sigma} \;=\; \mu \;+\; \frac{1}{\kappa}
    \log\!\left( \frac{1}{\lvert \Omega_c \rvert}\sum_{e} v_e\,
      e^{\kappa\,\left(R_{\mathrm{NRL},e} - \mu\right)} \right),\\
  \mu \;=\; \max_e R_{\mathrm{NRL},e},
  \qquad
  \lvert \Omega_c \rvert \;=\; \sum_e v_e ,
  \end{gathered}
\end{equation}
where $v_e$ is the measure of element $e$ in $\Omega_c$, $R_{\mathrm{NRL},e}$ is
the pointwise NRL Sum sampled in element $e$, and $\kappa > 0$, with units of
inverse response, sets the sharpness. Weighting by the element measure is what makes $f_{\sigma}$ converge under
mesh refinement. An aggregate that weights elements equally counts
discretization rather than material, and drifts as the mesh changes.

Normalizing by the total measure has a consequence the certification must
respect. Every exponent in~\eqref{eq:ks_aggregate} is non-positive, so
$f_{\sigma} \le \mu$ for every $\kappa$, with equality only when the field is
constant over $\Omega_c$. Keeping only the peak element $e^{*}$ in the sum
bounds the gap from the other side,
$\mu + \kappa^{-1}\log\bigl(v_{e^{*}}/\lvert\Omega_c\rvert\bigr) \le f_{\sigma}
\le \mu$. The aggregate approaches the peak from below as $\kappa$ grows and never
bounds it from above. A constraint $f_{\sigma} \le \sigma_{\mathrm{allow}}$
therefore bounds the aggregate, not the peak, and the peak $\mu$ may exceed the
allowable.

\paragraph{Complete dominance}
The pointwise form needs a stronger version of the regularization discussed
after~\eqref{eq:nrl_grad}. The scalar form drops the background term only as
it vanishes; the pointwise form must drop it earlier, for the following reason.

Where one mode $m$ dominates a point completely, the background radical
vanishes. The sensitivity to each background mode $i \neq m$,
$\partial R_{\mathrm{NRL}} / \partial R_i = R_i / \bigl(\sum_{k \neq m}
R_k^{2}\bigr)^{1/2}$, is then a ratio of two vanishing quantities, and it does
not vanish with them: it stays of order unity. A mode that carries no
participation would therefore receive a full-weight sensitivity, amplified
further by $\SDi = \Dai/\omega_i^{2}$, which grows without bound as
$\omega_i \to 0$.

To prevent this, the background is treated as absent wherever it falls below a
fixed fraction $\tau$ of the dominant term. The truncation is applied to the
criterion and to its derivative alike, so that the two describe the same
function. Its price is a jump in $R_{\mathrm{NRL}}$, of at most $\tau$ times
the dominant term, where the background crosses the threshold.

\paragraph{Sensitivity}
By~\eqref{eq:vm_mode}, the stress criterion depends on the design in three
ways: through the spectral factor $\lvert \SDi\, \Gamma_i \rvert$, the scalar
amplitude that scales the stress of mode $i$; through the mode shape
$\bphi_i$, which sets the invariant $s_i$; and directly, through the
interpolated stiffness $C(\bz)$ inside $s_i$. The first reuses the SRSS
results, while the other two require new terms. All three pass through the
same two sets of weights. The aggregate weights
\begin{equation}
  \label{eq:stress_chi}
  \chi_e \;=\; \frac{\partial f_{\sigma}}{\partial R_{\mathrm{NRL},e}}
  \;=\; \frac{v_e\, e^{\kappa(R_{\mathrm{NRL},e}-\mu)}}
             {\sum_{e'} v_{e'}\, e^{\kappa(R_{\mathrm{NRL},e'}-\mu)}}
\end{equation}
are non-negative and sum to one over $\Omega_c$ (the shift by $\mu$ cancels
in them). The pointwise combination weights
$c_{i,e} = \partial R_{\mathrm{NRL},e} / \partial R_{i,e}$ come
from~\eqref{eq:nrl_field}.

The spectral factor is the only place where the eigenvalue and the
participation factor enter. It multiplies $s_{i,e}$ in every element at once,
so the whole field responds to it through one number per mode, the
sensitivity of the criterion to the spectral factor,
\begin{equation}
  \label{eq:stress_A}
  A_i \;=\; \frac{\partial f_{\sigma}}{\partial \lvert \SDi\, \Gamma_i \rvert}
  \;=\; \sum_{e} \chi_e\, c_{i,e}\, s_{i,e} .
\end{equation}
Differentiating $\lvert \SDi\, \Gamma_i \rvert$ with respect to each of its
two factors then gives
\begin{equation}
  \label{eq:stress_coeffs}
  \frac{\partial f_{\sigma}}{\partial \SDi} = A_i\, \varsigma_i\, \Gamma_i ,
  \qquad
  \frac{\partial f_{\sigma}}{\partial \Gamma_i} = A_i\, \varsigma_i\, \SDi ,
  \qquad
  \varsigma_i = \operatorname{sign}(\SDi\, \Gamma_i) .
\end{equation}
These are the stress-criterion counterparts of $\alpha_i/(2f)$ and
$\beta_i/(2f)$ in~\eqref{eq:df_dz}, which are $\partial f/\partial \SDi$ and
$\partial f/\partial \Gamma_i$ for the SRSS criterion. The eigenvalue and
participation terms of Section~\ref{subsec:srss_sens} therefore carry over
unchanged once two substitutions are made in~\eqref{eq:df_dz}
and~\eqref{eq:full_sens}: $\alpha_i/(2f)$ becomes $A_i \varsigma_i \Gamma_i$,
and $\beta_i/(2f)$ becomes $A_i \varsigma_i \SDi$. The outer factor $1/(2f)$
is dropped, since it is the derivative of the SRSS square root and the
aggregate has no such root.

Since $\SDi > 0$, $\varsigma_i = \operatorname{sign}(\Gamma_i)$. The sign is
carried explicitly because $\Gamma_i$ takes either sign under the eigenvector
normalization, and $\lvert \cdot \rvert$ is not differentiable at the origin.
A sign flip of the eigenvector cancels, because $s_i$ is even in $\bphi_i$.

The mode shape needs a different adjoint load. The displacement criterion
contributes $\gamma_i W \bphi_i$, the derivative of a quadratic form in
$\bphi_i$. The stress criterion instead contributes
\begin{equation}
  \label{eq:stress_adjoint_load}
  \frac{\partial f_{\sigma}}{\partial \bphi_i}
    \;=\; \sum_{e} \chi_e\, c_{i,e}\, \lvert \SDi\, \Gamma_i \rvert \,
      \frac{\partial s_{i,e}}{\partial \bphi_i} ,
\end{equation}
taken at fixed $\Gamma_i$. This load depends on $\bphi_i$ through the
derivative of the von Mises invariant, not through a fixed weighting matrix
such as $W$. The dependence of $\Gamma_i$ on $\bphi_i$ is already counted in
the participation term, through~\eqref{eq:dGamma}.

Finally, the modal stress depends on the design variables directly, and not
only through the eigenpair, because the stiffness $C(\bz)$ is interpolated,
as in stress-based topology
optimization~\cite{duysinxBendsoe1998,chengGuo1997,le2010stress}. The total
derivative therefore acquires a term with no counterpart in the displacement
case,
\begin{equation}
  \label{eq:stress_explicit}
  \left.\frac{\partial f_{\sigma}}{\partial \bz}\right|_{\mathrm{explicit}}
    \;=\; \sum_i \sum_e \chi_e\, c_{i,e}\, \lvert \SDi\, \Gamma_i \rvert \,
      \frac{\partial s_{i,e}}{\partial \bz} .
\end{equation}
This term is not a small correction: on a stress variant of the verification
problem of Section~\ref{subsec:fd_check} (not shown) it supplies most of the
total derivative. Omitting it leaves a finite-difference error that is flat in the step size
rather than convergent. That is the signature of a missing term rather than of
a mistuned coefficient.

\section{Algorithmic implementation}
\label{sec:algorithm}

The DDAM criterion and its sensitivity are evaluated at every outer
iteration of the topology optimizer. Figure~\ref{fig:morphorm_pipeline}
shows the data flow per outer iteration, and Algorithm~\ref{alg:morphorm_ddam}
summarizes the corresponding workflow.

\begin{widefigure}[htbp]
\centering
\begin{tikzpicture}[
  font=\small,
  node distance=5mm,
  stage/.style = {rectangle, draw, rounded corners=2pt, align=center,
                  minimum height=9mm, inner xsep=4pt, inner ysep=3pt,
                  text width=52mm},
  loop/.style  = {rectangle, draw, rounded corners=2pt, align=center,
                  minimum height=10mm, inner xsep=3pt, inner ysep=3pt,
                  text width=30mm, fill=black!5},
  decision/.style = {diamond, draw, aspect=2.4, align=center, inner sep=1pt,
                     text width=18mm},
  arr/.style   = {-{Latex[length=2mm]}, thick},
]
\node[stage] (mesh)                       {Design $\bz^{(n)}$ on the fixed mesh; filtered densities $\tilde{\bz}=\mathbf{F}\bz^{(n)}$};
\node[stage, below=of mesh]   (assem)     {SIMP assembly: $\K(\bz^{(n)}),\,\M(\bz^{(n)})$};
\node[stage, below=of assem]  (reduce)    {Rigid reduction: $\K_r=\mathbf{T}\T\K\mathbf{T},\,\M_r=\mathbf{T}\T\M\mathbf{T}$};
\node[stage, below=of reduce] (eig)       {Eigensolver: $(\lambda_i,\mathbf{q}_i)_{i=1}^{N_{\!m}},\ \bphi_i=\mathbf{T}\mathbf{q}_i$};
\node[stage, below=of eig]    (modal)     {Modal data and retained set: $\Gamma_i,\bar m_i,\Dai,\SDi$; criterion with frozen argmax};
\coordinate[below=7mm of modal] (fanout);
\node[loop, below=7mm of fanout]                  (nelson) {Reduced Nelson adjoint\\$\bmu_{r,i}$ via~\eqref{eq:nelson_rigid}, load $v_i$ or~\eqref{eq:stress_adjoint_load}};
\node[loop, left=4mm of nelson]                   (fk)     {Fox--Kapoor\\$\lambda_i'$};
\node[loop, right=4mm of nelson]                  (kernel) {Element kernels\\$T^e_\mathrm{I},T^e_\mathrm{II},T^e_\mathrm{III}$ (AD)};
\node[draw, dashed, rounded corners=2pt,
      fit=(fk)(nelson)(kernel), inner xsep=3mm, inner ysep=2mm,
      label={[font=\scriptsize, yshift=-1mm]above:\itshape per mode $i = 1,\dots,N_{\!m}$}]
      (loopbox) {};
\coordinate[below=4mm of loopbox.south] (fanin);
\node[stage, below=4mm of fanin] (grad)   {Combine modes: \eqref{eq:full_sens}, \eqref{eq:nrl_grad}, or the stress weights and~\eqref{eq:stress_explicit}; floor gradient};
\node[stage, below=of grad]      (filter) {Filter transpose $\mathbf{F}\T$ \& MMA update};
\node[decision, below=of filter] (conv)   {Converged?};
\node[stage, below=of conv]      (done)   {Optimized $\bz^{*}$};
\draw[arr] (mesh)   -- (assem);
\draw[arr] (assem)  -- (reduce);
\draw[arr] (reduce) -- (eig);
\draw[arr] (eig)    -- (modal);
\draw[arr] (modal) -- (fanout);
\draw[arr] (fanout) -| (fk.north);
\draw[arr] (fanout) -- (nelson.north);
\draw[arr] (fanout) -| (kernel.north);
\draw[arr] (fk.south)     |- (fanin);
\draw[arr] (nelson.south) -- (fanin);
\draw[arr] (kernel.south) |- (fanin);
\draw[arr] (fanin) -- (grad);
\draw[arr] (grad)   -- (filter);
\draw[arr] (filter) -- (conv);
\draw[arr] (conv)   -- node[right, font=\scriptsize]{yes} (done);
\draw[arr] (conv.west) -- ++(-40mm,0)
           node[midway, above, font=\scriptsize]{no, $\bz^{(n+1)}$}
           |- (mesh.west);
\end{tikzpicture}
\caption{Per-iteration data flow of DDAM-driven topology optimization in
Morphorm Studio. The outer pipeline executes once per design iterate on a
fixed mesh: filtering, SIMP assembly, rigid-attachment reduction, eigensolver,
modal data and the criterion, the per-mode sensitivities, the combination of
the modes for the criterion in use, the filter transpose and the update, and
the convergence check. The design-independent tie $\mathbf{T}$
reduces $\K,\M$ to $\K_r,\M_r$ before the eigensolve, and the reduced Nelson
adjoint of~\eqref{eq:nelson_rigid} is solved in the tied space. The shaded
per-mode loop executes once per retained mode, and covers Fox--Kapoor, the
reduced Nelson adjoint and the AD-assembled element kernels. Because
$\mathbf{T}$ is constant the loop kernels are reduction-invariant, and they
supply the three additive contributions of~\eqref{eq:full_sens} unchanged.}
\label{fig:morphorm_pipeline}
\end{widefigure}

\begin{algorithm}[h!]
\KwIn{Initial topology $\bz^{(0)}$; allowable response
$R_{\mathrm{allow}}$ (when DDAM is used as a constraint); convergence
tolerance $\varepsilon_{\mathrm{tol}}$; max iterations $N_{\max}$.}
\KwOut{Optimized topology $\bz^{*}$; certifying NRL Sum response
$\Rtot^{*}$.}
\For{$n = 0, 1, \ldots, N_{\max}$}{
  Assemble $\K(\bz^{(n)})$ and $\M(\bz^{(n)})$ from the SIMP-interpolated
    element matrices, and form the rigid-attachment reduction
    $\K_r=\mathbf{T}\T\K\mathbf{T}$, $\M_r=\mathbf{T}\T\M\mathbf{T}$
    (Section~\ref{subsec:topo}; $\mathbf{T}$ design-independent)\;
  Solve the reduced generalized eigenproblem
    $\K_r\,\mathbf{q}_i=\lambda_i\,\M_r\,\mathbf{q}_i$ for the lowest
    $N_{\!m}$ pairs $(\lambda_i,\mathbf{q}_i)$ and recover
    $\bphi_i=\mathbf{T}\,\mathbf{q}_i$\;
  Evaluate $\Gamma_i$, $\bar m_i$, $\Dai$, and $\SDi$ from
    Section~\ref{subsec:modal_response}\;
  Evaluate the criterion $f^{(n)}$: the SRSS form~\eqref{eq:srss}, the
    NRL Sum~\eqref{eq:nrl_g}, or for stress the pointwise NRL
    Sum~\eqref{eq:nrl_field} and its aggregate~\eqref{eq:ks_aggregate}, with
    the dominant mode fixed by the frozen argmax; every demonstration of
    Section~\ref{sec:results} uses an NRL Sum\;
  \For{each retained mode $i$}{
    Evaluate the Fox--Kapoor expression~\eqref{eq:fk_lambda_prime} for
      $\lambda_i'$\;
    Solve the reduced Nelson adjoint system~\eqref{eq:nelson_rigid}, with
      the load $v_i$ of~\eqref{eq:nelson_v} or, for stress,
      \eqref{eq:stress_adjoint_load}, as a bordered saddle-point solve for
      $\bmu_{r,i}$, and expand $\bmu_i=\mathbf{T}\,\bmu_{r,i}$\;
    Evaluate $T^{e}_{\mathrm{I}}, T^{e}_{\mathrm{II}}, T^{e}_{\mathrm{III}}$
      from~\eqref{eq:T_I}--\eqref{eq:T_III} using $k_e'$, $m_e'$ supplied
      by forward-mode AD of the element kernels, and for stress the explicit
      term~\eqref{eq:stress_explicit}\;
  }
  Combine the modes: \eqref{eq:full_sens} for the SRSS form,
    \eqref{eq:nrl_grad} for the NRL Sum, the substitution of
    Section~\ref{subsec:nrl_stress} for stress; add the gradient of the
    frequency floor from~\eqref{eq:fk_lambda_prime}\;
  Map the gradient back through the filter transpose $\mathbf{F}\T$ of the
    density filter~\cite{bourdin2001filter} and update $\bz^{(n+1)}$
    (MMA~\cite{Svanberg_1987})\;
  \If{$\| \bz^{(n+1)} - \bz^{(n)} \| < \varepsilon_{\mathrm{tol}}$
       \textbf{and} every constraint holds}{
    \textbf{break}\;
  }
}
\caption{DDAM-driven topology optimization in Morphorm Studio. Every run
reported in Section~\ref{sec:results} stops at its iteration cap $N_{\max}$.}
\label{alg:morphorm_ddam}
\end{algorithm}

\paragraph{The design spectrum as a user module}
The Navy design acceleration is not coded in the solver. It is supplied by a
Python module that the input names, which returns, for a mode, the design
value $\Dai$ and its derivative $\partial \Dai / \partial \omega_i$ with
respect to the angular frequency. A module may also return the derivative
with respect to the effective modal mass; the solver then passes each mode's
effective mass $\bar m_i$ to the module and carries the mass-loading term
of~\eqref{eq:dSD_dz_full}. The solver applies the remaining chain rule,
$\partial\omega_i/\partial\lambda_i = 1/(2\omega_i)$ and
$\partial \bar m_i/\partial \Gamma_i = 2\Gamma_i$, itself. Ship type, mounting
location, direction and the tabulated coefficients therefore live in the
module, and the formulation of Section~\ref{sec:sensitivity} applies unchanged
to a different Navy spectrum: only the module changes. The demonstrations of
Section~\ref{sec:results} use modules implementing the NRL-1396 surface-ship,
deck-mounted elastic design values~\cite{nrl1396}, at a fixed reference weight
and, for the variant of Section~\ref{subsec:demo1_mass_loaded} (Mass-loaded
spectrum), mode by mode.

\paragraph{AD-assembled $k_e'$ and $m_e'$}
The element stiffness and mass matrices are coded as templated kernels. Their
arguments are the element pseudo-density $z_e$, the element geometry, and the
constitutive parameters. Forward-mode AD over the control template parameter $z_e$ yields
the SIMP-derivative entries needed by~\eqref{eq:T_I}--\eqref{eq:T_III},
exact to round-off. It removes hand-coded derivative expressions, and
the maintenance burden they carry when new element types or constitutive
models are introduced. Forward mode suits the element level because each kernel has a single
control input $z_e$ and many output entries; the global gradient is then
assembled by the adjoints of Section~\ref{subsec:autodiff}.

\paragraph{Mode selection}
The criterion identifies the mode it acts on in one of two ways, both local
to the current design point. Fixed-index criteria select by frequency ordering, the $i$-th smallest
converged eigenvalue. The NRL Sum selects its dominant mode by the response
magnitude $g_i$, recomputed at each evaluation as described in
Section~\ref{subsec:nrl_exact}. Neither uses a modal-assurance-criterion
(MAC)~\cite{allemang2003mac} pairing between consecutive iterates, so modes are not
tracked from one iterate to the next. When the
retained set changes between iterates, the criterion changes by the
contribution of the mode that enters or leaves. A mode removed by the
participation floor of Section~\ref{sec:results} carries little effective
mass, but one added to reach the coverage target need not.
Section~\ref{subsec:crossings} (Mode crossings) discusses what mode tracking
would and would not change.

\section{Numerical demonstration}
\label{sec:results}

We demonstrate the workflow on three surface-ship problems on one geometry,
the equipment frame of Figure~\ref{fig:setup_cabinet}. Each demonstration
after the first changes a single element of the one before, so the three form
a controlled sequence rather than three unrelated examples:
\begin{itemize}
  \item \emph{Demonstration~1} certifies the fore-and-aft interface
        displacement, and establishes the workflow with a displacement
        acceptance measure.
  \item \emph{Demonstration~2} changes the acceptance measure alone, to
        von~Mises stress, in the same fore-and-aft direction. The direction is
        fore-and-aft rather than vertical, for a reason given in
        Section~\ref{subsec:demo2}.
  \item \emph{Demonstration~3} certifies the same frame against all three
        orthogonal directions. The bound on the most severe direction is
        imposed as one constraint per direction, so it adds no hard-max layer
        beyond the pointwise dominant-mode selector of the NRL
        Sum~\eqref{eq:nrl_field}.
\end{itemize}
Demonstrations~2 and~3 share geometry, mesh, material and allowable, so the
difference between them isolates what the combined-direction requirement
costs. The criterion is used as a \emph{constraint} on a volume-minimizing problem in
all three. Section~\ref{sec:limitations} explains why the objective form of
Section~\ref{subsec:topo} is not demonstrated.

All three problems model the structure as isotropic, linear-elastic mild
structural steel ($E = \SI{200}{GPa}$, $\nu = 0.30$,
$\rho = \SI{7850}{kg/m^3}$) and represent the supported equipment the same way,
as described below. Its minimum specified yield supplies the
allowable of Demonstrations~2 and~3, which are certified on stress.
Demonstration~1 is certified on a displacement and carries no yield limit. A mild grade is a deliberate choice for the stress demonstrations. Its yield
keeps the allowable active rather than leaving a margin the optimizer never
approaches. The supported equipment is represented by a rigid, non-structural lumped mass
clamped to the mounting interface. Its mass therefore enters $\M(\bz)$ and the
effective modal mass~\eqref{eq:effective_mass}, but is excluded from the SIMP
design domain.

The formulation retains the mass-loading coupling. The effective modal mass
varies with the mode shapes as the topology evolves, and $\Dai$ could vary
with it. That is the $\partial \Dai/\partial \Gamma_i$ term
of~\eqref{eq:dSD_dz_full}. The sensitivity of the displacement criterion
carries this term; that of the stress criterion does not yet
(Section~\ref{subsec:mass_loading} (Mass loading)).

Demonstrations~2 and~3 are certified on stress, so they evaluate the effective
weight in~\eqref{eq:nrl1396_A0V0} at a fixed reference, the weight of the
supported equipment. The design acceleration then depends on frequency alone,
and the term vanishes. Demonstration~1 uses the same fixed weight, so that all
three demonstrations share one spectrum. A variant of Demonstration~1
evaluates the weight mode by mode and so exercises the term
(Section~\ref{subsec:demo1_mass_loaded} (Mass-loaded spectrum)).

Every demonstration and both companion studies share these settings. We use
the modified SIMP interpolation~\eqref{eq:simp} with $p=3$ and
$\epsilon_{\min}=10^{-6}$ on both the stiffness and the structural mass, and the MMA update~\cite{Svanberg_1987} with a move limit
of $0.1$. The density filter
is the kernel filter~\cite{bourdin2001filter,brunsTortorelli2001}, without
projection, of radius $2.5$ times the average element edge length. Eigenvectors are mass-normalized
(Remark~\ref{rem:massnorm}). The NAVSEA guidance~\cite{navsea_0908,navsea_t9070} sets the coverage requirement: the
retained modes must reach \SI{80}{\percent} cumulative effective mass in the shock
direction. The runs meet it through a mode count rather than a frequency
cutoff. The eigensolver starts with six modes and adds modes, up to twelve,
until the solved set reaches \SI{80}{\percent}. Every solved mode is retained
except those below $10^{-3}$ of the total effective mass, and that floor is
never applied if it would take coverage below \SI{80}{\percent}. The design acceleration for each mode is the lesser of the
acceleration- and velocity-limited branches,
\begin{equation}
  \label{eq:Da_branches}
  \Dai \;=\; \max\Bigl(6\,g,\ \min\bigl(A_{a,i}\, g,\ \omega_i\, V_{a,i}\bigr)\Bigr),
\end{equation}
where $A_{a,i}$ and $V_{a,i}$ are the NRL-1396 acceleration and velocity
design values for the selected ship type, mounting location, direction, and
elastic analysis, $A_{a,i}$ a multiple of the gravitational acceleration $g$
and $V_{a,i}$ converted from in/s to consistent units. They are evaluated at an
effective weight $W_a$ in kips, fixed at the weight of the supported
equipment, except in the variant of Section~\ref{subsec:demo1_mass_loaded} (Mass-loaded spectrum),
which uses the modal effective weight
$W_{a,i}=\bar m_i\, g$~\cite{nrl1396,navsea_0908,navsea_t9070}. On the velocity branch
$\SDi = V_{a,i}/\omega_i$, so
$\partial \SDi / \partial \lambda_i = -V_{a,i}/(2\omega_i^{3})$; on the
acceleration branch and on the floor $\Dai$ is constant, so
$\partial \SDi / \partial \lambda_i = -\Dai/\omega_i^{4}$.

The branch switch in~\eqref{eq:Da_branches} between the velocity- and
acceleration-limited regimes is the source of the piecewise non-smoothness
discussed in Section~\ref{sec:limitations}. Its location moves with the modal
mass as the topology changes. Table~\ref{table:demo_summary} collects
the headline results across the three problems.

\subsection{Verification of the analytic sensitivity}
\label{subsec:fd_check}

\begin{figure}[htbp]
  \centering
  \includegraphics[width=0.62\linewidth]{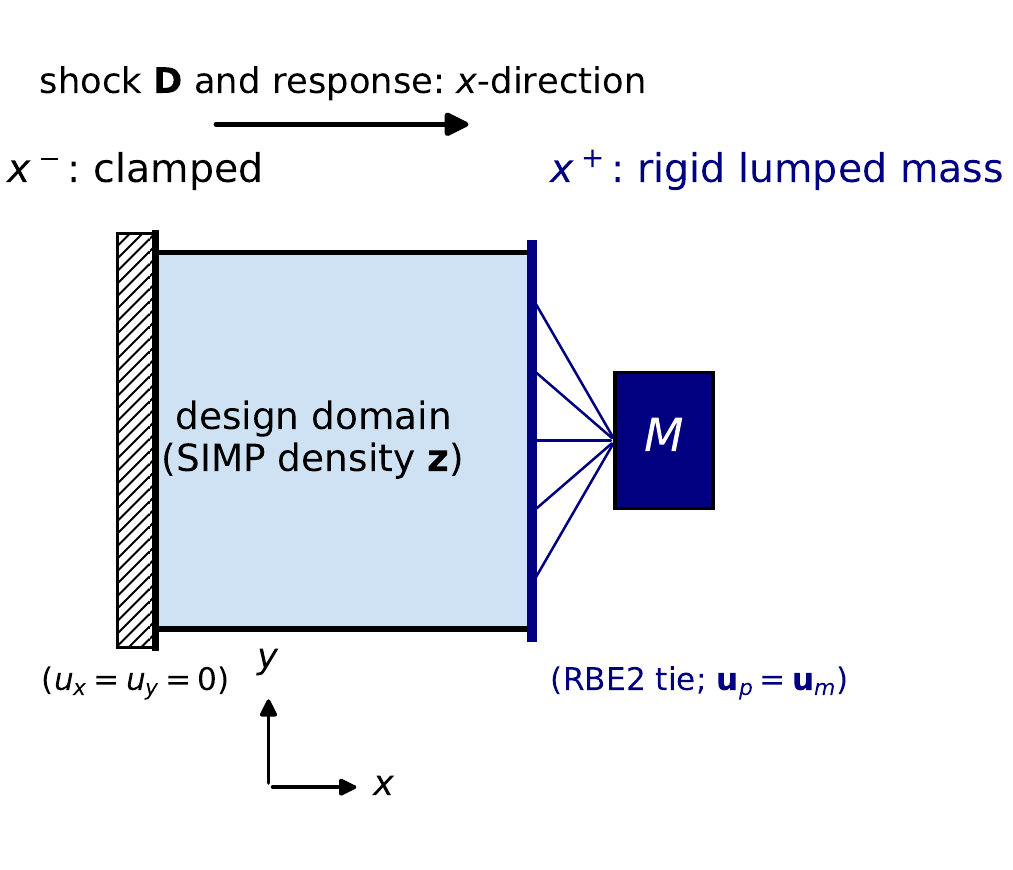}
  \caption{Verification problem for the finite-difference check of the analytic
  sensitivity. A square design domain of SIMP density
  $\mathbf{z}$ is clamped at $x^-$ and carries a rigid lumped mass $M$ tied
  across $x^+$. The shock influence vector $\mathbf{D}$ and the measured
  response are in the $x$-direction. This is the problem behind the V-curve of
  Figure~\ref{fig:fd_vcurve}.}
  \label{fig:fd_problem}
\end{figure}

We verify the assembled $\partial f / \partial \bz$ of~\eqref{eq:full_sens} by
finite differences on the small representative problem of
Figure~\ref{fig:fd_problem}. It is a square design domain clamped on its left edge ($x^-$, $u_x=u_y=0$). A
rigid lumped mass is tied across its right edge ($x^+$, the translational RBE2
tie). The shock influence vector
$\mathbf{D}$ and the measured response are in the $x$-direction.

This exercises the full SIMP-assembly, eigensolve, and rigid-reduced
Nelson-adjoint pipeline. The check validates the gradient assembly itself, and
is independent of the larger demonstration meshes.

A subtle point
specific to modal sensitivity is that the verification cannot be performed
at the criterion-evaluation level alone. The DDAM criterion receives the eigenpair
$(\lambda_i, \bphi_i)$ from the eigensolve. With the eigenpair held fixed, the
explicit map $\bz \mapsto f$ inside the criterion does not depend on $\bz$. A
finite-difference probe that perturbs $\bz$ without repeating the eigensolve
therefore has an identically zero numerator and passes vacuously.

The correct probe routes the perturbation through the full implicit
eigensolve closure. For each perturbed design $\bz \pm h\, \mathbf{e}_e$, with $\mathbf{e}_e$ the
canonical basis vector on element $e$, the entire pipeline of
Algorithm~\ref{alg:morphorm_ddam} is re-executed. That covers SIMP-interpolated
assembly of $\K(\bz), \M(\bz)$, the eigensolve of~\eqref{eq:gep}, eigenvector
mass normalization, modal post-processing, and criterion evaluation. The fourth-order central
difference
\begin{equation}
  \label{eq:fd_4pt}
  \begin{aligned}
  \widehat{\frac{\partial f}{\partial z_e}}
  \;=\; \frac{1}{12\, h}\Bigl[\,
  &f(\bz - 2h\,\mathbf{e}_e) - 8\,f(\bz - h\,\mathbf{e}_e)\\
  &+ 8\,f(\bz + h\,\mathbf{e}_e) - f(\bz + 2h\,\mathbf{e}_e)\,\Bigr]
  \end{aligned}
\end{equation}
is compared against the analytic element entry assembled from~\eqref{eq:full_sens}, repeated over a logarithmic sweep
$h \in \{10^{-1}, 10^{-2}, \dots, 10^{-10}\}$. The expected outcome is the canonical V-shaped error
curve~\cite{nocedal2006numerical}:
truncation-dominated for large $h$, round-off-dominated for small $h$. The
minimum-error window at intermediate $h$ indicates the achievable agreement
between analytic and finite-difference estimates of the modal sensitivity.

An error in any intermediate stage shows up as an inflated or non-V-shaped
curve. Three such failures are eigensolve non-convergence on a
near-singular intermediate iterate, rigid-body-mode contamination of the
Nelson adjoint, and boundary-condition inconsistency between the eigensolver
and the Nelson solve, all discussed in Appendix~B. The FD sweep therefore also
checks the safeguards of that appendix.

A second subtlety governs the \emph{direction} of the finite-difference
probe. The difference~\eqref{eq:fd_4pt} perturbs one element density $z_e$
at a time, and this coordinate-wise probing is essential rather than
incidental. Under a spatially uniform density perturbation the
participation-mass and mode-shape contributions cancel,
$\sum_e T^{e}_{\mathrm{II}} + \sum_e T^{e}_{\mathrm{III}} = 0$, leaving only
the eigenvalue term $\sum_e T^{e}_{\mathrm{I}}$. An aggregate probe along
that uniform direction then compares two near-cancelling quantities and is
uninformative. A random direction need not cancel exactly, but can still mask
a localized error in $T^{e}_{\mathrm{II}}$ or $T^{e}_{\mathrm{III}}$.
Perturbing a single element at a time breaks the cancellation and tests
every assembled gradient entry independently.

Finally, the achievable floor reports agreement between the
\emph{discretized} analytic and finite-difference gradients. It is therefore
sensitive to any quadrature inconsistency between the criterion value and its
sensitivity. The mode-shape term $T^{e}_{\mathrm{III}}$ involves element
integrals such as $\int_e \rho\, |\bphi_i|^{2}\, dV$, whose integrand is of
higher polynomial degree than the stiffness integrals. If they are evaluated
with a Gauss rule that under-integrates that degree while the value functional
uses the consistent finite-element mass operator, the floor of the V-curve
rises by a fixed amount which refining $h$ cannot lower. Driving the floor to round-off requires that the sensitivity
integrals use the same quadrature, or equivalently the same assembled
consistent mass operator, as the value.

\begin{figure}[htbp]
  \centering
  \includegraphics[width=0.88\linewidth]{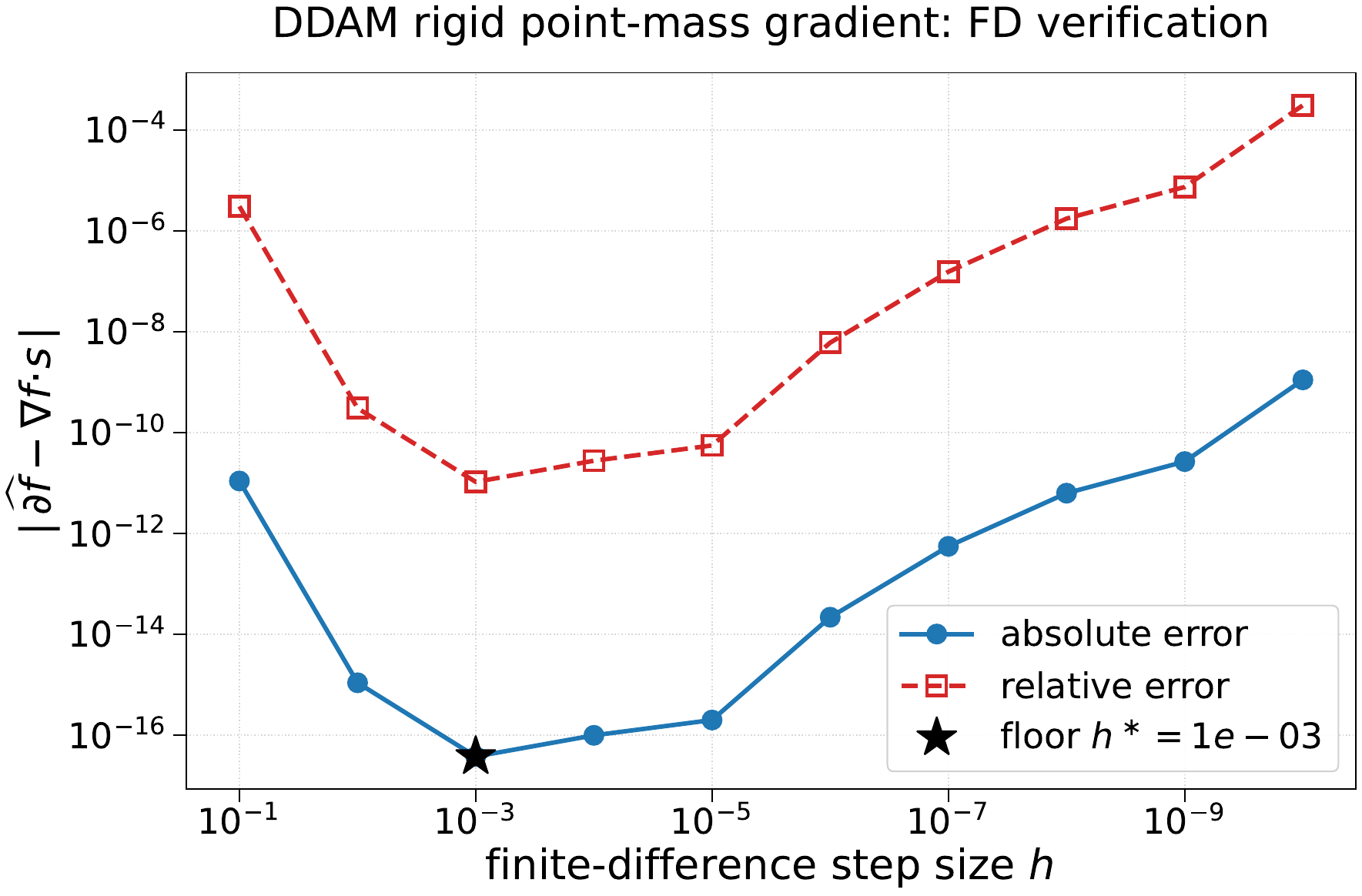}
  \caption{Finite-difference verification of the rigid point-mass DDAM
  gradient on the clamped square domain carrying a rigid lumped-mass attachment. The
  absolute and relative errors between the four-point central
  difference~\eqref{eq:fd_4pt} and the analytic sensitivity~\eqref{eq:full_sens}
  are swept over $h\in\{10^{-1},\dots,10^{-10}\}$, for one representative
  perturbation direction. The curve traces the canonical V:
  $O(h^{4})$ truncation-dominated at large $h$, $O(\varepsilon/h)$
  round-off-dominated at small $h$, with the floor at $h^{*}=10^{-3}$.}
  \label{fig:fd_vcurve}
\end{figure}

Figure~\ref{fig:fd_vcurve} shows the sweep for the rigid point-mass DDAM
gradient. The error reaches its floor at $h^{*}=10^{-3}$. There the estimate and the analytic gradient agree to a relative error of
$1.0\times10^{-11}$, absolute $3.8\times10^{-17}$. That is the round-off floor
of the discretized computation. The error rises monotonically on either side of
$h^{*}$, as $O(h^{4})$ truncation for $h>h^{*}$ and as $O(\varepsilon/h)$
round-off for $h<h^{*}$. The absence of a flat plateau above machine precision confirms that the
assembled gradient carries no systematic bias. In particular,
the rigid attachment's eigenvector sensitivity is correctly evaluated in the
constraint-reduced space of Section~\ref{subsec:rigid_sens} (a full-space Nelson
adjoint would inflate the floor).

\subsection{Demonstration~1: equipment frame, fore-and-aft interface displacement}
\label{subsec:demo1}

\begin{figure}[htbp]
  \centering
  \includegraphics[width=0.511\linewidth]{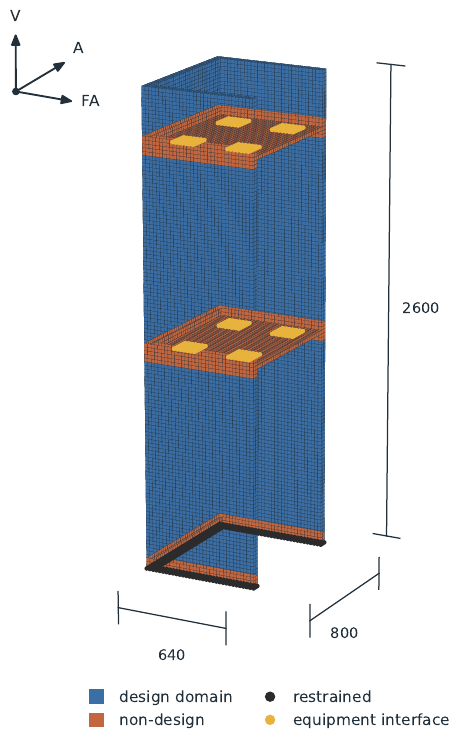}
  \caption{The model, shared by all three demonstrations. Two side frames and a
  back panel form the design domain (dark). The shelves, the collars where they
  meet the frames, and a skin at the bolted footprint are held non-design
  (light). Each shelf carries its equipment on four pads through a rigid
  attachment, and the footprint is restrained at the deck. The axes at top
  left give the three shock directions: vertical (V), athwartship (A) and
  fore-and-aft (FA). One model is
  certified three times, so what separates the demonstrations is the criterion
  and the directions it combines, nothing else. Dimensions in \si{mm}; the
  \SI{640}{mm} is the design depth, the cabinet envelope being \SI{900}{mm}
  deep with the front \SI{260}{mm} left open for access.}
  \label{fig:setup_cabinet}
\end{figure}

\paragraph{Setup}
Figure~\ref{fig:setup_cabinet} shows the model. It is the same equipment frame
all three demonstrations use, and the same one Section~\ref{subsec:compliance}
redesigns by a conventional criterion. Two side frames and a back panel form
the design domain. The shelves, the collars where they meet the frames, and a
skin at the bolted footprint are non-design. Each of the two shelves carries a
\SI{2700}{kg} unit on four pads, tied by the rigid attachment of
Section~\ref{subsec:topo}. The footprint is restrained at the deck. The
frame, its provenance, its material and its allowable are described in full
with Demonstration~2 (Section~\ref{subsec:demo2}).

The certifying shock here is a single \emph{fore-and-aft} UNDEX event, and the
certified quantity is the interface displacement rather than a stress. The
influence vector $\mathbf{D}$ in~\eqref{eq:participation} selects the
fore-and-aft translational degrees of freedom. The response is read over the
upper shelf's mounting pads by the area-average
operator~\eqref{eq:surface_weights}.

That reading deserves a word, because the alternative is available and wrong.
A response taken as a raw sum over the interface nodes has no continuum limit,
as Section~\ref{subsec:srss} notes. On this mesh it differs from the area
average by a factor of \num{14}. The rigid attachment makes the response
uniform across the pads, so the area average and the pointwise value agree.
That does not make the nodal sum agree with either.

\paragraph{Spectrum and criterion}
The NRL-1396 surface-ship, \emph{deck-mounted}, elastic spectrum is used with
the fore-and-aft design coefficient. For this category the unclassified
design-value reference
equations~\cite{nrl1396} are
\begin{equation}
  \label{eq:nrl1396_A0V0}
  \begin{aligned}
  A_0 &= 10\,\frac{(37.5 + W_a)(12 + W_a)}{(6 + W_a)^2}\;[\mathrm{g}],\\
  V_0 &= 30\,\frac{(12 + W_a)}{(6 + W_a)}\;[\text{in/s}],
  \end{aligned}
\end{equation}
with $W_a$ the effective weight in kips, evaluated at the weight of the
supported equipment as stated at the start of this section. The directional
elastic design coefficients that scale $A_0$ and $V_0$ into $A_{a,i}$ and
$V_{a,i}$ are taken as unity in every direction;
Section~\ref{subsec:limits} (Design coefficients) gives the reason and what
the choice controls.

DDAM is used as a \emph{constraint}. We minimize volume subject to two constraints. The NRL Sum of the fore-and-aft
interface response must not exceed an allowable, and the frequency floor of
Section~\ref{subsec:topo} applies,
\begin{equation}
  \label{eq:demo1_problem}
  \begin{aligned}
  &\underset{\bz}{\text{minimize}}\;\; v(\bz)
  \quad\text{s.t.}\quad
  f_{\mathrm{NRL}}(\Lambda(\bz),\Phi(\bz)) \le R_{\mathrm{allow}},\\
  &\lambda_1(\bz) \ge \omega_{\min}^{2},
  \;\; z_{\min}\le z_e \le 1,
  \end{aligned}
\end{equation}
with the interface response measured by the area-average
operator~\eqref{eq:surface_weights} over the upper shelf's pads. The allowable is
$R_{\mathrm{allow}} = \SI{6.3651}{mm}$, which is $1.2$ times the response of the
fully solid domain. Setting it this way places the constraint inactive at the starting point and
active at the optimum. That is what a demonstration of the sensitivities
requires. It is a modeling choice for this problem rather than a qualification
value for a particular component. The floor is \SI{25}{Hz}, that is,
$\omega_{\min} = 2\pi \times \SI{25}{Hz} \approx \SI{157.1}{rad/s}$, the same
as in Demonstrations~2 and~3 and below the \SI{37.87}{Hz} fundamental of the
solid domain.

\paragraph{Discretization}
The domain is meshed with trilinear hexahedra (HEX8) of nominal size
$h=\SI{20}{mm}$. It has \num{28160} elements and
\num{43893} nodes, or \num{131679} displacement degrees of freedom. The
shelves, their collars and a skin at the bolted footprint are held non-design,
which protects the restraint and the two mass attachments from erosion. The
mesh, the restraint and the attachments are those of
Demonstrations~2 and~3 without modification.

The baseline is the fully solid domain, whose fundamental frequency is
\SI{37.8743}{Hz} and whose fore-and-aft interface NRL Sum is
\SI{5.3042}{mm}. Mode retention follows the rule stated at the start of this section, as in
the other two demonstrations. Six modes are solved and the count
raised toward the \SI{80}{\percent} coverage target. Every solved mode is then
retained, subject to the participation floor. The fundamental carries
\SI{80.0}{\percent} of the fore-and-aft effective mass at the solid baseline.

\begin{figure}[htbp]
  \centering
  \includegraphics[width=0.773\linewidth]{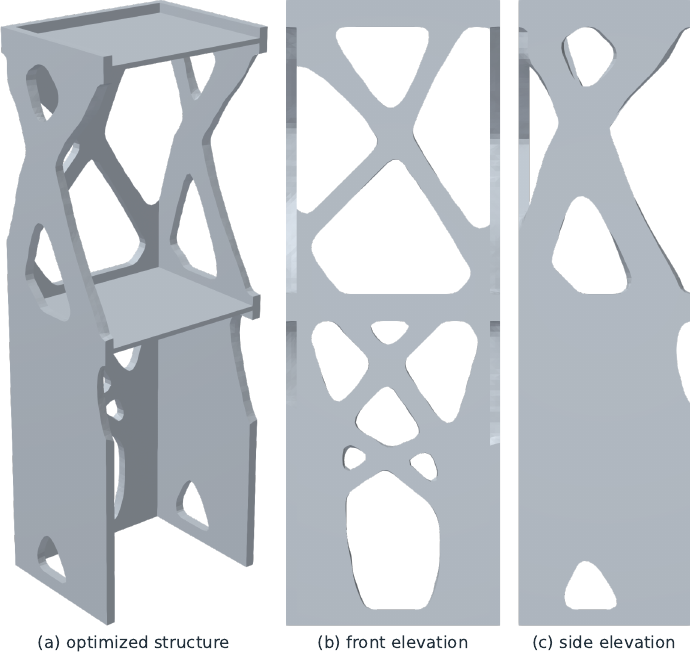}
  \caption{Demonstration~1 optimized structure at iteration~199, shown at the
  \num{0.5} level of the density field. (a)~The full frame, (b)~the front
  elevation and (c)~the side elevation. Camera and scale match
  Figures~\ref{fig:demo2_topology} and~\ref{fig:demo3_topology}, so the three
  designs compare directly.}
  \label{fig:demo1_topology}
\end{figure}

\paragraph{Results}
Figure~\ref{fig:demo1_topology} shows the optimized structure at
iteration~199, the last iteration satisfying both constraints.
Table~\ref{tab:demo1_results} compares it with the solid baseline. The
optimizer removes \SI{43.2}{\percent} of the material. The interface response
rises from \SI{5.3042}{mm} to the allowable and meets it to within
\SI{0.0004}{\percent}. The fundamental settles \SI{0.0024}{\percent} above the
\SI{25}{Hz} floor. Both constraints are therefore active at the optimum, as
they are in Demonstrations~2 and~3 on the same frame.

\begin{figure}[htbp]
  \centering
  \includegraphics[width=0.689\linewidth]{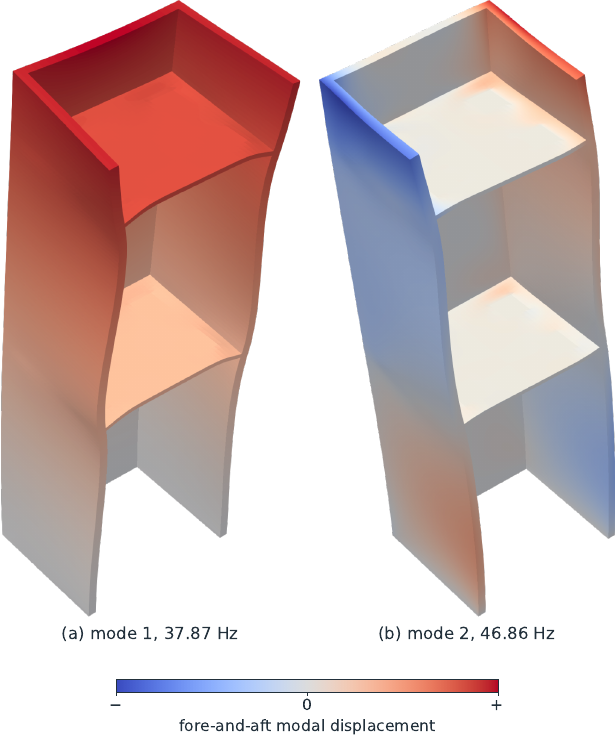}
  \caption{The two lowest modes of the solid frame shared by all three
  demonstrations. They are drawn at equal peak deflection and colored by
  fore-and-aft modal displacement, each panel on its own scale. (a)~is symmetric
  about the plane containing the shock and carries the response,
  $\Gamma = \num{2.3945}$. (b)~is antisymmetric, the two side frames moving in
  opposite directions, and carries none of it: $\Gamma = \num{-0.0000}$ to four
  decimals, as parity requires. What (b) does carry is the athwartship
  response, $\Gamma_{\mathrm{A}} = \num{2.4481}$, which
  Section~\ref{subsec:demo3} needs. The two modes
  are a fore-and-aft and an athwartship translation respectively. The pair is
  \SI{8.99}{Hz} apart. Compare the optimized pairs of
  Figures~\ref{fig:demo1_mode_optimized}, \ref{fig:demo2_mode_optimized}
  and~\ref{fig:demo3_mode_optimized}.}
  \label{fig:frame_mode_baseline}
\end{figure}

\begin{table}[htbp]
  \centering
  \small
  \caption{Demonstration~1. The optimized column is the design at
  iteration~199, the last iteration satisfying both constraints. The limit
  column gives the allowable $R_{\mathrm{allow}}$, set at \num{1.2} times the
  solid response, and the frequency floor.}
  \label{tab:demo1_results}
  \begin{adjustbox}{max width=\linewidth}
\begin{tabular}{lccc}
    \toprule
     & Solid baseline & Optimized & Limit \\
    \midrule
    Volume fraction                                   & 1.000  & 0.568   & \\
    Fore-and-aft interface NRL Sum, \si{mm}           & 5.3042 & 6.3651  & 6.3651 \\
    Fundamental frequency, \si{Hz}                    & 37.87  & 25.0006 & 25.00 \\
    Governing mode, fore-and-aft mass, \si{\percent}  & 80.0   & 68.2    & \\
    \bottomrule
  \end{tabular}
\end{adjustbox}
\end{table}

\begin{figure}[htbp]
  \centering
  \includegraphics[width=0.689\linewidth]{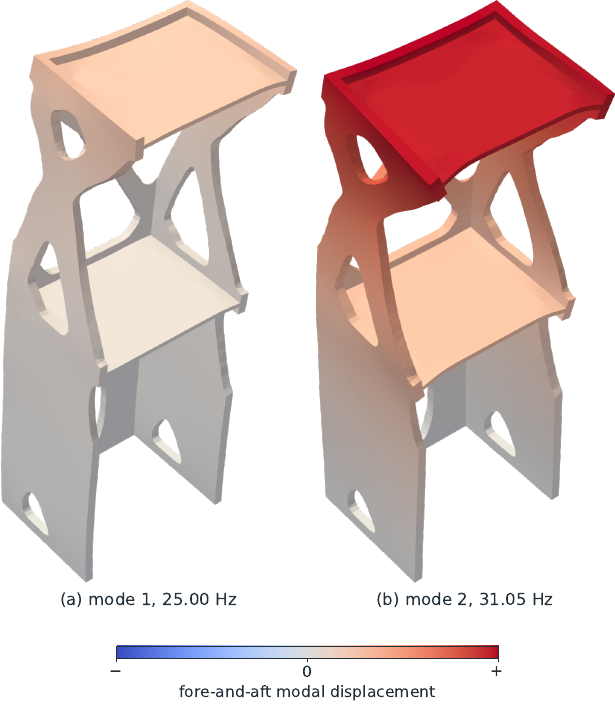}
  \caption{The first two modes of the Demonstration~1 optimum, drawn on the
  deformed shape. (a)~Mode~1, at \SI{25.00}{Hz}, holds the frequency floor.
  Its displacement is \SI{95}{\percent} athwartship, and it carries
  \SI{3.0}{\percent} of the fore-and-aft effective mass. (b)~Mode~2, at
  \SI{31.05}{Hz}, carries \SI{68.2}{\percent} and governs the certified
  response. Both panels share one color scale for the fore-and-aft component,
  so the color compares fore-and-aft content between them. Each deflection is
  drawn to the same visible amplitude, because mass-normalized amplitudes carry
  no physical meaning. The modes of the solid frame are those of
  Figure~\ref{fig:frame_mode_baseline}.}
  \label{fig:demo1_mode_optimized}
\end{figure}

Why both constraints bind is visible in the modes. Table~\ref{tab:demo1_modal}
gives the six modes of the basis at both ends of the run. At the solid baseline
the fundamental carries \SI{80.0}{\percent} of the fore-and-aft effective mass.
Figure~\ref{fig:frame_mode_baseline} shows it beside the second mode, an
athwartship translation. At the optimum the fundamental carries
\SI{3.0}{\percent}. It has become an athwartship sway like the second mode of the solid frame,
its displacement \SI{95}{\percent} athwartship. The
certified response is carried instead by the second mode, at \SI{31.05}{Hz},
with \SI{68.2}{\percent} of the fore-and-aft mass.
Figure~\ref{fig:demo1_mode_optimized} draws the two modes. The floor is
therefore held by a mode the certified direction barely sees. The optimizer
cannot trade one constraint against the other through a single mode, and both
remain active.

\begin{table}[htbp]
  \centering
  \small
  \caption{Demonstration~1, fore-and-aft (FA) effective modal mass as a percentage
  of the total, for the six modes of the basis. The largest entry in each
  column is set in bold. On the solid frame the fundamental carries the
  fore-and-aft direction. At the optimum it carries almost none of it: the
  fundamental is an athwartship sway, and the second mode carries the
  certified response.}
  \label{tab:demo1_modal}
  \begin{tabular}{crrcrr}
    \toprule
    & \multicolumn{2}{c}{Solid baseline} & & \multicolumn{2}{c}{Optimized} \\
    \cmidrule{2-3}\cmidrule{5-6}
    Mode & $f$ [\si{Hz}] & FA & & $f$ [\si{Hz}] & FA \\
    \midrule
    1 & 37.87  & \textbf{80.0} & & 25.00  & 3.0  \\
    2 & 46.87  & 0.0           & & 31.05  & \textbf{68.2} \\
    3 & 135.22 & 0.0           & & 63.68  & 0.8  \\
    4 & 138.64 & 0.9           & & 79.63  & 1.9  \\
    5 & 148.87 & 13.9          & & 101.90 & 21.9 \\
    6 & 153.22 & 1.0           & & 129.86 & 0.3  \\
    \bottomrule
  \end{tabular}
\end{table}

\begin{figure}[htbp]
  \centering
  \includegraphics[width=0.82\linewidth]{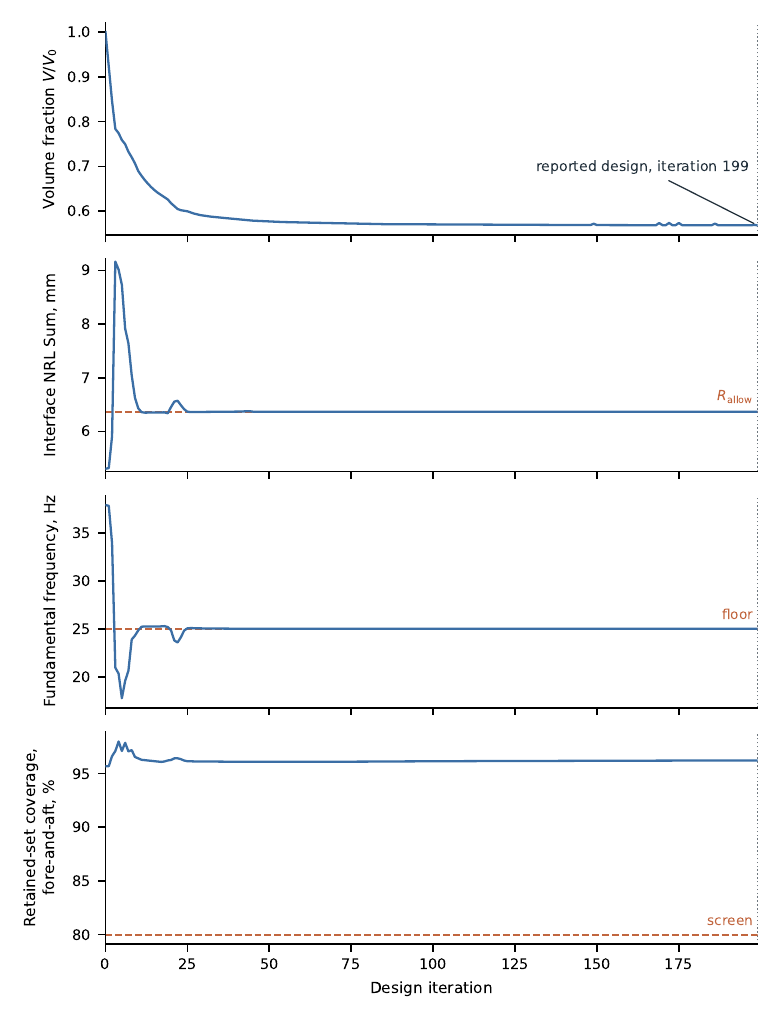}
  \caption{Demonstration~1 convergence history. Dashed lines mark the
  allowable $R_{\mathrm{allow}}$, the frequency floor and the
  \SI{80}{\percent} coverage target. The dotted vertical line marks the
  reported design at iteration~199. Both constraints are violated in the
  transient around the mode crossing at iteration~3, and both are active from
  about iteration~24 onward. The frequency panel plots the fundamental, whose
  identity changes at the crossing: fore-and-aft sway before it, athwartship
  sway after.}
  \label{fig:demo1_convergence}
\end{figure}

The two modes cross in a single step early in the run, between design
iterations~2 and~3. The fundamental falls from \SI{33.91}{Hz} to
\SI{20.96}{Hz}, and its fore-and-aft effective mass from \SI{81.6}{\percent}
to zero. Figure~\ref{fig:demo1_convergence} shows the transient around the
crossing. The response peaks at \SI{9.16}{mm} at iteration~3,
\SI{44}{\percent} over the allowable. The fundamental reaches its lowest value,
\SI{17.78}{Hz}, at iteration~5. Both are within \SI{1}{\percent} of their
limits from iteration~24 onward. The objective changes by
\SI{0.062}{\percent} over the final forty iterations. At the optimum the
retained set covers \SI{96.2}{\percent} of the fore-and-aft effective mass,
against the \SI{80}{\percent} target.

\paragraph{Mass-loaded spectrum}
\label{subsec:demo1_mass_loaded}
A variant of Demonstration~1 lifts the fixed reference weight. The problem is
unchanged except for the spectrum. Each mode now sees the design values
of~\eqref{eq:nrl1396_A0V0} at its own effective weight, $W_{a,i} =
\Gamma_i^2$, converted to kips. The gradient then carries the second term
of~\eqref{eq:dSD_dz_full}. The allowable stays at \SI{6.3651}{mm}, so both runs
answer the same requirement.

The solid baseline response rises from \SI{5.3042}{mm} to \SI{5.3637}{mm}. The
fundamental carries \SI{5730}{kg} of effective mass against the
\SI{5400}{kg} reference, the two \SI{2700}{kg} units, so its shock is slightly
milder. The higher modes carry less than \SI{1000}{kg} each and see up to 3.3 times the reference acceleration. Their added
response outweighs the relief on the fundamental.

We report the last iterate, iteration~200. The optimizer removes
\SI{39.6}{\percent} of the material, against \SI{43.2}{\percent} with the
fixed weight. The response meets the allowable to within \SI{0.002}{\percent},
and the fundamental sits \SI{0.009}{\percent} above the floor. Both constraints
are again active. The heavier optimum follows from the sign of the coupling.
Removing material lowers the effective weight of the mode that carries the
response, and a lighter mode sees a larger design acceleration. The allowable is therefore reached with
more material in place.

\begin{figure}[htbp]
  \centering
  \includegraphics[width=0.773\linewidth]{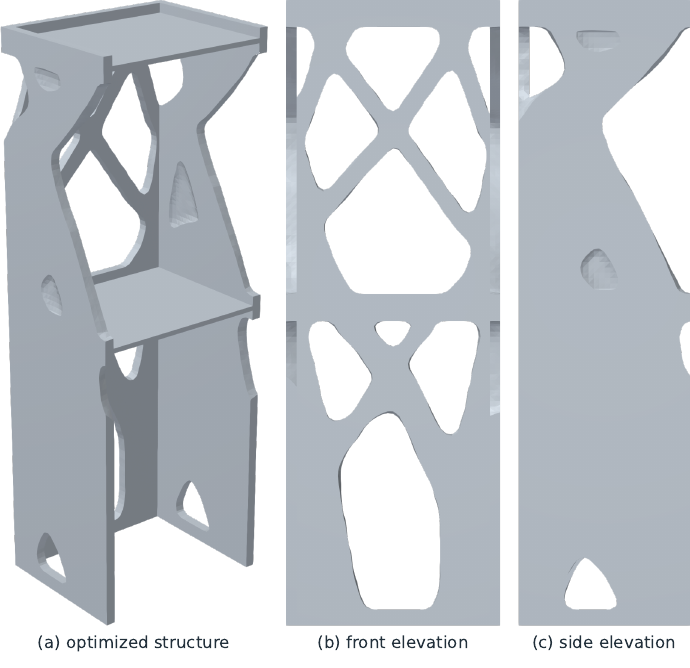}
  \caption{Demonstration~1 with the effective weight taken mode by mode, the
  optimized structure at iteration~200, shown at the \num{0.5} level of the
  density field. (a)~The full frame, (b)~the front elevation and (c)~the side
  elevation. Camera and scale match Figure~\ref{fig:demo1_topology}, the same
  problem with the fixed reference weight. The mass-loaded design keeps
  \SI{60.4}{\percent} of the solid volume, against \SI{56.8}{\percent},
  and the difference sits mainly in the side frames.}
  \label{fig:demo1b_topology}
\end{figure}

Figure~\ref{fig:demo1b_topology} shows the mass-loaded design, drawn exactly
as Figure~\ref{fig:demo1_topology}. The extra material stays in the side
frames, which carry the fore-and-aft response. Their large openings close, and
only a few small holes remain. In the front elevation the upper-bay bracing remains,
and the lower bay's small openings give way to fewer, larger ones.

\subsection{Demonstration~2: equipment frame, fore-and-aft stress}
\label{subsec:demo2}

\paragraph{Setup}
The model is the one Section~\ref{subsec:demo1} introduces and
Figure~\ref{fig:setup_cabinet} shows. It is a hard-mounted equipment frame of
the kind used to carry shipboard electronic and control units. Demonstration~2
changes one thing about it: the certified quantity becomes a stress rather
than the interface displacement. The configuration follows the four-leg, three-shelf assembly published as a
DDAM benchmark by the Naval Surface Warfare Center~\cite{koehler2018ddam}. Its
peak response under fore-and-aft shock is reported at or above the yield of
a mild structural steel. That is the property this problem is chosen for. The allowable is genuinely
active, rather than a margin so large that the design is unconstrained.

The published model~\cite{koehler2018ddam} is built from shells with welded
connections. The present formulation is a displacement-only continuum discretization, so
the frame is reinterpreted rather than reproduced. The equipment layout, the
supported masses, the mounting class and the allowable are taken from the
reference assembly. The leg and bracing region is treated as a solid design domain in which the
optimizer places material. The shelves and the mounting interfaces are
non-design.

The shelf count and heights are part of what is reinterpreted rather than
carried over. They were chosen from modal solves of the solid frame. The aim
was a fore-and-aft fundamental below the crossover frequency
$\omega^{*} = A_a/V_a$ of the deck-mounted spectrum, where its acceleration-
and velocity-limited branches meet. Below the crossover the mode lies on the
velocity-limited branch of~\eqref{eq:Da_branches}, where the design
acceleration varies with frequency. The frequency-dependent terms of the
sensitivity are then active rather than zero.

Six configurations were compared, varying the shelf count, the shelf heights
and the envelope height. The fundamental moved with the height of the
supported mass, not with the number of shelves; two evenly spaced shelves were
even stiffer athwartship than three. The configuration adopted has two
shelves, at \SI{1190}{mm} and \SI{2290}{mm}, in a \SI{2.6}{m} envelope. It
places the fore-and-aft fundamental at \num{0.69} of the crossover frequency.
A taller envelope would lower the fundamental further, but would not be a
plausible shipboard cabinet. Because the shelf layout and the discretization both depart from the reference
assembly, the reported values are not comparable with the reference model,
and no agreement with it is claimed.

The frame occupies an envelope of
$\SI{800}{mm}\times\SI{900}{mm}\times\SI{2600}{mm}$ in the athwartship,
fore-and-aft and vertical directions. Two side frames and a back panel form the design domain, each \SI{40}{mm}
thick. The side frames extend \SI{640}{mm} of the depth, so that the front
\SI{260}{mm} remains open for access. Two shelves at \SI{1190}{mm} and \SI{2290}{mm} each carry
\SI{2700}{kg} of equipment on four pads, tied to the pads by the rigid
attachment of Section~\ref{subsec:topo}.

The material is a mild structural steel, $E = \SI{200}{GPa}$, $\nu = 0.30$ and
$\rho = \SI{7850}{kg/m^3}$, with minimum specified yield
$\sigma_y = \SI{235}{MPa}$ supplying the allowable. The allowable is taken from the published assembly the configuration is drawn
from~\cite{koehler2018ddam}, as Section~\ref{sec:limitations} records. The elastic properties are
nominal for the class rather than certified for a particular heat. No standard grade
designation is claimed. Minimum specified yield falls with section thickness in the structural steel
standards~\cite{en10025_2}. An allowable tied to a certified grade at the \SI{40}{mm} frame
thickness would therefore sit below the nominal value used here.
What the demonstrations exercise is the formulation, not a qualification.

The certifying shock for this demonstration is a single
\emph{fore-and-aft} event. The choice of direction is dictated by the structure rather than by the
spectrum. With the unit design coefficient of Section~\ref{subsec:limits} (Design coefficients),
every direction sees the same reference spectrum. A modal analysis of the frame places its vertical response in a regime where
the method has nothing to demonstrate. The mode carrying the vertical mass is a bending mode of the shelves rather
than of the frame. It sits at a frequency well above the branch switch
of~\eqref{eq:Da_branches}. There the design value
reduces to the acceleration-limited $A_a(W_a)$ and no longer depends on
frequency. The frequency-dependent terms of the
sensitivity derived in Section~\ref{sec:sensitivity} are then dormant. The fore-and-aft response of the same frame lies on the velocity branch, where
those terms are active. It is also the direction in which the reference
assembly reports its peak. Section~\ref{sec:limitations} returns to this
point.

\paragraph{Spectrum and criterion}
The NRL-1396 surface-ship, \emph{deck-mounted}, elastic spectra
of~\eqref{eq:nrl1396_A0V0} are used with the fore-and-aft design coefficient. The
certifying quantity is the pointwise NRL Sum~\eqref{eq:nrl_field} of the von~Mises modal
stress~\eqref{eq:Ri_stress}. It is formed mode by mode and then combined
across modes, which is the order the certification procedure
prescribes~\cite{navsea_t9070}. The pointwise NRL Sum field is then reduced to a single scalar by a
measure-weighted aggregate over the design domain, which is the region
$\Omega_c$ here: the side frames and the back panel, without the non-design
shelves, collars and footprint skin. The reported quantity then converges
under mesh refinement.

The measure weighting is what secures that convergence. It carries the price
Section~\ref{subsec:nrl_stress} derives: the aggregate approaches the peak from
below rather than bounding it from above. The constraint below bounds the
aggregate, and the peak of the field may lie above the allowable.

DDAM is used as a \emph{constraint}: minimize volume subject to the
fore-and-aft NRL Sum stress not exceeding the elastic allowable,
\begin{equation}
  \label{eq:demo2_problem}
  \begin{aligned}
  &\underset{\bz}{\text{minimize}}\;\; v(\bz)
  \quad\text{s.t.}\quad
  f_{\mathrm{NRL}}^{(\mathrm{FA})}(\bz) \le \sigma_{\mathrm{allow}},\\
  &\lambda_1(\bz) \ge \omega_{\min}^{2},
  \;\; z_{\min}\le z_e \le 1,
  \end{aligned}
\end{equation}
with $\sigma_{\mathrm{allow}} = \sigma_y$. A floor of \SI{25}{Hz},
$\omega_{\min} \approx \SI{157.1}{rad/s}$, is imposed on the fundamental frequency,
below the \SI{37.87}{Hz} of the solid domain and the same as in
Demonstration~1. The floor is a connectivity
guard~\cite{liShiXia2017,coolAageSigmund2025,bonaccorsi2025connectivity}
rather than a qualification value: it prevents the optimizer from reaching a
disconnected design whose lowest modes are local to a fragment. A guard need not
bind, but on this frame it does: at the optimum, the floor and the shock
criterion are both active, as the results below show.

\paragraph{Discretization}
The domain is meshed with trilinear hexahedra (HEX8) of nominal size
$h=\SI{20}{mm}$. This gives \num{28160} elements, \num{43893} nodes and
\num{131679} displacement degrees of freedom. Three regions are held non-design, so that the optimizer cannot erode a
boundary condition or a load introduction. They are a \SI{60}{mm} skin at the
bolted footprint, a \SI{50}{mm} collar in the frames on either side of each
shelf, and the shelves themselves. The
remaining \SI{1.87e8}{mm^3} is design domain, whose solid mass is \num{0.27}
of the supported equipment mass.

\begin{figure}[htbp]
  \centering
  \includegraphics[width=0.773\linewidth]{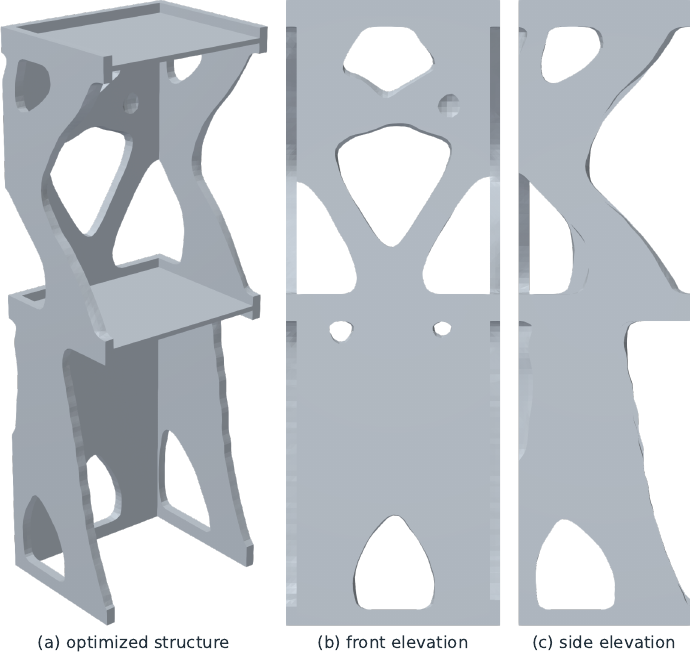}
  \caption{Demonstration~2 optimized structure at iteration~174, shown at the
  \num{0.5} level of the density field. The non-design shelves, collars and
  footprint skin are drawn with the design domain, so the structure is seen as
  it is supported. (a)~The full frame, (b)~the front elevation, looking
  along the fore-and-aft shock direction, and (c)~the side elevation, looking
  along the athwartship axis. The elevations are drawn without perspective and
  at one scale. Material is
  concentrated low, where the shear from the two supported masses is largest.
  The back web is solid across its full width in the lower bay. It is retained
  over about half of it in the upper, \num{1.000} against \num{0.488} by mean
  density.}
  \label{fig:demo2_topology}
\end{figure}

\begin{figure}[htbp]
  \centering
  \includegraphics[width=0.689\linewidth]{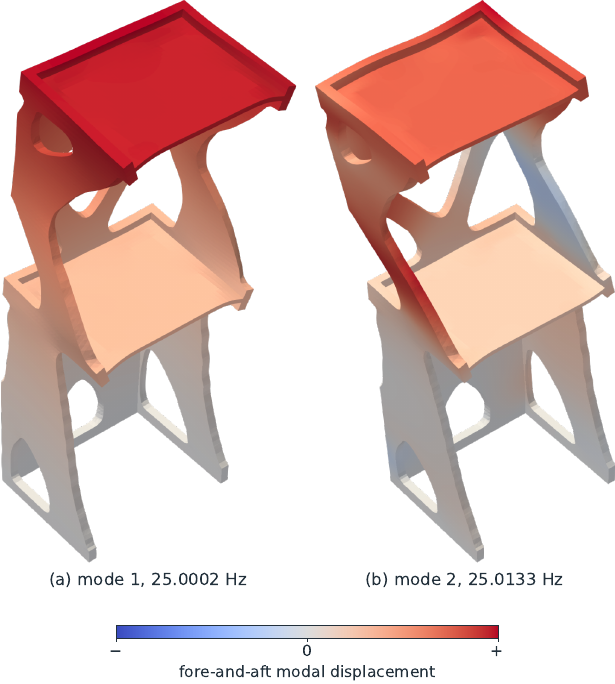}
  \caption{The same two modes of the optimized design, drawn and colored as in
  Figure~\ref{fig:frame_mode_baseline}. Both have been driven onto the
  frequency floor, and are now \SI{0.013}{Hz} apart rather than \SI{8.99}{Hz}.
  The mode that is antisymmetric on the solid frame no longer carries nothing,
  $\Gamma = \num{0.1601}$ against \num{-0.0000} there. The near
  degeneracy is produced by the optimization rather than inherited from the
  geometry. It is why the retained basis is unstable between design
  iterations.}
  \label{fig:demo2_mode_optimized}
\end{figure}

\begin{figure}[htbp]
  \centering
  \includegraphics[width=0.82\linewidth]{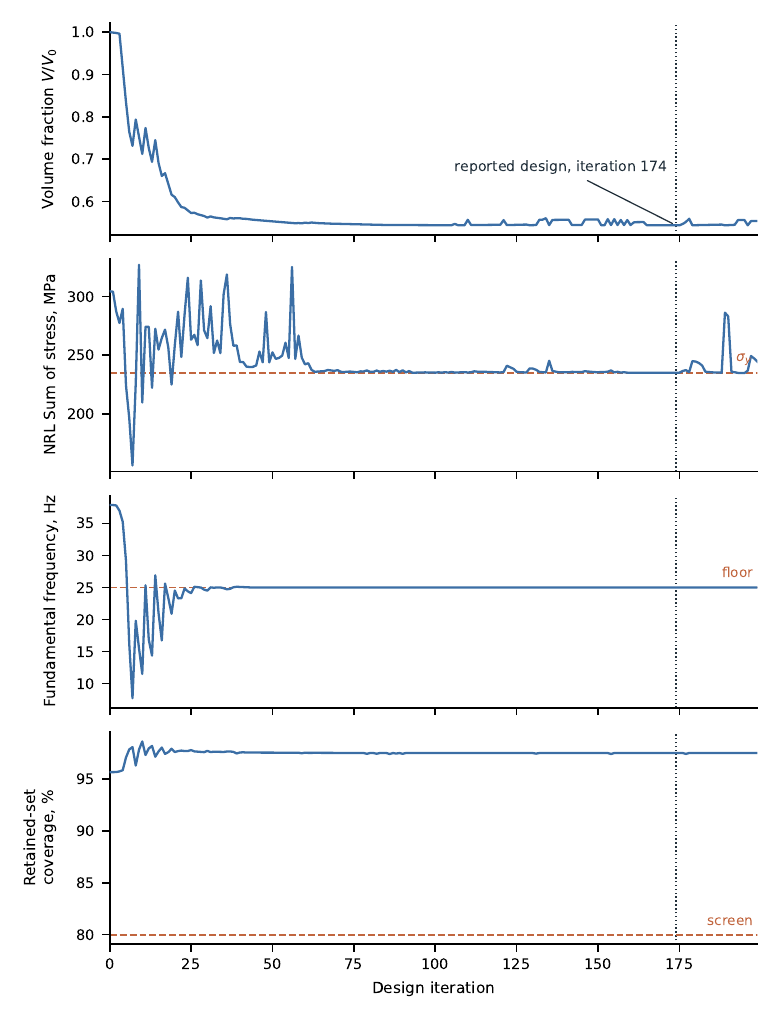}
  \caption{Demonstration~2 convergence. The dotted line marks iteration~174,
  the reported design. The response settles onto the allowable and then
  oscillates across it rather than converging onto it. The fundamental locks
  onto the floor after an excursion to \SI{7.7}{Hz} at iteration~7, from which
  the optimizer recovers unaided. The retained set holds its coverage near
  \SI{97.5}{\percent} throughout, well clear of the \SI{80}{\percent} target.
  The reported response is therefore not an artifact of a thinning modal
  basis.}
  \label{fig:demo2_convergence}
\end{figure}

\paragraph{Results}
Table~\ref{tab:demo2_results} reports the lightest design that satisfies both
constraints, at iteration~174, and Figure~\ref{fig:demo2_topology} the structure
it arrived at. Figures~\ref{fig:frame_mode_baseline}
and~\ref{fig:demo2_mode_optimized} show the mode pair that governs it before
and after, and Figure~\ref{fig:demo2_convergence} the history that produced
it. The optimizer removes \SI{45.6}{\percent} of the
material, driving the aggregated response from \SI{304.44}{MPa} on the solid
baseline to the allowable. It meets that allowable to within
\SI{0.011}{\percent}.

Both constraints are active at the optimum, as in Demonstration~1. The response
sits at \SI{234.975}{MPa} against \SI{235}{MPa} and the fundamental at
\SI{25.0002}{Hz} against a floor of \SI{25}{Hz}. Both constraints are active, so the design is determined by the two
together.

The optimizer does not settle onto that point smoothly, as
Figure~\ref{fig:demo2_convergence} shows. It approaches the feasible boundary and then oscillates across it. The reported design is the lightest feasible iterate. Iterates~193 to~195 return
to the feasible side later, carrying more material.

The cause is visible in Figures~\ref{fig:frame_mode_baseline}
and~\ref{fig:demo2_mode_optimized}. The two lowest modes lie \SI{8.99}{Hz}
apart on the solid frame, and are driven together onto the floor until they
lie \SI{0.013}{Hz} apart. Symmetric and antisymmetric on the solid frame,
they are not separated by the response either, since parity gives the
antisymmetric member no fore-and-aft participation there. The basis the criterion is formed
from is therefore not stable between design iterations.

At that separation the pair is numerically repeated, which is not a regime the
present sensitivity treatment covers. Section~\ref{subsec:crossings} (Mode crossings) states
what that costs and what would resolve it. The frequency history also records
an excursion to \SI{7.7}{Hz} early in the run, from which the optimizer
recovers unaided. That is the configuration the floor exists to guard
against.

The structure that results is a channel: two side frames carrying the shelf
loads into the footprint, closed at the back by a web. The closure is not
uniform. Measured as a mean density over the design band, the web is solid across its
full width in the lower bay. It is retained over about half of it in the upper bay,
\num{1.000} against \num{0.488}. Figure~\ref{fig:demo2_topology}(b) shows that
as a solid lower panel thinning into a pair of flanges above.

\begin{figure}[htbp]
  \centering
  \includegraphics[width=0.689\linewidth]{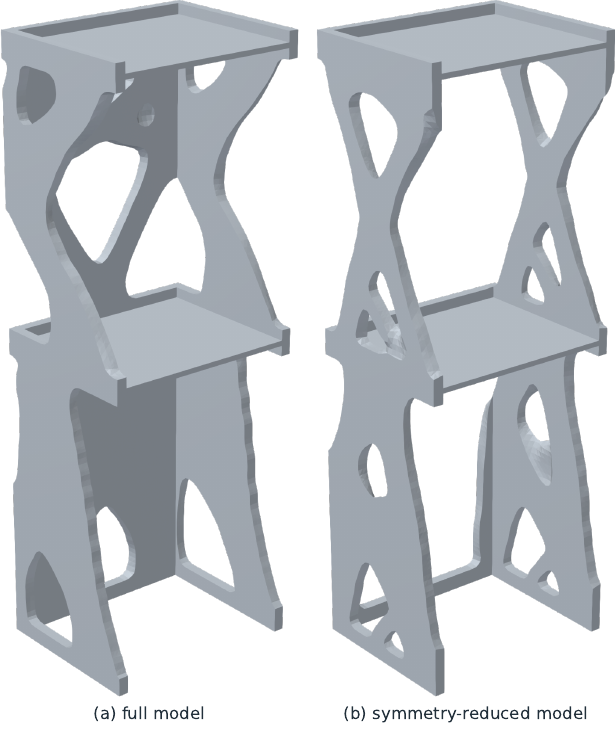}
  \caption{The effect of halving the model on its plane of symmetry.
  (a)~is the design of Table~\ref{tab:demo2_results}, $v/v_0 = \num{0.544}$.
  (b)~is obtained on the reduced model, $v/v_0 = \num{0.425}$, and is shown
  reflected; both panels are at one scale and one viewpoint. The reduced model
  removes the back web almost entirely, retaining a mean density of \num{0.116}
  there against \num{0.751}. That design is \SI{21.8}{\percent} lighter and is
  not admissible: evaluated on the unreduced model its fundamental is
  \SI{3.70}{Hz} against a floor of \SI{25}{Hz}.}
  \label{fig:demo2_comparison}
\end{figure}

The web is not carrying the fore-and-aft response, which the side frames take
in bending. It is holding the torsional modes above the frequency floor, and
the demand for torsional stiffness follows the shear.
Figure~\ref{fig:demo2_comparison} shows what the design becomes when those
modes are removed from the model. The paragraph below shows why that design
cannot be used.

\begin{table}[!htbp]
\begin{center}
\small
\caption{Demonstration~2. The optimized column is the design at
iteration~174, the lightest iteration satisfying both constraints. Both limits
are attained, as in Demonstration~1: the response to within
\SI{0.011}{\percent} and the frequency floor to within \SI{0.001}{\percent}.}
\label{tab:demo2_results}
\begin{adjustbox}{max width=\linewidth}
\begin{tabular}{lccc}
\toprule
 & Solid baseline & Optimized & Limit \\
\midrule
Volume fraction & 1.000 & 0.544 & \\
Fore-and-aft NRL Sum of stress, \si{MPa} & 304.44 & 234.975 & 235.00 \\
Fundamental frequency, \si{Hz} & 37.87 & 25.0002 & 25.00 \\
\bottomrule
\end{tabular}
\end{adjustbox}
\end{center}
\end{table}

\paragraph{Symmetry reduction}
The frame is mirror-symmetric about the plane containing the shock direction,
and the loading and the supported masses are symmetric with it. Halving the
model and imposing a symmetry condition on the cut plane is therefore the
standard economy~\cite{cookMalkusPleshaWitt2002}. Here it halves the degrees of freedom.

For the certifying criterion the reduction is exact. Write $y$ for the
coordinate normal to the plane of symmetry. A mode antisymmetric about that plane has a fore-and-aft displacement
component odd in $y$. The influence vector $\mathbf{D}$ is even in $y$, so the
participation factor $\Gamma_i$ of~\eqref{eq:participation} integrates to
zero. Antisymmetric modes therefore carry no fore-and-aft effective mass and
cannot enter the NRL Sum~\eqref{eq:nrl_sum}. They would never be retained
by a selection on effective mass. Discarding them costs the shock criterion nothing.

The frequency constraint is a different quantity. It bounds the lowest eigenvalue of the structure, irrespective of the mode
shape that attains it. It is therefore not filtered by the parity argument
above. The same
reduction is exact for one constraint in the problem and unsound for the other.

The full model is not held to that symmetry, and its optimum does not keep it.
The initial design is mirror-symmetric to round-off, its density differing
from its mirror image by at most \num{3e-10}. The difference then grows by
about an order of magnitude per iteration over the first ten iterations. By iteration~99 the design has settled into an asymmetric layout:
\num{1520} of its \num{43893} nodes, \SI{3.5}{\percent}, are solid on one side
of the plane and void on the other (\num{1516} at the reported design of iteration~174). This is why the mode that is antisymmetric on the solid frame
acquires participation in Figure~\ref{fig:demo2_mode_optimized}. The full-model
optimum is therefore not a design the reduced model can represent.

The consequence is not small. Optimizing the reduced model to the same allowable and the same frequency
bound gives $v/v_0 = \num{0.4251}$, against $\num{0.5438}$ for the full model.
That design is \SI{21.8}{\percent} lighter, reports a fundamental of
\SI{25.0004}{Hz}, and appears to satisfy every constraint.

The saving is concentrated in the back panel, the web that closes the open
channel section and supplies its torsional stiffness. Over the design band the
reduced model retains a mean density of \num{0.116} there against \num{0.751},
and removes the web entirely over the inner \SI{250}{mm}.
Figure~\ref{fig:demo2_comparison} places the two designs side by side.

Evaluating that same design on the unreduced model, where both mode families
are present, gives the numbers in Table~\ref{tab:symmetry_reduction}. The NRL
Sum is reproduced to within \SI{0.02}{\percent}, as the parity argument
requires. The fundamental is not. It falls to \SI{3.70}{Hz}, a factor of \num{6.8}
below the bound, and two modes now lie below the floor. Those two modes carry
participation factors of order $10^{-12}$ and $10^{-14}$, zero to working
precision, which identifies them as the antisymmetric family the reduction
discarded. They are the modes that collapse when the web is removed, and they
are invisible to the model that removed it.

\begin{table}[htbp]
  \centering
  \caption{The design obtained from the symmetry-reduced model, as that model
  reports it and as the unreduced model measures it. The criterion that is
  filtered by parity is reproduced; the one that is not is violated by a factor
  of \num{6.8}.}
  \label{tab:symmetry_reduction}
  \begin{adjustbox}{max width=\linewidth}
\begin{tabular}{lcc}
    \toprule
    & Reduced model & Unreduced model \\
    \midrule
    NRL Sum of stress, \si{MPa} & \num{234.984} & \num{234.940} \\
    Fundamental frequency, \si{Hz} & \num{25.0004} & \num{3.698} \\
    \bottomrule
  \end{tabular}
\end{adjustbox}
\end{table}

The reduced design is therefore not a lighter solution but an inadmissible one
and the reduction is what conceals that. A symmetry reduction is sound for a response filtered by the same symmetry.
It is unsound for a constraint on the raw spectrum. A formulation combining the two, as this one does, admits the
reduction for one and not the other. The
demonstrations that follow use the unreduced model throughout.

\subsection{Demonstration~3: equipment frame, combined-direction shock}
\label{subsec:demo3}

\paragraph{Setup}
The third problem is the frame of Section~\ref{subsec:demo2}, unchanged in
geometry, material, mesh and allowable. It is certified instead against shock
from an \emph{unknown} attack orientation. The component must survive shock in each of
the vertical, athwartship and fore-and-aft directions. Each direction~$d$ has its
own influence vector $\mathbf{D}^{(d)}$ and therefore its own modal
participation factors, design accelerations, and per-direction NRL Sum.

Holding everything but the certified directions fixed makes the comparison
with Section~\ref{subsec:demo2} a controlled one. The two demonstrations
differ in the certified directions and in nothing else. Demonstration~2 certifies
against fore-and-aft alone; this one adds the vertical and athwartship
directions to the same design problem. Two outcomes are informative and they
are informative for different reasons. If the optimum is unchanged, the
fore-and-aft direction governs throughout and the single-direction design was
already the combined-direction design. If it moves, the design that satisfies one direction does not satisfy three.
That is the case for treating the directional maximum as a constraint in its
own right, rather than certifying each direction after the fact.

The vertical direction is not expected to govern this frame, for the reason
given in Section~\ref{subsec:demo2}. Its vertical response sits on the
acceleration-limited branch, where the design value is set by the modal weight
alone.

\paragraph{Spectrum and criterion}
The three directional accelerations follow from $A_0$ and $V_0$
of~\eqref{eq:nrl1396_A0V0} through~\eqref{eq:Da_branches}, with the unit
design coefficient of Section~\ref{subsec:demo1}. The three directions
therefore share a spectrum. The governing direction is decided by the
structure's modal response rather than by a directional table. 
For each direction we form the pointwise NRL Sum~\eqref{eq:nrl_field} of the von~Mises modal
stress~\eqref{eq:Ri_stress}. The component is certified against the most
severe direction,
\begin{equation}
  \label{eq:demo3_combined}
  f_{\mathrm{NRL}}^{\mathrm{comb}}(\bz)
  \;=\; \max_{d \in \{\mathrm{V},\,\mathrm{A},\,\mathrm{FA}\}}
        f_{\mathrm{NRL}}^{(d)}(\bz).
\end{equation}
The outer maximum in~\eqref{eq:demo3_combined} need not be differentiated.
Bounding the maximum of the three directions is the same as bounding each of
them, $f_{\mathrm{NRL}}^{(d)}(\bz) \le \sigma_{\mathrm{allow}}$ for
$d \in \{\mathrm{V},\mathrm{A},\mathrm{FA}\}$, and that is how the problem is
posed to the optimizer: three constraints, each with its own gradient, and the
optimizer itself decides which are active. The directional combination
therefore introduces no new sensitivity machinery and no second hard-max
layer. Nor does it need three eigenproblems: a single modal basis is extracted
and reused across all three directional influence vectors. Whether the maximum is the combination
of record for this component class is a separate question, and
Section~\ref{sec:limitations} takes it up.

DDAM is again used as a \emph{constraint}: minimize volume subject to the
combined-direction NRL Sum stress not exceeding the elastic allowable,
\begin{equation}
  \label{eq:demo3_problem}
  \begin{aligned}
  &\underset{\bz}{\text{minimize}}\;\; v(\bz)
  \quad\text{s.t.}\quad
  f_{\mathrm{NRL}}^{\mathrm{comb}}(\bz) \le \sigma_{\mathrm{allow}},\\
  &\lambda_1(\bz) \ge \omega_{\min}^{2},
  \;\; z_{\min}\le z_e \le 1,
  \end{aligned}
\end{equation}
with $\sigma_{\mathrm{allow}} = \sigma_y$ as in
Section~\ref{subsec:demo2}. No additional factor of safety is applied. DDAM places its conservatism on the
load side, the design spectrum being a worst-case envelope rather than a
best-estimate load~\cite{navsea_0908}. Deflating the resistance would
therefore double-count margin. The yield limit bounds the measure-weighted
aggregate of the von~Mises NRL Sum, which lies at or below its peak. We
therefore certify against $\sigma_y$ and report the reserve factor
$\sigma_y / f_{\mathrm{NRL}}^{\mathrm{comb}}$ at the optimum, as in
Table~\ref{table:demo_summary}. That quantity is not a certified
margin. Section~\ref{sec:limitations} states what formal certification
additionally requires of the stress measure.

\begin{figure}[htbp]
  \centering
  \includegraphics[width=0.773\linewidth]{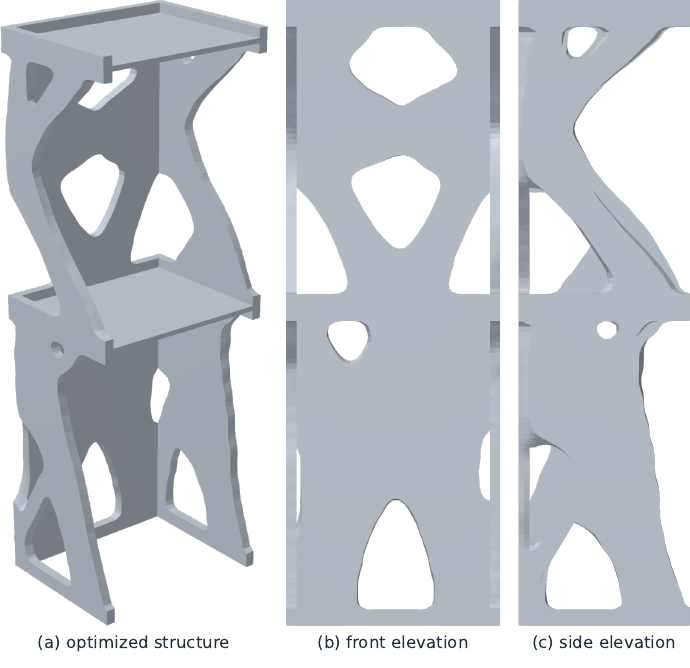}
  \caption{Demonstration~3 optimized structure at iteration~199, shown at the
  \num{0.5} level of the density field. It is drawn from the viewpoint, at the
  scale and in the color of Figure~\ref{fig:demo2_topology}. The
  three-direction design can then be read against the single-direction one. The
  non-design shelves, collars and footprint skin are drawn with the design
  domain, so the structure is seen as it is supported. (a)~The full frame,
  (b)~the front elevation, looking along the fore-and-aft direction, and
  (c)~the side elevation, looking along the athwartship axis. The elevations
  are drawn without perspective. The architecture is that of Demonstration~2. A pair of side
  frames carries the shelf loads into the footprint, and a web closes the
  channel. The combined-direction requirement redistributes material within
  that architecture rather than replacing it.}
  \label{fig:demo3_topology}
\end{figure}

\begin{figure}[htbp]
  \centering
  \includegraphics[width=0.82\linewidth]{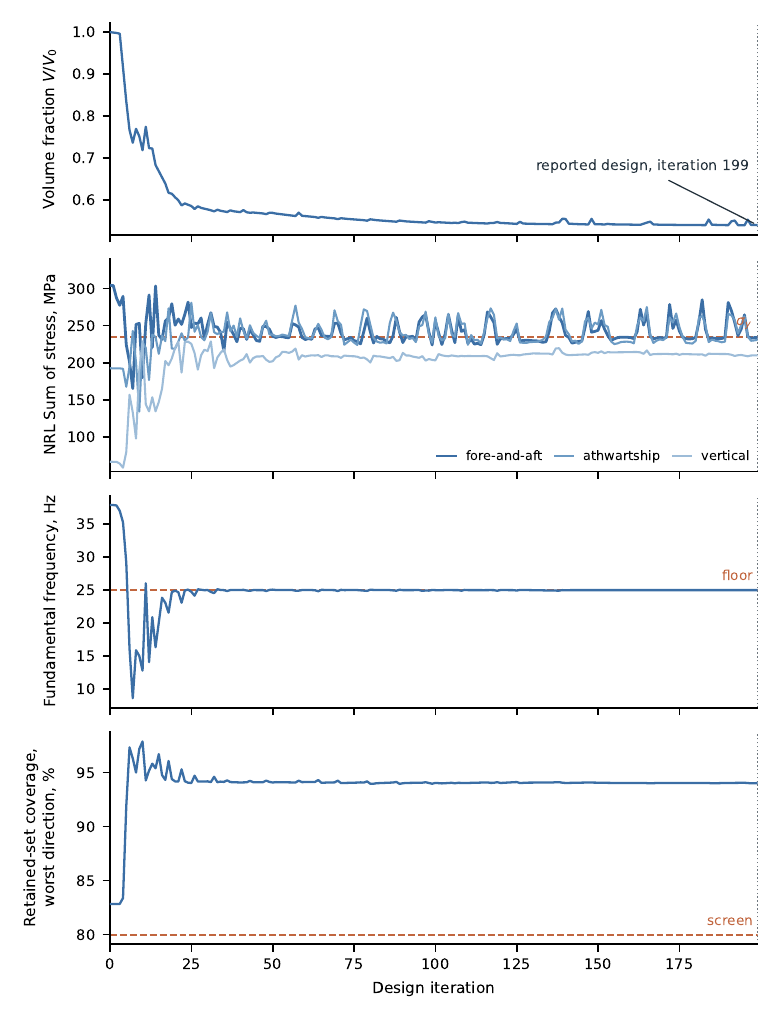}
  \caption{Demonstration~3 convergence. The dotted line marks iteration~199,
  the reported design and also the final one. The three directional responses
  are drawn against their shared allowable. Fore-and-aft governs at both ends,
  athwartship overtakes it on about half of the iterates, and the vertical climbs from far below without
  ever reaching the limit. The fourth panel reports the coverage of the
  retained set in its worst direction, which is the vertical.}
  \label{fig:demo3_convergence}
\end{figure}

\begin{figure}[htbp]
  \centering
  \includegraphics[width=0.689\linewidth]{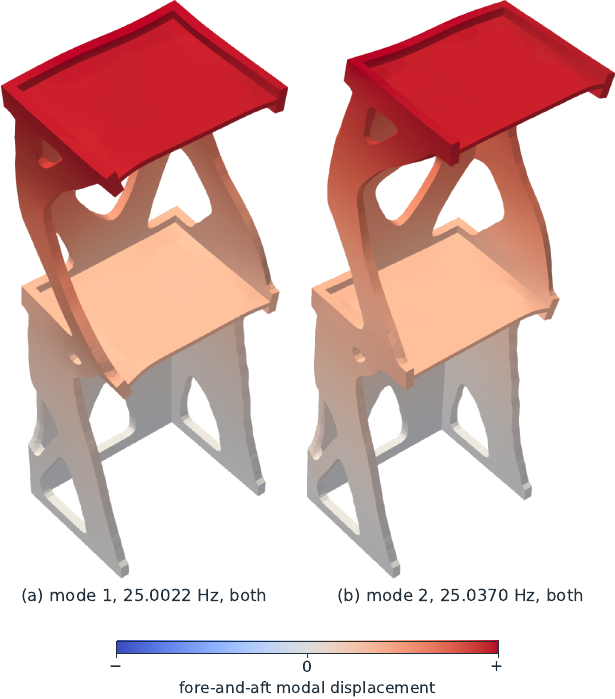}
  \caption{The two lowest modes of the design optimized against all three
  directions. They are drawn and colored as in
  Figure~\ref{fig:frame_mode_baseline}, whose panels are this frame before
  optimization. Both have been driven onto the frequency floor and now lie
  \SI{0.035}{Hz} apart rather than \SI{8.99}{Hz}, and neither belongs to a
  direction any longer. The participation factors are
  $\Gamma_{\mathrm{FA}} = \num{-1.7221}$, $\Gamma_{\mathrm{A}} = \num{+1.2777}$
  in (a), and $\Gamma_{\mathrm{FA}} = \num{+1.3603}$,
  $\Gamma_{\mathrm{A}} = \num{+1.6215}$ in (b). The cancellation that made the
  second mode carry no fore-and-aft response at the baseline is gone, which the
  color shows directly. This pair carries the two directions that oscillate
  across the allowable in Figure~\ref{fig:demo3_convergence}.}
  \label{fig:demo3_mode_optimized}
\end{figure}

\paragraph{Results}
Table~\ref{tab:demo3_results} reports the design at the last iteration that
satisfies all four constraints, and Table~\ref{tab:demo3_modal} the
per-direction modal content before and after.
Figure~\ref{fig:demo3_topology} shows the structure it arrived at,
Figure~\ref{fig:demo3_mode_optimized} the mode pair that governs it, and
Figure~\ref{fig:demo3_convergence} the history that produced it. The frame is the one of Section~\ref{subsec:demo2}, so its
modes before optimization are those already shown in
Figure~\ref{fig:frame_mode_baseline}. The optimizer removes \SI{46.1}{\percent} of the material, driving the
combined-direction response of~\eqref{eq:demo3_combined} from
\SI{304.44}{MPa} on the solid baseline to \SI{234.04}{MPa}. That meets the
allowable to within \SI{0.41}{\percent}. The fundamental settles on its floor at
\SI{25.0022}{Hz}, as in Demonstration~2.

Fore-and-aft governs the combined maximum at the solid baseline and still
governs it at the optimum, consistent with the expectation of the setup
that vertical would not govern. The vertical direction sits on the acceleration-limited branch, where the design value is set by the modal
weight alone. The prediction is therefore confirmed at both ends of the
run, but not along it: the athwartship response exceeds the fore-and-aft one
on \num{96} of the \num{200} iterates, first at iteration~7 and last at
iteration~196, by up to \SI{77.8}{MPa}, and on \num{19} of the last fifty. At
the reported design fore-and-aft leads, \SI{234.04}{MPa} against
\SI{230.94}{MPa}.

That fore-and-aft governs at the reported design does not make the other two
inert. The margins they carry at the solid baseline are consumed almost
entirely. The athwartship response rises from \SI{192.49}{MPa} to \SI{230.94}{MPa}, and
the vertical from \SI{66.39}{MPa} to \SI{209.87}{MPa}, the latter more than
tripling. At the optimum all three directions therefore lie within
\SI{10.7}{\percent} of the allowable, and two of them within
\SI{1.8}{\percent}. A design certified on fore-and-aft alone is free to spend
the vertical and athwartship reserves. This one is not, and the directional
maximum is what stops it. The three constraints are not redundant even though
one of them is the only active member of the set at the reported design: the
athwartship one is active on many of the iterates along the way.

The mechanism is visible in the modal content of
Table~\ref{tab:demo3_modal} and in
Figures~\ref{fig:frame_mode_baseline} and~\ref{fig:demo3_mode_optimized}. On the solid frame the three directions are carried by three distinct modes.
The fundamental at \SI{37.87}{Hz} holds \SI{80.0}{\percent} of the
fore-and-aft effective mass, and the second mode at \SI{46.87}{Hz} holds
\SI{83.6}{\percent} of the athwartship. The fourth, at \SI{138.64}{Hz}, holds
\SI{74.8}{\percent} of the vertical. Each direction has
a dominant mode of its own, and the NRL Sum in each is close to a
single-mode response. At the optimum that separation is gone. The two lowest modes have been driven onto the frequency floor and lie
\SI{0.035}{Hz} apart, and neither belongs to one direction any longer. The
first carries \SI{48.4}{\percent} of the fore-and-aft effective mass and
\SI{26.6}{\percent} of the athwartship, the second \SI{30.2}{\percent} and
\SI{42.9}{\percent}. The
participation factors state the same thing more sharply than the fractions do.
At the baseline the cross terms vanish, $\Gamma_{\mathrm{A}} = \num{-0.0000}$
in the first mode and $\Gamma_{\mathrm{FA}} = \num{-0.0000}$ in the second.
Each mode is a translation in one direction and carries none of the other. At
the optimum every one of the four is of order unity. This is the same
near-degenerate pair that Section~\ref{subsec:demo2} identifies as the source
of the oscillation there. The directional requirement does not remove it. 
It mixes two directions into the pair rather than one.
The vertical direction is left with no dominant mode at all. Its effective
mass is spread over four modes between \SI{64}{Hz} and \SI{121}{Hz}, of which
the largest holds \SI{32.2}{\percent}.

The optimizer does not settle onto that point smoothly, and
Figure~\ref{fig:demo3_convergence} shows that the roughness is not shared
equally between the three directions. The fore-and-aft response lies above the allowable on \num{113} of the
\num{200} iterates and the athwartship on \num{104}. Both oscillate across the
limit rather than approaching it from below, and over the last fifty iterates
they span \SI{55}{MPa} and \SI{50}{MPa} respectively. The vertical response never exceeds the allowable
at any iterate and moves within \SI{8}{MPa} over the same fifty. The two directions that chatter are exactly the two carried by the
near-degenerate pair of Table~\ref{tab:demo3_modal}. The one that does not is
the one with no dominant mode. That identifies the pair rather than the
directional maximum as the source.

The frequency history repeats Demonstration~2's behavior closely. There is an early excursion to \SI{8.65}{Hz} at iteration~7, from which the
optimizer recovers unaided. The fundamental reaches the floor by
iteration~29 and then stays on it only on average: it dips below it on
\num{63} later iterates, the last at iteration~196, by at most
\SI{0.45}{Hz} (\SI{24.55}{Hz}). The retained set covers \SI{82.8}{\percent} of
the effective mass in its worst direction on the solid frame, above
\SI{91.8}{\percent} from iteration~5 on and \SI{94.0}{\percent} at the
reported design, clear of the \SI{80}{\percent} target throughout. The reported
response is therefore not an artifact of a thinning modal basis.

Against Demonstration~2 the comparison must be read with care. The two
problems share geometry, mesh, material, allowable and frequency floor. 
Demonstration~3 also adds two constraints to that design space. Its feasible set is
therefore a strict subset of Demonstration~2's and its optimum cannot be the
lighter of the two. The reported designs do not respect that ordering:
$v/v_0 = \num{0.5394}$ here against \num{0.5438} in
Section~\ref{subsec:demo2}. Neither figure is a converged optimum. Both runs were stopped by an iteration cap rather than by a convergence test.
In this one the reported design is the final iterate, with the objective still
falling when the run ended. The ordering is evidence of that, and the
quantity it bounds is Demonstration~2's optimum, which must lie at or below
\num{0.5394}. The comparison does establish one thing, and it does not depend on either run
being converged. The combined-direction requirement is satisfiable on this frame at a volume
fraction near that of the single-direction problem. It is satisfied by
consuming the off-axis margins rather than by adding material.

\begin{table}[htbp]
  \centering
  \small
  \caption{Demonstration~3. The optimized column is the design at
  iteration~199, the last iteration satisfying all four constraints. The
  certifying quantity is the combined-direction maximum
  of~\eqref{eq:demo3_combined}, attained by the fore-and-aft direction at both
  the baseline and the optimum.}
  \label{tab:demo3_results}
  \begin{adjustbox}{max width=\linewidth}
\begin{tabular}{lccc}
    \toprule
     & Solid baseline & Optimized & Limit \\
    \midrule
    Volume fraction                             & 1.000  & 0.539    & \\
    Fore-and-aft NRL Sum of stress, \si{MPa}    & 304.44 & 234.04   & 235.00 \\
    Athwartship NRL Sum of stress, \si{MPa}     & 192.49 & 230.94   & 235.00 \\
    Vertical NRL Sum of stress, \si{MPa}        & 66.39  & 209.87   & 235.00 \\
    Combined-direction NRL Sum, \si{MPa}        & 304.44 & 234.04   & 235.00 \\
    Fundamental frequency, \si{Hz}              & 37.87  & 25.0022  & 25.00 \\
    \bottomrule
  \end{tabular}
\end{adjustbox}
\end{table}

\begin{table}[htbp]
  \centering
  \small
  \caption{Demonstration~3, effective modal mass by direction as a percentage
  of the total, for the six modes of the shared basis. Directions are
  fore-and-aft (FA), athwartship (A) and vertical (V); the largest entry in
  each direction is set in bold. On the solid frame each direction is carried
  by a distinct dominant mode. At the optimum the two lowest modes sit
  \SI{0.035}{Hz} apart on the frequency floor, and each carries both horizontal
  directions. The vertical direction retains no dominant mode, its largest
  share being \SI{32.2}{\percent}.}
  \label{tab:demo3_modal}
  \begin{adjustbox}{max width=\linewidth}
\begin{tabular}{crrrrcrrrr}
    \toprule
    & \multicolumn{4}{c}{Solid baseline} & & \multicolumn{4}{c}{Optimized} \\
    \cmidrule{2-5}\cmidrule{7-10}
    Mode & $f$ [\si{Hz}] & FA & A & V & &
           $f$ [\si{Hz}] & FA & A & V \\
    \midrule
    1 & 37.87  & \textbf{80.0} & 0.0  & 0.1  & & 25.00  & \textbf{48.4} & 26.6 & 0.0  \\
    2 & 46.87  & 0.0  & \textbf{83.6} & 0.0  & & 25.04  & 30.2 & \textbf{42.9} & 0.0  \\
    3 & 135.22 & 0.0  & 12.5 & 0.0  & & 64.23  & 2.5  & 13.6 & 28.0 \\
    4 & 138.64 & 0.9  & 0.0  & \textbf{74.8} & & 66.69  & 2.3  & 14.4 & 21.6 \\
    5 & 148.87 & 13.9 & 0.0  & 0.7  & & 75.80  & 14.2 & 0.0  & 12.2 \\
    6 & 153.22 & 1.0  & 0.0  & 7.3  & & 121.47 & 0.1  & 0.0  & \textbf{32.2} \\
    \bottomrule
  \end{tabular}
\end{adjustbox}
\end{table}

\subsection{Companion study: what a conventional criterion produces}
\label{subsec:compliance}

\paragraph{The counterfactual}
Section~\ref{sec:intro} argues that a component designed against conventional
structural criteria may fail DDAM afterwards, and that the rework is
expensive. That argument motivates the whole method and is so far untested
here. This companion study tests it on the frame of Section~\ref{subsec:demo2}.

\paragraph{The design problem}
The frame is designed the way a structural engineer would design it, by
minimizing compliance under a static equivalent of the shock, the classical
problem of topology optimization~\cite{bendsoe1989,bendsoeSigmund2003}. The equipment
is applied as a fore-and-aft inertial load on the mounting pads, and the
volume is \emph{constrained} rather than minimized,
\begin{equation}
  \label{eq:compliance_problem}
  \underset{\bz}{\text{minimize}}\;\; \mathbf{f}\T \bu(\bz)
  \quad\text{s.t.}\quad
  v(\bz) \le \num{0.5438}\, v_0,
  \;\; z_{\min}\le z_e \le 1 ,
\end{equation}
with $\K(\bz)\,\bu = \mathbf{f}$. The volume bound is the value Demonstration~2 reached. The two designs
therefore carry the same material, and the comparison is one of layout
alone. Mesh, material, non-design
regions, penalization, filter and move limit are those of
Section~\ref{subsec:demo2}.

The magnitude of $\mathbf{f}$ does not affect the outcome. Scaling the load by $\alpha$ scales the compliance by $\alpha^{2}$ and leaves
the minimizer of~\eqref{eq:compliance_problem} unchanged. The resulting layout
is therefore a property of the load direction alone.

Everything the shock criterion supplies is absent from~\eqref{eq:compliance_problem}. There is no spectrum, no modal combination, no
frequency floor, and no direction other than the one the load acts in. The
design is then evaluated against the criterion of
Section~\ref{subsec:demo3} without further optimization.

\paragraph{Results}
The compliance run also stopped at its iteration cap, \num{100} iterations,
but unlike Demonstrations~2 and~3 it had settled before reaching it. Its
objective moves \SI{0.014}{\percent} over the final ten iterations and the
volume constraint is met to seven decimals. The design compared below is
therefore effectively stationary at exactly Demonstration~2's mass, and the
comparison needs no allowance for an oscillating design or a mass difference.

\begin{figure}[htbp]
  \centering
  \includegraphics[width=0.773\linewidth]{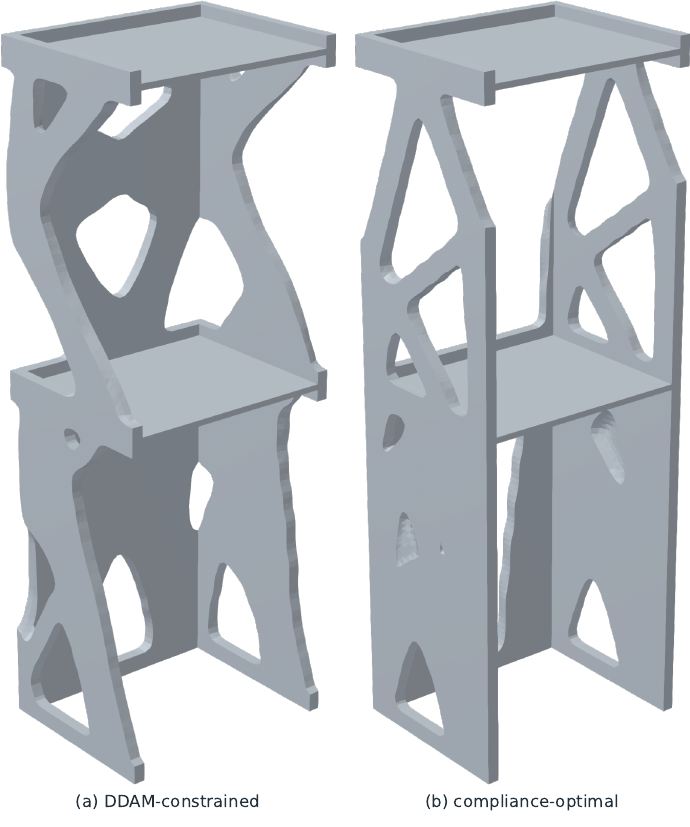}
  \caption{The same frame at nearly the same mass ($v/v_0 = \num{0.5394}$
  and \num{0.5438}), designed two ways, shown at the
  \num{0.5} level of the density field. (a)~is the design of
  Section~\ref{subsec:demo3}, certified against three directions.
  (b)~minimizes compliance under a fore-and-aft load at the volume of
  Demonstration~2. Both
  panels are at one camera and one scale, and at those of
  Figure~\ref{fig:demo3_topology}, so the difference is the design rather than
  the framing. The compliance design is a planar braced frame, its diagonals
  lying in the plane of the load it was given. It is stiff fore-and-aft and
  soft athwartship, and athwartship then governs its certification.}
  \label{fig:compliance_comparison}
\end{figure}

Table~\ref{tab:compliance_results} reports what the criterion says about it,
and Figure~\ref{fig:compliance_comparison} places the two designs side by
side. The response exceeds the allowable in all three directions, by a factor
of \num{2.75} in the worst of them. The fundamental frequency is
\SI{5.234}{Hz}, a factor of \num{4.8} below the floor of
Section~\ref{subsec:demo2}.

That last number governs how the result should be read. A design whose
fundamental lies so far beneath the floor is the quasi-disconnected
configuration of Section~\ref{subsec:topo}, which the floor exists to exclude.
Its computed responses therefore lie outside the modeling assumptions under
which they are evaluated. The finding is not that the design misses the
allowable by a stated margin. It is that a conventional criterion, given the
same material, produced a structure the certification procedure cannot qualify
at all.

\paragraph{The failure is directional}
The modal content of Table~\ref{tab:compliance_modal} locates it precisely.
The lowest \emph{fore-and-aft} mode of the compliance design lies at
\SI{28.160}{Hz}, above the floor. In the direction the load acted, the design
is adequate. The fundamental is instead a pure \emph{athwartship} mode at
\SI{5.234}{Hz} carrying \SI{53.4}{\percent} of the athwartship effective mass,
with a second athwartship mode at \SI{25.576}{Hz}.

The spectral displacement of~\eqref{eq:srss} varies as $\Dai/\omega_i^{2}$, so
a mode this low carrying half the directional mass dominates the NRL Sum in
that direction. That accounts for most of the difference between
\SI{645.15}{MPa} athwartship and \SI{334.55}{MPa} fore-and-aft. Minimizing
compliance under a single-direction load bought stiffness in that direction
and paid for it in the transverse one.

Figure~\ref{fig:compliance_comparison} shows the same thing as geometry. The
compliance design is a planar braced frame, its diagonals lying in the plane
of the load it was given. The certified design of Section~\ref{subsec:demo3} carries load through curved
closed sections instead. It closes the channel with the web that
Section~\ref{subsec:demo2} shows is holding the torsional modes above the
floor.

This is the argument for the combined-direction requirement of
Section~\ref{subsec:demo3}, made from the other side. A design process that
sees one direction produces a structure whose certification is decided by
another.

\begin{widetable}[htbp]
  \centering
  \small
  \caption{The compliance-optimal design certified against the criterion of
  Section~\ref{subsec:demo3}, at the mass Demonstration~2 reached, beside the
  two DDAM-constrained designs. The allowable is $\sigma_y$ and the frequency
  floor is \SI{25}{Hz}. Demonstration~2 is certified fore-and-aft only, so its
  off-axis responses are not reported.}
  \label{tab:compliance_results}
  \begin{adjustbox}{max width=\linewidth}
  \begin{tabular}{lcccc}
    \toprule
    & Compliance-optimal & Demonstration~2 & Demonstration~3 & Limit \\
    \midrule
    Volume fraction                          & 0.5438 & 0.5438 & 0.5394 & \\
    Fundamental frequency, \si{Hz}           & \textbf{5.234} & 25.0002 & 25.0022 & 25.00 \\
    Fore-and-aft NRL Sum, \si{MPa}           & 334.55 & 234.98 & 234.04 & 235.00 \\
    Athwartship NRL Sum, \si{MPa}            & \textbf{645.15} & & 230.94 & 235.00 \\
    Vertical NRL Sum, \si{MPa}               & 246.42 & & 209.87 & 235.00 \\
    Combined-direction NRL Sum, \si{MPa}     & \textbf{645.15} & & 234.04 & 235.00 \\
    \bottomrule
  \end{tabular}
  \end{adjustbox}
\end{widetable}

\begin{table}[htbp]
  \centering
  \small
  \caption{Effective modal mass by direction for the compliance-optimal
  design, as a percentage of the total. The design is stiff in the direction the
  compliance load acted and soft in the one it did not. The lowest fore-and-aft
  mode clears the floor, while the fundamental is an athwartship mode at a
  fifth of it.}
  \label{tab:compliance_modal}
  \begin{tabular}{crrrr}
    \toprule
    Mode & $f$ [\si{Hz}] & FA & A & V \\
    \midrule
    1 & \textbf{5.234}  & 0.0  & \textbf{53.4} & 0.0  \\
    2 & 25.576 & 0.0  & 42.8 & 0.0  \\
    3 & \textbf{28.160} & \textbf{72.6} & 0.0  & 0.0  \\
    4 & 70.100 & 0.0  & 0.0  & 50.0 \\
    5 & 97.808 & 23.3 & 0.0  & 2.0  \\
    6 & 114.943 & 0.7 & 0.0  & 28.6 \\
    \bottomrule
  \end{tabular}
\end{table}

\subsection{Companion study: the cost of separating the modes}
\label{subsec:demo3_gap}

\paragraph{What the collapse costs}
Demonstrations~2 and~3 both end with their two lowest modes on the frequency
floor. Section~\ref{subsec:crossings} (Mode crossings) treats that as a limitation. This
companion study asks whether it can be avoided, and what avoiding it costs.

The reason to want it avoided is specific to the NRL Sum. For an exactly
repeated pair the eigensolver may return any rotation of the basis, and for a
near-degenerate pair the basis is determined but rotates rapidly with the
design. Only quantities of the subspace are invariant under that rotation,
such as the sum of squares $\Gamma_1^{2} + \Gamma_2^{2}$ and the coherent
field $\Gamma_1 \bphi_1 + \Gamma_2 \bphi_2$. The NRL Sum of a pair,
$\lvert R_1 \rvert + \lvert R_2 \rvert$, is not, and neither is any other
combination formed mode by mode. The ratio
$(\lvert R_1 \rvert + \lvert R_2 \rvert)/(R_1^{2} + R_2^{2})^{1/2}$ is $1$ when
one mode carries everything and $\sqrt{2}$ when the split is even. The run of Section~\ref{subsec:demo3} has \num{114} iterates on which the pair lies
closer than \SI{0.1}{Hz}. Over those the ratio runs from \num{1.036} to
\num{1.4142}, reaching the bound exactly. The reported response therefore
moves through a \SI{36.5}{\percent} band for reasons the structure does not
fix. The larger carrier of a direction swaps between the two modes on
\num{46} of those iterates.

\paragraph{A criterion on the cluster rather than on a mode}
The quantity to constrain is the one that survives the rotation. Let
$\Phi_c$ collect the $n_c$ lowest eigenvectors as columns, normalized so that
$\Phi_c\T \M \Phi_c = \mathbf{I}$, and define the cluster mean
\begin{equation}
  \label{eq:cluster_mean}
  f_{\mathrm{cl}}(\bz) \;=\; \frac{1}{n_c}\sum_{r=1}^{n_c} \lambda_r(\bz)
    \;=\; \frac{1}{n_c}\operatorname{tr}\bigl(\Phi_c\T \K \Phi_c\bigr).
\end{equation}
Any other $\M$-orthonormal basis of the same subspace is $\Phi_c \mathbf{Q}$
with $\mathbf{Q}$ an $n_c \times n_c$ orthogonal matrix, and the trace
in~\eqref{eq:cluster_mean} is unchanged by that substitution. The criterion
therefore depends on the invariant subspace and not on the basis the
eigensolver returned. Constraining a symmetric function of a repeated cluster,
rather than an eigenvalue within it, is the established remedy for this class
of non-smoothness~\cite{seyranian_lund_olhoff_1994,overtonWomersley1993,duOlhoff2007}. It is smooth wherever the cluster stays separated from
$\lambda_{n_c+1}$, including at the coalescence that makes the individual
$\lambda_r$ non-smooth.

Its sensitivity needs nothing beyond Section~\ref{subsec:fox_kapoor}. At a
coalescence the individual $\lambda_r$ are not differentiable, so the mean is
not differentiated term by term there. Its derivative is
\begin{equation}
  \label{eq:cluster_mean_sens}
  \frac{\partial f_{\mathrm{cl}}}{\partial \bz}
    \;=\; \frac{1}{n_c}\sum_{r=1}^{n_c}
      \bphi_r\T \bigl(\K' - \lambda_r \M'\bigr)\bphi_r ,
\end{equation}
which at an exact coalescence, $\lambda_r = \lambda$, is
$n_c^{-1}\operatorname{tr}\bigl(\Phi_c\T (\K' - \lambda\M')\Phi_c\bigr)$, the trace
of the $n_c \times n_c$ Fox--Kapoor matrix of the subspace. The trace does not
depend on the basis, so summing the Fox--Kapoor diagonals~\eqref{eq:fk_lambda_prime}
in whatever basis the eigensolver returns is exact. As with the directional maximum of
Section~\ref{subsec:demo3}, the cluster-mean criterion introduces no new
sensitivity machinery.

\paragraph{The gap follows from the floor}
A constraint on $\lambda_2$, alone or as $\lambda_2 - (1+\delta)\lambda_1 \ge 0$,
is not differentiable at the coalesced iterates where the run spends its time,
while the cluster mean is. We impose the separation indirectly instead. The
floor on $\lambda_1$ shares the non-smoothness at coalescence. The gap runs
keep it, and once the pair is held apart, it no longer coalesces. The frequency 
floor~\eqref{eq:frequency_floor} is active at the optimum of all three demonstrations. 
Thus, $\lambda_1 = \omega_{\min}^{2}$ at the optimum. Constraining the cluster 
mean at $n_c=2$ then bounds the second eigenvalue,
\begin{equation}
  \label{eq:gap_from_mean}
  \frac{\lambda_1 + \lambda_2}{2} \;\ge\; \underline{f}_{\mathrm{cl}}
  \quad\Longrightarrow\quad
  \frac{\lambda_2}{\lambda_1} \;\ge\; \frac{2\,\underline{f}_{\mathrm{cl}}}{\lambda_1} - 1 ,
\end{equation}
and the choice $\underline{f}_{\mathrm{cl}} = \omega_{\min}^{2}\,(2+\delta)/2$ gives
$\lambda_2 \ge (1+\delta)\,\lambda_1$ when $\lambda_1$ sits on the floor. If
$\lambda_1 = \omega_{\min}^{2}(1+\theta)$, the bound is
$\lambda_2/\lambda_1 \ge (1+\delta-\theta)/(1+\theta)
\approx 1 + \delta - (2+\delta)\,\theta$. The design problem of
Section~\ref{subsec:demo3} gains one constraint,
\begin{equation}
  \label{eq:demo3_gap_problem}
  \begin{aligned}
  &\underset{\bz}{\text{minimize}}\;\; v(\bz)
  \quad\text{s.t.}\quad
  f_{\mathrm{NRL}}^{\mathrm{comb}}(\bz) \le \sigma_{\mathrm{allow}},\\
  &\lambda_1(\bz) \ge \omega_{\min}^{2},
  \;\; f_{\mathrm{cl}}(\bz) \ge \underline{f}_{\mathrm{cl}},
  \;\; z_{\min}\le z_e \le 1 .
  \end{aligned}
\end{equation}

The separation in~\eqref{eq:gap_from_mean} is implied rather than imposed, and
the implication is only as good as its premise. Should $\lambda_1$ rise above
the floor, the mean can be satisfied without separating the pair. We therefore
report the fundamental alongside the separation achieved, rather than assuming
the construction held.

Designing a structure to prescribed modal properties rather than to an
extremum of them is an established aim in its own
right~\cite{Maeda2006,duOlhoff2007}. Two values of $\delta$ are run. The solid frame separates the pair by
$\lambda_2/\lambda_1 = \num{1.531}$. Setting $\delta = 0.5$ therefore asks the optimizer to hold roughly the
separation the geometry supplies unaided. Setting $\delta = 0.2$ lets the
excess $\lambda_2/\lambda_1 - 1$ close to about two fifths of its solid-frame
value of \num{0.531}. All other settings are those of
Section~\ref{subsec:demo3}.

\paragraph{Results}
Both runs completed \num{200} iterations. Every number below is read at the
last iterate of its run. For Section~\ref{subsec:demo3} the last iterate is
also the last feasible one, so the three runs are compared on a common
footing. The problem setup is that of Section~\ref{subsec:demo3} unchanged,
including the mesh, material, allowable and filter.

The premise of~\eqref{eq:gap_from_mean} held: the frequency floor is active
in both runs. The fundamental sits \SI{0.0178}{\percent} above the floor at
$\delta = 0.2$ and \SI{0.0017}{\percent} above it at $\delta = 0.5$, and
neither run falls below the floor after its sixteenth iterate. With $\lambda_1$
on the floor, \eqref{eq:gap_from_mean} bounds the separation, and both runs
attain it almost exactly. At the last iterate, $\lambda_2/\lambda_1$ is
\num{1.2005} for a requested \num{1.2}, and \num{1.5001} for a requested
\num{1.5}. Both exceed the bounds that~\eqref{eq:gap_from_mean} gives for
these fundamentals, \num{1.1996} and \num{1.5000}.

The pair separates as intended. Median separation over the run is
\SI{2.3882}{Hz} at $\delta = 0.2$ and \SI{5.6193}{Hz} at $\delta = 0.5$. The
median in Section~\ref{subsec:demo3} is \SI{0.0749}{Hz}. Neither run has a
single iterate on which the pair lies closer than \SI{0.1}{Hz}, against
\num{114} such iterates there. The rotation band of the preceding paragraphs
therefore never opens. The larger carrier of a direction does not swap at any
iterate of either run, against \num{46} swaps there.

The cost in volume is small. Table~\ref{tab:demo3_gap} collects the comparison. The spread in
$v/v_0$ across the three runs is under two percentage points, while the separation
they hold differs by a factor of \num{161}. Both gap runs end marginally over
the allowable, by \SI{1.18}{MPa} at $\delta = 0.2$ and by \SI{0.056}{MPa} at
$\delta = 0.5$. The Section~\ref{subsec:demo3} run ends \SI{0.956}{MPa} under
it.

That last comparison is closer than it appears. Over its final fifty
iterates, the Section~\ref{subsec:demo3} run moves the response of the direction
that governs its reported design, fore-and-aft, by \SI{10.93}{MPa} per iterate
on average. It crosses the allowable
nine times in that window. Which side of the limit its two-hundredth design
falls on is therefore not something the run controls. Both gap runs move the
response of their own governing direction, athwartship, by about
\SI{0.16}{MPa} per iterate over the same window. At
$\delta = 0.5$ it does not cross the allowable at all. The frequency
floor behaves the same way. The Section~\ref{subsec:demo3} run sits below it
on \num{19} of its final fifty iterates, as late as iterate \num{196}.
Neither gap run falls below it after iterate \num{16}.

The separation constraint does not buy a lighter frame. It buys one whose
reported performance is settled rather than read off an oscillation. That is
the answer to the design question this companion study was built to ask. Spacing the
modes is close to free in mass. What it purchases is a settled design, and a
reported response that means what it says; both runs end within
\SI{0.5}{\percent} of the allowable, on the infeasible side. Both gap runs still stopped at their
\num{200}-iteration cap; settled here means that the response no longer
oscillates, not that a convergence test was met.

\begin{widetable}[!htbp]
\begin{center}
\small
\begin{tabular}{lccc}
\toprule
& Section~\ref{subsec:demo3} & $\delta = 0.2$ & $\delta = 0.5$ \\
\midrule
$v/v_0$                                        & \num{0.5394}  & \num{0.5313}  & \num{0.5494}  \\
Largest directional response                   & \SI{234.04}{MPa} & \SI{236.18}{MPa} & \SI{235.06}{MPa} \\
Excess over $\sigma_{\mathrm{allow}}$          & \SI[retain-explicit-plus]{-0.956}{MPa}
  & \SI[retain-explicit-plus]{+1.182}{MPa} & \SI[retain-explicit-plus]{+0.056}{MPa} \\
Fundamental above floor                        & \SI{0.0086}{\percent} & \SI{0.0178}{\percent} & \SI{0.0017}{\percent} \\
$\lambda_2/\lambda_1$                          & \num{1.0028}  & \num{1.2005}  & \num{1.5001}  \\
Pair separation                                & \SI{0.0348}{Hz} & \SI{2.3926}{Hz} & \SI{5.6193}{Hz} \\
Median pair separation                         & \SI{0.0749}{Hz} & \SI{2.3882}{Hz} & \SI{5.6193}{Hz} \\
Iterates with pair $<\SI{0.1}{Hz}$             & \num{114}     & \num{0}       & \num{0}       \\
Carrier swaps                                  & \num{46}      & \num{0}       & \num{0}       \\
Mean step of the governing direction, last \num{50} & \SI{10.93}{MPa} & \SI{0.16}{MPa} & \SI{0.16}{MPa} \\
Allowable crossings, last \num{50}             & \num{9}       & \num{2}       & \num{0}       \\
Sub-floor iterates, last \num{50}              & \num{19}      & \num{0}       & \num{0}       \\
\bottomrule
\end{tabular}
\end{center}
\caption{The cost of separating the first two modes. All three runs are the
Demonstration~3 problem at \num{200} iterations, differing only in the cluster-mean
constraint of~\eqref{eq:demo3_gap_problem}. Quantities in the upper block
are read at the last iterate of each run; those in the lower block describe
the final fifty. A positive margin is a response above the allowable. The
largest directional response is fore-and-aft in Demonstration~3 and
athwartship in both gap runs. The
separation ratio $\lambda_2/\lambda_1$ reproduces the requested $1 + \delta$
to four decimal places, confirming the construction
of~\eqref{eq:gap_from_mean}.}
\label{tab:demo3_gap}
\end{widetable}

\begin{sidewaystable}
\begin{center}
\small
\setlength{\tabcolsep}{4pt}
\begin{tabular}{lcccccccc}
\toprule
& Ship / mount / dir. & Form & DoFs
& Baseline & Optimized & $R_{\mathrm{allow}}$ & Reserve & Mass red. \\
& & & & NRL Sum & NRL Sum & & factor & vs.\ base \\
\midrule
Demo~1 & surf. / deck / FA        & con.\ (disp.)  & $\num{1.32e5}$ & \SI{5.3042}{mm} & \SI{6.3651}{mm} & \SI{6.3651}{mm} & \num{1.0000} & \SI{43.2}{\percent} \\
Demo~1b & surf. / deck / FA       & con.\ (disp.)  & $\num{1.32e5}$ & \SI{5.3637}{mm} & \SI{6.3650}{mm} & \SI{6.3651}{mm} & \num{1.0000} & \SI{39.6}{\percent} \\
Demo~2 & surf. / deck / FA        & con.\ (stress) & $\num{1.32e5}$ & \SI{304.44}{MPa} & \SI{234.98}{MPa} & $\sigma_y$     & \num{1.0001} & \SI{45.6}{\percent} \\
Demo~3 & surf. / deck / V+A+FA    & con.\ (stress) & $\num{1.32e5}$ & \SI{304.44}{MPa} & \SI{234.04}{MPa} & $\sigma_y$     & \num{1.0041} & \SI{46.1}{\percent} \\
\bottomrule
\end{tabular}
\end{center}
\caption{Headline results for the demonstration problems. Each uses the DDAM criterion as a constraint (con.) on a volume-minimizing
problem. ``Form'' records whether that criterion is a displacement (disp.) or
a stress criterion. Directions: V vertical, A athwartship, FA fore-and-aft.
All four rows are the same model, mesh and material. Demonstration~1
certifies an interface displacement and Demonstration~2 a stress, in the same
direction; Demonstration~3 adds the two remaining directions.
Demonstrations~1, 2 and~3 each differ from the previous one in a single
element. Demo~1b is
the variant of Demonstration~1 with the effective weight taken mode by mode (Section~\ref{subsec:demo1_mass_loaded} (Mass-loaded spectrum)). The reserve factor
is the allowable over the optimized NRL Sum,
$R_{\mathrm{allow}}/f_{\mathrm{NRL}}$, so that unity is an exactly attained
limit.}
\label{table:demo_summary}
\end{sidewaystable}

\section{Limitations}
\label{sec:limitations}

The formulation and its demonstrations have the limitations of scope set out
below. They concern the Navy inputs and combination rules, the stress
measure and the benchmark, the role of the criterion, and the modal
assumptions.

\paragraph{Design coefficients are taken as unity}
\label{subsec:limits}
The elastic design coefficients scale the reference values $A_0$ and $V_0$
of~\eqref{eq:nrl1396_A0V0}. They give the directional values $A_{a,i}$ and
$V_{a,i}$. They are tabulated~\cite{nrl1396,navsea_t9070} by ship type,
mounting location and direction, and are not used here. The coefficient is taken as unity in every direction,
so the published reference equations are applied unmodified.

The choice affects more than the reported magnitudes. Because the tabulated
coefficients reduce the design values, the responses here are an upper bound
and the resulting designs are conservative. The coefficient also decides \emph{whether the shock criterion binds at
all}. For the frame of Section~\ref{subsec:demo2}, a coefficient of $0.4$
places the response far below the allowable. The frequency floor alone then
decides the design, and the shock criterion never becomes active. At unity the criterion does bind, and
both constraints are active at the optimum, as
Table~\ref{tab:demo2_results} shows. The reported margins, governing constraints and masses are therefore
specific to this choice. A certification against the governing tables would
rerun the optimization with those coefficients rather than rescale these
results.

The directional structure of Demonstration~3 is unaffected in kind. With a shared coefficient the governing direction is decided by the
structure's modal response rather than by the table, which is the mechanism
the demonstration is meant to show. A directional table would reweight the
three responses before the maximum is taken.

\paragraph{The directional combination rule}
Demonstration~3 certifies against the worst of the three orthogonal directions,
\eqref{eq:demo3_combined}. Whether that maximum is the combination of record for this component class is
left open here, for the same reason the design coefficients are taken as
unity: the governing tables are not used.

The formulation covers the alternative without extension. If the rule is
instead an SRSS over the three directions, as some non-Navy seismic practice
does~\cite{nrcRG192}, the substitution is immediate. That
form is smooth, and its sensitivity is given by the SRSS coefficients of~\eqref{eq:srss_coef_alpha}--\eqref{eq:srss_coef_gamma} applied at the
directional level rather than the modal level. It removes a hard-max layer
rather than adding one.

The choice of rule does change the answer. At the design of
Section~\ref{subsec:demo3}, the three directional responses combine by SRSS to
\SI{390.07}{MPa}, against \SI{234.04}{MPa} for the maximum, or
\SI{66.0}{\percent} above the allowable. The solid baseline combines to
\SI{366.26}{MPa}, also above it. With unit design coefficients, the SRSS rule
therefore places even the solid frame above the allowable; whether a feasible
design exists under that rule was not tested.

Two points put this in proportion. The unit coefficients overstate the design
values (Design coefficients, above), so under the tabulated coefficients every
response here would be lower, whichever rule applies. The formulation also
handles either rule, and a change of rule changes the constraint, not the
method. The combination rule is a modeling input that can decide whether
the problem is feasible, and it should be fixed from the governing
requirements before these results are used for certification.

\paragraph{The stress measure differs from the certification measure}
Demonstrations~2 and~3 bound the measure-weighted aggregate of the von~Mises
NRL Sum by the material's yield strength. Formal certification uses a
different measure. It resolves the modal stress into the NAVSEA general membrane, local membrane, and
membrane-plus-bending categories, each carrying its own
allowable~\cite[Ch.~6]{navsea_t9070}.

The single-yield allowable used here is the practical fallback for solid
three-dimensional parts. In those the through-thickness stress linearization
the categories presuppose is ill-defined, there being no shell section over
which to separate membrane from bending.

Because the aggregate lies below the peak, the bound does not cap the largest
local response, and the designs have not been checked against the category
criteria. They meet the simpler measure, and the reserve factors quoted in
Table~\ref{table:demo_summary} are margins against it rather than certified
margins.

\paragraph{A reinterpreted benchmark} The equipment frame is a continuum
reinterpretation of a published shell model with welded
connections~\cite{koehler2018ddam}, as Section~\ref{subsec:demo2} describes.
What is solved is a foundation design problem posed on that configuration,
not a reproduction of that model, and the reported values are not intended
for direct comparison with it.

\paragraph{The criterion as an objective} Section~\ref{subsec:topo} presents
the DDAM criterion in both roles, as an objective and as a constraint. All
three demonstrations use it as a constraint on a volume-minimizing problem.
The choice follows from the physics. Minimizing the response drives the retained frequencies \emph{upward}. Once a
mode crosses the branch point of~\eqref{eq:Da_branches} onto the
acceleration-limited plateau, its design value stops depending on frequency.
Then $\partial \Dai / \partial \omega_i$ vanishes, and the eigenvalue path
through the chain rule of Section~\ref{sec:sensitivity} goes quiet.

Minimizing volume does the opposite. Material leaves, the frequencies fall,
and the design remains on the velocity-limited branch, where that dependence
is live along the whole trajectory. The objective form is therefore the one that suppresses the
coupling this paper derives, and the constraint form is the one that exercises
it.

\paragraph{A certified direction on the plateau} The same argument applies
within Demonstration~3. Its vertical component sits on the acceleration
plateau, for the reason Section~\ref{subsec:demo2} gives for certifying
Demonstration~2 fore-and-aft, so its frequency dependence is inert; the two
horizontal components remain on the velocity branch, where the coupling this
paper derives is live.

\paragraph{Mode crossings and clustered eigenvalues}
\label{subsec:crossings}
The SRSS criterion~\eqref{eq:srss} is everywhere differentiable in $\bz$
\emph{between} mode crossings. The modal index $i$ is not a smooth function of
$\bz$ near a crossing, and modes are not tracked across iterates (Section~\ref{sec:algorithm}, Mode selection). A
modal-assurance-criterion (MAC)~\cite{allemang2003mac} pairing between
consecutive iterates would mitigate, though not eliminate, the gradient
discontinuities that fixed-index criteria incur at crossings. It would not
alter the NRL Sum, whose selector is response-based rather than shape-based by
construction.

A sustained cluster is the harder case. Demonstrations~2 and~3 both exhibit one,
rather than the isolated crossings the preceding paragraph addresses.
Demonstration~1 shows the isolated kind, a single crossing at iteration~3, and
its lowest pair ends \SI{6}{Hz} apart. The
measurements here are Demonstration~2's. Its two lowest modes are \SI{8.99}{Hz} apart on the solid frame. They are driven together onto the frequency floor until they lie
\SI{0.013}{Hz} apart. There they remain for the rest of the run
(Figures~\ref{fig:frame_mode_baseline} and~\ref{fig:demo2_mode_optimized}). At
that separation the pair is numerically repeated, and the derivative of an
individual eigenvector within a repeated subspace is not defined unless the
eigenvalue derivatives separate the
pair~\cite{seyranian_lund_olhoff_1994,millsCurran1988,dailey1989,linMottersheadNg2020}. What is
differentiable is the subspace, not its basis. The consequence is measurable
in the design history. The retained set changes on \num{47} of \num{200} iterations. The criterion
moves by as much as \SI{117}{MPa} between consecutive designs, under a move
limit of \num{0.1}. About half of the large jumps occur with the retained set unchanged. That is
the eigenvectors themselves exchanging character, rather than the selection
rule switching between them.

In the demonstrations the optimizer works across the resulting
non-smoothness and still reaches feasible designs. The price is some
oscillation about the constraint boundary, so each reported design is a
feasible iterate rather than a converged one: the lightest feasible iterate in
Demonstration~2 and the last in Demonstration~3.

A subspace-aware criterion sidesteps the issue, depending on the eigenvalue
cluster rather than on individual modes; the companion study of
Section~\ref{subsec:demo3_gap} tests it. That criterion is not applied to the
reported designs of Sections~\ref{subsec:demo2} and~\ref{subsec:demo3}, which
are reported as run.

\paragraph{Tabulated shock spectra} The Navy-supplied design acceleration $\Dai = G(\bar m_i, \omega_i)$
in~\eqref{eq:Dai} is a piecewise-defined tabulated
function~\cite{navsea_0908}. Its derivative $\partial \Dai / \partial \omega_i$
in~\eqref{eq:dSD_dz_full} is therefore piecewise continuous. We compute these derivatives by analytic differentiation of the
piecewise expressions within each branch. Written as in~\eqref{eq:Da_branches}, the design value carries two kinks: the
\SI{6}{\g} floor~\cite{navsea_t9070}, and the crossover $\omega^{*} = A_a/V_a$ at which the
velocity-limited and acceleration-limited branches meet. The derivative is
$V_a$ strictly inside the velocity-limited branch, and zero on both the floor
and the acceleration plateau.

At either kink we take the zero-slope side. That choice is an element of the generalized gradient~\cite{clarke1983} at the kink, which is
the whole interval $[0, V_a]$ there. It is therefore admissible rather than
merely convenient. It is also the
conservative element, in the sense that at the crossover it reports no
frequency sensitivity where the branch above supplies none.

The selection is consequential only on a set of designs of measure zero, and
in the present demonstrations it was never reached. For the supported weight of the equipment frame the crossover falls at
\SI{56.52}{Hz}. It is the same in all three directions, because the
directional coefficients cancel in the ratio $A_a/V_a$. Across the \num{1200} mode evaluations of the three-direction run, the closest
any computed mode came to that frequency was \SI{0.45}{Hz}. The retained modes
sit clearly on one side or the other. The pair on the frequency floor near \SI{25}{Hz} is on the velocity-limited
branch. The modes carrying the vertical response between \SI{64}{Hz} and
\SI{121}{Hz} are on the plateau.

\paragraph{Mass loading of the design spectrum}
\label{subsec:mass_loading}
The NRL-1396 design values fall as the modal effective weight $W_a$ rises. The
design acceleration therefore depends on the participation factor through the
effective modal mass, and~\eqref{eq:dSD_dz_full} carries that dependence as
its second term. The sensitivity of the displacement criterion includes
that term. The variant of Demonstration~1 in
Section~\ref{subsec:demo1_mass_loaded} (Mass-loaded spectrum) exercises it.

The term was verified with the finite-difference probe of
Section~\ref{subsec:fd_check}, on a three-dimensional test problem with an
analytic mass-dependent spectrum. The analytic gradient agrees to a relative
error of $5\times10^{-12}$ or better for both the SRSS criterion and the NRL
Sum. A spectrum that omits its mass derivative misses by \SI{7.7}{\percent},
so the probe resolves the term.

The stress criterion does not yet carry it, which is why every demonstration
uses the fixed reference weight of Section~\ref{sec:results}. The effect of
the fixed reference has no fixed sign. For a mode whose effective weight exceeds the reference it
overstates the shock, and for a lighter mode it understates it. On the frame of
Demonstration~1 the net effect is a heavier optimum, \SI{39.6}{\percent}
removed against \SI{43.2}{\percent}.

\paragraph{Linearity assumption} DDAM assumes a linear modal response of
the equipment; non-linear material response, large rotations, and contact
are not represented. For components that approach the non-linear regime
under the design shock, DDAM provides a screening criterion
only~\cite{navsea_t9070}, and a follow-on transient analysis of the shock
response~\cite{sigristBroc2023} is the natural complement.

\section{Conclusions}
\label{sec:conclusions}

We have presented a complete differentiable formulation of the Dynamic Design
Analysis Method. It allows the Navy-mandated shock criterion to be used as a
design driver inside large-scale density-based topology optimization.

The sensitivity expression combines two mechanisms. Forward-mode automatic
differentiation supplies the element kernels, and physics-level adjoints
assemble the gradient: Fox--Kapoor for the eigenvalue sensitivity, Nelson for
the eigenvector-sensitivity inner product. The chain rule is derived for a spectral displacement that depends jointly on
the natural frequency and the modal participation factor. The frequency path is
implemented and verified. The path through the participation factor is
implemented and verified for the displacement criterion. For the stress
criterion it is derived only, and those demonstrations evaluate the spectrum
at a fixed reference weight. The AD-plus-adjoint construction removes the maintenance burden of hand-coded
derivative expressions. It also keeps the cost of the gradient, relative to one analysis,
independent of the number of design variables.

The formulation was exercised on three shipboard problems, arranged so that
each adds one element to the one before. On all three the criterion governs
the design rather than merely permitting it. Two companion studies then
isolate single questions those problems raise. One asks what a conventional
criterion produces at the same mass. The other asks what it costs to keep the
two lowest modes apart.

The optimizer removed \SI{43.2}{\percent} of the material from an equipment
frame certified on the fore-and-aft NRL Sum of interface displacement. It
removed \SI{45.6}{\percent} from the same frame certified on the fore-and-aft
NRL Sum of von~Mises stress. It removed \SI{46.1}{\percent} from the same frame
certified against the combined-direction maximum over three orthogonal shock
directions. In each case the shock
criterion is active at the optimum, attained to within \SI{0.0004}{\percent},
\SI{0.011}{\percent} and \SI{0.41}{\percent} of the allowable respectively.
The reported designs therefore sit on the acceptance boundary rather than
inside it.

Three findings emerged that are properties of the problem rather than of the
implementation. First, the frequency floor and the shock criterion determine
the frame design jointly: both are active at every optimum, and a formulation
carrying only one of them admits designs the other rejects.

Second, the combined-direction requirement is satisfied not by adding material
but by consuming the margins the off-axis directions carry at the baseline.
Those rise from \SI{192.49}{MPa} and \SI{66.39}{MPa} to \SI{230.94}{MPa} and
\SI{209.87}{MPa} while the governing direction is unchanged. Certifying one
direction and checking the others afterwards therefore need not reproduce the
design.

Third, in Demonstrations~2 and~3 the optimizer drives the two lowest modes onto
the frequency floor together, leaving them numerically repeated. The resulting
non-smoothness makes those two problems oscillate across the acceptance
boundary rather than converge onto it. Their reported designs are feasible
iterates rather than converged optima. Demonstration~1 avoids the coalescence.
A different mode reaches its floor, and its lowest pair ends \SI{6}{Hz} apart.
Section~\ref{subsec:crossings} (Mode crossings) states what a treatment of repeated eigenvalues
would change.

That treatment is no longer hypothetical. Section~\ref{subsec:demo3_gap}
constrains the mean of the two lowest eigenvalues, which is invariant to the
basis the eigensolver returns inside a repeated cluster. Holding the second
eigenvalue \SI{20}{\percent} above the first removes the degeneracy outright,
while the run ends within \SI{0.5}{\percent} of the allowable. No iterate has the
pair within \SI{0.1}{Hz}, against \num{114}, and the carrier never swaps,
against \num{46}. The oscillation goes with it. Over the final fifty iterates
the governing response moves \SI{0.16}{MPa} per iterate rather than
\SI{10.93}{MPa}, while the volume fraction differs by under two percentage points.
Separating the modes is therefore close to free. What it buys is a settled design rather
than a lighter one.

What the workflow delivers is a design that satisfies the DDAM acceptance criterion at
the optimum, not a completed qualification. Formal certification additionally
requires resolving the modal stress into the NAVSEA stress categories
(Section~\ref{sec:limitations}). The reinterpreted benchmark of
Section~\ref{sec:limitations} is also not comparable with its published
source.

Section~\ref{subsec:compliance} shows what the design-then-check pattern
costs. The same frame was minimized for compliance under a fore-and-aft load,
at nearly the same mass. It reaches a combined-direction NRL Sum of
\SI{645.15}{MPa} against an allowable of \SI{235.00}{MPa}. Its fundamental
falls to \SI{5.234}{Hz} against a floor of \SI{25}{Hz}. The design is braced
in the plane of the load it was given and soft in the direction it was not. A
criterion that never saw the shock therefore produces a component whose
fundamental lies outside the range the certification can qualify, and whose
nominal response exceeds the allowable by a factor of~\num{2.75}.

The workflow brings the Navy-mandated shock criterion~\cite{navsea_t9070} inside the optimization
loop, so that shock response is driven rather than checked after the fact. It
replaces the iterative design-then-check pattern of current
practice~\cite{jengDeLimaPagaldipti,koehler2018ddam}.

\section*{Code and data availability}
The software used to produce the results reported in this work is proprietary.
Neither the source code nor the executables can be shared. The Navy design
spectrum enters the software through a Python interface rather than being
coded in it (Section~\ref{sec:algorithm}): a user supplies the design
acceleration for a mode and its derivatives, so the same formulation applies
to different Navy spectra, ship types, mounting locations and directions
without changes to the solver. The numerical experiments and methodology are
described in sufficient detail in this paper to convey the underlying
approach. The author welcomes opportunities to collaborate
and partner with interested parties, who are encouraged to contact the
corresponding author.

\section*{Appendix A. Notation}
\label{sec:notation}

For reference, Tables~\ref{table:notation} and~\ref{table:notation2}
collect the principal symbols used in the paper. The first covers the design,
the modal problem and the shock spectrum; the second the criteria, the
sensitivity and the study parameters.

\begin{notationtable}
\begin{center}
\small
\renewcommand{\arraystretch}{1.25}
\begin{tabular}{ll}
\toprule
Symbol & Meaning \\
\midrule

\multicolumn{2}{l}{\emph{Design and discretization}} \\
$\bz,\ z_e$ & Topology (pseudo-density) design vector and element density \\
$N_e$ & Number of elements \\
$z_{\min},\ p$ & Density lower bound and SIMP penalization exponent \\
$\mathcal{I},\ \epsilon_{\min}$ & Modified SIMP interpolation and its minimum ersatz fraction \\
$\mathbf{F},\ \tilde{\bz}$ & Linear density filter and filtered densities \\
$E,\ \nu,\ \rho_0,\ \mathbf{C}_0$ & Solid Young's modulus, Poisson's ratio, density, stiffness \\
$v(\bz),\ v_0,\ v_e$ & Design volume, its solid reference value, and element measure \\
$v_f$ & Volume-fraction target \\
$\Omega_c$ & Region the stress aggregate runs over (the design domain here) \\
\midrule
\multicolumn{2}{l}{\emph{Modal problem}} \\
$\K(\bz),\ \M(\bz)$ & Stiffness and mass matrices \\
$\M_{\mathrm{s}},\ \M_{\mathrm{ns}}$ & Structural and non-structural (equipment) mass \\
$\mathbf{E}_p,\ m_p,\ \mathcal{S}$ & Nodal translational selector, nodal mass, interface node set \\
$m_{\mathrm{eq}}$ & Equipment mass on one interface \\
$p_0$ & Master node of the rigid tie \\
$\K^{\mathrm{BC}},\ \M^{\mathrm{BC}}$ & Dirichlet-eliminated operators \\
$\mathbf{T},\ \K_r,\ \M_r,\ \mathbf{q}_i$ & Rigid tie, reduced operators, reduced eigenvector \\
$\lambda_i,\ \omega_i$ & Eigenvalue, natural frequency; $\lambda_i = \omega_i^{2}$ \\
$N_{\!m}$ & Number of modes solved \\
$\bphi_i$ & Mass-normalized eigenvector of mode $i$ \\
$m_i^{*}$ & Generalized mass, $\bphi_i\T \M \bphi_i$ \\
$\bar m_i$ & Effective modal mass of mode $i$ \\
$\mathbf{D}$ & Shock-direction influence vector \\
$d \in \{\mathrm{V},\mathrm{A},\mathrm{FA}\}$ & Shock direction: vertical, athwartship, fore-and-aft \\
$\Gamma_i$ & Modal participation factor of mode $i$ \\
$\Phi_{\mathrm{rb}},\ \mathbf{P}$ & Near-null-space basis and its $\M$-orthogonal projector \\
$\Phi_c,\ \mathbf{Q}$ & Cluster eigenvectors and an orthogonal change of basis \\
\midrule
\multicolumn{2}{l}{\emph{Navy shock spectrum}} \\
$\Dai(\bar m_i, \omega_i)$ & Navy-supplied design acceleration for mode $i$ \\
$G$ & Tabulated map from $(\bar m_i, \omega_i)$ to $\Dai$ \\
$\SDi = \Dai / \omega_i^{2}$ & Spectral displacement for mode $i$ \\
$A_0,\ V_0$ & NRL-1396 reference acceleration and velocity design values \\
$A_{a,i},\ V_{a,i}$ & Their directional counterparts, after the elastic coefficients \\
$W_{a,i}$ & Modal effective weight $\bar m_i\, g$, in kips \\
$\omega^{*} = A_a/V_a$ & Crossover between the velocity- and acceleration-limited branches \\
\bottomrule
\end{tabular}
\end{center}
\caption{Principal symbols used in the paper, part 1 of 2: the design variables, the modal problem and the Navy shock spectrum. Volumes are lower case, $v(\bz)$, $v_0$ and $v_e$, so that the upper-case $V_0$ of~\eqref{eq:nrl1396_A0V0} is unambiguously the NRL-1396 velocity design value.}
\label{table:notation}
\end{notationtable}

\begin{notationtable}
\begin{center}
\small
\renewcommand{\arraystretch}{1.15}
\begin{tabular}{ll}
\toprule
Symbol & Meaning \\
\midrule

\multicolumn{2}{l}{\emph{Response and criteria}} \\
$R_i,\ \Rmax,\ \Rtot$ & Modal, peak, and NRL Sum total responses \\
$X_i,\ \boldsymbol{\sigma}_i$ & Peak modal displacement and stress \\
$B,\ C$ & Shape-function derivatives and (interpolated) material stiffness \\
$W$ & Response weighting operator; the area average~\eqref{eq:surface_weights} over an interface $\Sigma$ \\
$N_j,\ a_j,\ A$ & Interface shape function, nodal weight, area \\
$\mathrm{B}_i$ & Weighted square $\bphi_i\T W \bphi_i$ of mode $i$ \\
$f(\Lambda, \Phi)$ & SRSS DDAM criterion \\
$g_i$ & Per-mode squared response entering the NRL Sum \\
$f_{\mathrm{NRL}}$ & Exact NRL Sum criterion,~\eqref{eq:nrl_g} or~\eqref{eq:nrl_field} \\
$m,\ w_m,\ w_{\mathrm{bg}}$ & Dominant mode index and its dominant/background weights \\
$f_{\mathrm{NRL}}^{\mathrm{comb}}$ & Combined-direction maximum,~\eqref{eq:demo3_combined} \\
$s_i,\ s_{i,e}$ & von~Mises invariant of the modal stress of mode $i$,~\eqref{eq:vm_mode} \\
$R_{\mathrm{NRL}}(x),\ m(x)$ & Pointwise NRL Sum and its dominant-mode field,~\eqref{eq:nrl_field} \\
$f_{\sigma},\ \mu,\ \kappa$ & Measure-weighted aggregate, its peak, and the sharpness \\
$\chi_e,\ c_{i,e}$ & Aggregate and pointwise-combination weights \\
$\tau$ & Background truncation threshold of the stress criterion \\
$A_i,\ \varsigma_i$ & Sensitivity of $f_{\sigma}$ to the spectral factor $\lvert \SDi\Gamma_i \rvert$, and its sign,~\eqref{eq:stress_A} \\
$\epsilon$ & Radical regularization in~\eqref{eq:nrl_g} \\
$f_{\mathrm{cl}},\ n_c$ & Cluster mean of the $n_c$ lowest eigenvalues, and the cluster size \\
\midrule
\multicolumn{2}{l}{\emph{Sensitivity}} \\
$\alpha_i,\ \beta_i,\ \gamma_i$ & SRSS chain-rule coefficients \\
$\hat\beta_i$ & Participation coefficient under a mass-dependent spectrum,~\eqref{eq:beta_hat} \\
$\bmu_i,\ \xi_i$ & Nelson adjoint vector for mode $i$ and its bordered multiplier \\
$\widetilde{\bphi}_i',\ c_i$ & Nelson particular solution and null-space coefficient \\
$v_i,\ \eta_i$ & Eigenvector-sensitivity functional and its null-space scalar \\
$T^{e}_{\mathrm{I}},\,T^{e}_{\mathrm{II}},\,T^{e}_{\mathrm{III}}$ & Per-element sensitivity contributions \\
$k_e',\ m_e'$ & Element-level stiffness and mass derivatives (from AD) \\
$h,\ \mathbf{e}_e,\ \varepsilon$ & Finite-difference step, unit vector of element $e$, machine precision \\
$\varepsilon_{\mathrm{tol}},\ N_{\max}$ & Stopping tolerance and iteration cap \\
\midrule
\multicolumn{2}{l}{\emph{Limits and study parameters}} \\
$R_{\mathrm{allow}}$ & Allowable NRL Sum response \\
$\sigma_{\mathrm{allow}},\ \sigma_y$ & Allowable and minimum specified yield stress \\
$\omega_{\min}$ & Fundamental-frequency floor of~\eqref{eq:frequency_floor}, in \si{rad/s} \\
$\underline{f}_{\mathrm{cl}},\ \delta$ & Cluster-mean bound and the separation it implies,~\eqref{eq:gap_from_mean} \\
$\mathbf{f},\ \bu$ & Static equivalent load and its displacement,~\eqref{eq:compliance_problem} \\
\bottomrule
\end{tabular}
\end{center}
\caption{Principal symbols used in the paper, part 2 of 2: the response and its criteria, the sensitivity, and the limits and study parameters.}
\label{table:notation2}
\end{notationtable}

\section*{Appendix B. Numerical treatment of the Nelson solve}
\label{sec:appendix_nelson}

As written, the adjoint system~\eqref{eq:nelson_adjoint} is solved on the
complement of the null space of the singular operator $(\K - \lambda_i \M)$. For a fully constrained structure with a
well-separated spectrum, the projected right-hand side is consistent by
construction and the system is non-singular. In density-based topology
optimization, two issues arise repeatedly at intermediate iterates and
require explicit treatment.

\paragraph{Near-rigid-body and quasi-disconnected modes} Under the SIMP
interpolation~\eqref{eq:simp} the stiffness of low-density material falls to
$\epsilon_{\min}$ times its solid value. Intermediate optimization iterates routinely contain large low-density
regions that behave as quasi-disconnected substructures. These produce
additional near-zero eigenvalues, which contaminate the Krylov
iteration~\cite{saad2003iterative} applied to~\eqref{eq:nelson_adjoint}. For
any mode above the fundamental, $\K - \lambda_i \M$ is also indefinite, so the
iteration must be one for symmetric indefinite systems rather than conjugate
gradients.
Let $\Phi_{\mathrm{rb}}$ denote a basis for the numerical near-null space of
$\K$ at the current design, and define the projector
\begin{equation}
  \label{eq:rb_projector}
  \mathbf{P} \;=\; \mathbf{I} \;-\; \M\, \Phi_{\mathrm{rb}}\,
    \bigl( \Phi_{\mathrm{rb}}\T \M\, \Phi_{\mathrm{rb}} \bigr)^{-1}
    \Phi_{\mathrm{rb}}\T,
\end{equation}
which is idempotent and satisfies $\mathbf{P}\, \M \Phi_{\mathrm{rb}} = \mathbf{0}$
and $\Phi_{\mathrm{rb}}\T \mathbf{P} = \mathbf{0}$. It is oblique; its
transpose is the $\M$-orthogonal projector onto the $\M$-orthogonal
complement of $\Phi_{\mathrm{rb}}$.
The adjoint solve is then carried out in three complementary stages:
\begin{enumerate}
  \item \emph{Right-hand-side projection.} Replace the right-hand side
    $\mathbf{b}_i := v_i - (v_i\T \bphi_i)\, \M \bphi_i$ of~\eqref{eq:nelson_adjoint} by $\mathbf{P}\, \mathbf{b}_i$, removing the
    near-null-space content of the seed.
  \item \emph{Krylov-level deflation.} Apply $\mathbf{P}\,(\K - \lambda_i \M)$
    rather than $(\K - \lambda_i \M)$ throughout the iteration. This is
    equivalent to solving on the $\M$-orthogonal complement of
    $\Phi_{\mathrm{rb}}$, and it prevents accumulation of near-null-space
    components in the Krylov basis.
  \item \emph{Mass-orthogonality correction.} After convergence, enforce
    $\bmu_i\T \M \bphi_i = 0$ by
    $\bmu_i \leftarrow \bmu_i - (\bmu_i\T \M \bphi_i)\, \bphi_i$, removing the
    arbitrary $c_i \bphi_i$ component admissible by~\eqref{eq:nelson_general}.
\end{enumerate}
Stage~1 removes the projection of the seed onto $\Phi_{\mathrm{rb}}$, and
Stage~2 prevents the Krylov subspace from reintroducing it during iteration.
Together they drop the contribution of the near-null modes to $\bmu_i$, which
is not zero in general: they trade it for a well-conditioned solve, and the
finite-difference check of Section~\ref{subsec:fd_check} is what shows the
trade to be acceptable on the problems reported. Stage~3 is required rather
than a clean-up: the simplified form~\eqref{eq:nelson_pi_simplified} holds
only for the mass-orthogonal adjoint. Dropping any one stage causes the
finite-difference gradient check of Section~\ref{subsec:fd_check} to fail
intermittently at intermediate iterates, even when the final-design solve
passes.

\paragraph{Dirichlet boundary-condition consistency} The Nelson solve must
employ exactly the same Dirichlet enforcement as the eigensolver, applied
symmetrically to both $\K$ and $\M$. Constraints are imposed symmetrically:
the constrained rows and columns are cleared, a large value is placed on the
diagonal of $\K^{\mathrm{BC}}$ and a unit on the diagonal of $\M^{\mathrm{BC}}$. The unit on
$\M^{\mathrm{BC}}$ keeps $\M^{\mathrm{BC}}$ positive definite, which the
eigensolver requires as an inner-product weight. The large diagonal pushes the resulting spurious eigenpairs, one per
constrained degree of freedom, several decades above any structural mode. It is filtered out by requesting the bottom $N_{\!m}$ of the
spectrum. The same constrained operators $\K^{\mathrm{BC}}$, $\M^{\mathrm{BC}}$
are then reused verbatim in~\eqref{eq:nelson_adjoint}. Any inconsistency causes $\bmu_i$ to acquire nonzero components at constrained
degrees of freedom. Applying the constraint to one matrix only, or using an
asymmetric row-only elimination, are the two ways this happens. The
element-level inner products
$\bmu_i^{e\mathsf{T}} (k_e' - \lambda_i\, m_e')\, \bphi_i^{e}$ that assemble
$T^{e}_{\mathrm{III}}$ in~\eqref{eq:T_III} are then corrupted.

\paragraph{Participation factor and the physical mass operator} The
Dirichlet-eliminated operators are the correct ones for the eigensolve and
the Nelson solve. The participation factor~\eqref{eq:participation} is the
single quantity that must instead be formed with the physical
(pre-elimination) mass operator $\M$. Mass normalization is insensitive to
the choice, since the eigenvector vanishes at constrained degrees of freedom
and hence $\bphi_i\T \M^{\mathrm{BC}} \bphi_i = \bphi_i\T \M \bphi_i$. The numerator $\bphi_i\T \M \mathbf{D}$, however, does depend on it. Symmetric
elimination zeros the (free, constrained) blocks of $\M^{\mathrm{BC}}$.
Forming $\M^{\mathrm{BC}} \mathbf{D}$ therefore drops the inertial coupling
between the free degrees of freedom and the constrained directions that
$\mathbf{D}$ excites. That systematically under-predicts $\Gamma_i$. Retaining the unconstrained $\M$ in~\eqref{eq:participation},
while using $\M^{\mathrm{BC}}$ for the solves, keeps $\Gamma_i$ faithful to
its definition. It then propagates correctly into the coefficients $\beta_i$
and $\gamma_i$ of~\eqref{eq:nelson_v}. The term $\hat\beta_i \M \mathbf{D}$ of
$v_i$ has entries at constrained degrees of freedom, and they must be cleared
before the solve with $\M^{\mathrm{BC}}$.

\paragraph{Free-free structures and indefinite shift-invert} Some DDAM components
are evaluated in a fully unsupported configuration, e.g., equipment whose
mounting interface has not yet been added to the design domain. The
structure then admits an exact six-dimensional rigid-body null space
$\Phi_{\mathrm{rb}}$ (three translations and three rotations linearized about
the centroid, for three-dimensional elasticity). We construct $\Phi_{\mathrm{rb}}$ analytically from the mesh rather than
extracting it from the eigensolve. The eigensolver does not guarantee a
canonical orthogonalization within the deflated subspace, and would return an
arbitrary basis. The analytical basis feeds the projector $\mathbf{P}$ of~\eqref{eq:rb_projector} directly.
In a fully free configuration the shock direction vector $\mathbf{D}$ is
itself a rigid-body mode, so every flexible mode has
$\Gamma_i = \bphi_i\T \M \mathbf{D} = 0$ and the DDAM criterion is degenerate;
the machinery matters for the near-null modes of intermediate designs. None of
the demonstrations is free-free.

For the eigensolve itself, a shift-invert spectral transformation with shift
$\sigma > 0$ is required because $\K$ is singular. The resulting shifted
operator $\K - \sigma \M$ is symmetric \emph{indefinite}: on any
$\mathbf{v} \in \mathrm{span}(\Phi_{\mathrm{rb}})$,
\begin{equation}
  \label{eq:indefinite_si}
  \mathbf{v}\T (\K - \sigma \M)\, \mathbf{v}
  \;=\; -\sigma\, \mathbf{v}\T \M\, \mathbf{v} \;<\; 0,
\end{equation}
while on the flexible subspace the form is positive only when $\sigma$ lies
below the first flexible eigenvalue. Sparse-Cholesky factorizations of $\K - \sigma \M$ therefore fail;
a sparse-LU or symmetric-indefinite factorization must be used to apply the
shift-invert step. The same three-stage projection scheme~\eqref{eq:rb_projector} then deflates
$\Phi_{\mathrm{rb}}$ from the Krylov basis of the eigensolve and from the
Nelson adjoint. Both then return only structural information.

\section*{References}

\bibliographystyle{elsarticle-num}
\bibliography{ddam}

\end{document}